\documentclass[reprint, amsmath,amssymb, preprintnumbers,aps,nofootinbib,superscriptaddress,usenames,dvipsnames]{revtex4-1}
\usepackage{graphicx}				% Use pdf, png, jpg, or eps§ with pdflatex; use eps in DVI mode
\usepackage{amssymb}
\usepackage{amsmath, amsthm, amssymb}
\usepackage{dcolumn}
\usepackage{bm}
\usepackage{slashed}
\usepackage{tikz}
\usepackage{rotating}
\usepackage{hyperref}
\usepackage{array}
\usepackage{color}
\usepackage{multirow}
\usepackage{xcolor} 
\usepackage[T1]{fontenc}
\usepackage{tocloft}
\usepackage[normalem]{ulem}
\usepackage{subcaption}
\usepackage{enumitem}
\usepackage{extarrows}
\usepackage{mathtools}
\usepackage{lipsum}
\usepackage{bbold}
\usepackage{mathrsfs}
\usepackage{lmodern}
\usepackage{blindtext}
\usepackage{lipsum}

\newcommand{\nn}{\nonumber}

\newcommand{\be}{\begin{equation}}
\newcommand{\ee}{\end{equation}}
\newcommand{\bea}{\begin{eqnarray}}
\newcommand{\eea}{\end{eqnarray}}

\newcommand{\as}{\alpha_s}

\newcommand{\Gcusp}{\Gamma_{\rm cusp}}

\newcommand{\cO}{\mathcal{O}}

\newcommand{\cG}{\mathcal{G}}

\newcommand{\wt}[1]{\widetilde{#1}}

\renewcommand{\sec}[1]{Sec.~\ref{sec:#1}}

\newcommand{\fig}[1]{Fig.~\ref{fig:#1}}

\newcommand{\comment}[1]{}

\begin{document}

\preprint{LA-UR-23-26138}
\preprint{SI-HEP-2023-24}
\preprint{P3H-23-079}

%%%%%%%%%%%%%%%%%%%%%%%%%%%%%%%%%%%%%%%%%%%%%%%%%%%%%%%%

\title{Effects of Renormalon Scheme and Perturbative Scale Choices \\[0.1em] 
on  Determinations of the Strong Coupling from $e^+e^-$ Event Shapes}

\author{Guido Bell}
\email{bell@physik.uni-siegen.de}
\affiliation{Theoretische Physik 1, Center for Particle Physics Siegen, Universit\"at Siegen, \\Walter-Flex-Strasse 3, 57068 Siegen, Germany}
\author{Christopher Lee}
\email{clee@lanl.gov}
\affiliation{Theoretical Division, Los Alamos National Laboratory, P.O. Box 1663, MS B283, \\Los Alamos, NM  87545, USA}
\author{Yiannis Makris}
\affiliation{Theoretical Division, Los Alamos National Laboratory, P.O. Box 1663, MS B283, \\Los Alamos, NM  87545, USA}
\affiliation{INFN Sezione di Pavia, via Bassi 6, I-27100 Pavia, Italy}
\author{Jim Talbert} 
\email{talbert@lanl.gov}
\affiliation{DAMTP, University of Cambridge, Wilberforce Road, Cambridge, CB3 0WA, United Kingdom}
\affiliation{Theoretical Division, Los Alamos National Laboratory, P.O. Box 1663, MS B283, \\Los Alamos, NM  87545, USA}
\author{Bin Yan} 
\email{yanbin@ihep.ac.cn}
\affiliation{Theoretical Division, Los Alamos National Laboratory, P.O. Box 1663, MS B283, \\Los Alamos, NM  87545, USA}
\affiliation{Institute of High Energy Physics, Chinese Academy of Sciences, Beijing 100049, China}

%%%%%%%%%%%%%%%%%%%%%%%%%%%%%%%%%%%%%%%%%%%%%%%%%%%%%%%%

\begin{abstract}
We study the role of renormalon cancellation schemes and perturbative scale choices in extractions of the strong coupling constant $\alpha_s(m_Z)$ and the leading non-perturbative shift parameter $\Omega_1$ from resummed predictions of the $e^+e^-$ event shape thrust.  We calculate the thrust distribution to N$^{3}$LL$^\prime$ resummed accuracy in Soft-Collinear Effective Theory (SCET) matched to the fixed-order $\mathcal{O}(\alpha_s^2)$ prediction, and perform a new high-statistics computation of the $\mathcal{O}(\alpha_s^3)$ matching in \texttt{EERAD3}, although we do not include the latter in our final $\alpha_s$ fits due to some observed systematics that require further investigation. We are primarily interested in testing the phenomenological impact sourced from varying amongst three renormalon cancellation schemes and two sets of perturbative scale profile choices. We then perform a global fit to available data spanning center-of-mass energies between 35-207 GeV in each scenario. Relevant subsets of our results are consistent with prior SCET-based extractions of $\alpha_s(m_Z)$, but we are also led to a number of novel observations. Notably, we find that the combined effect of altering the renormalon cancellation scheme and profile parameters can lead to few-percent-level impacts on the extracted values in the $\alpha_s-\Omega_1$ plane, indicating a potentially important systematic theory uncertainty that should be accounted for. We also observe that fits performed over windows dominated by dijet events are typically of a higher quality than those that extend into the far tails of the distributions, possibly motivating future fits focused more heavily in this region. Finally, we discuss how different estimates of the three-loop soft matching coefficient $c_{\tilde{S}}^3$ can also lead to measurable changes in the fitted $\lbrace \alpha_s, \Omega_1 \rbrace$ values.
\end{abstract}
\maketitle

%%%%%%%%%%%%%%%%%%%%%%%%%%%%%%%%%%%%%%%%%%%%%%%%%%%%%%%%
%%%%%%%%%%%%%%%%%%%%%%%%%%%%%%%%%%%%%%%%%%%%%%%%%%%%%%%
\section{Introduction}
\label{sec:INTRO}

Electron-positron event shapes are amongst the oldest and most established observables for testing the predictions of perturbative QCD \cite{Dasgupta:2003iq}.  They typically measure geometric configurations of hadronic final-state momentum distributions in detectors which, thanks to the absence of initial-state color-charged particles, are free of many uncertainties associated to hadron-hadron collisions.  They are therefore generally assumed to be theoretically clean observables, permitting high-order perturbative calculations as well as precision extractions of the fundamental parameter of massless QCD---the strong coupling constant $\alpha_s$.

It is to this end that multiple experimental and theoretical collaborations have pursued $e^+e^-$ event shapes, utilizing available data from the Large Electron-Positron collider (LEP) and other experiments.  Some of these results 
\cite{Bethke:2009ehn,Dissertori:2009ik,Dissertori:2009qa,Abbate:2010xh,OPAL:2011aa,Schieck:2012mp,Gehrmann:2012sc,Hoang:2015hka,Kardos:2018kqj,Verbytskyi:2019zhh}
contribute along with analogous studies utilizing lattice simulations, $\tau$-decays, PDF fits, heavy quarkonia decays, etc. to the 2022 Particle Data Group (PDG) world average, which is found to be $\alpha_s(m_Z) = 0.1179 \pm 0.0009$ \cite{Workman:2022ynf}.  Thanks to the desirable properties of $e^+e^-$ event shapes and the associated precision of theoretical predictions now available---which exist up to the N$^3$LL$'$ resummed perturbative accuracy using Soft-Collinear Effective Theory (SCET) \cite{Bauer:2000ew,Bauer:2000yr,Bauer:2001yt,Beneke:2002ph} and $\mathcal{O}(\as^3)$ accuracy in fixed-order QCD \cite{Gehrmann-DeRidder:2007vsv,Weinzierl:2009ms}---determinations based on observables such as thrust \cite{Abbate:2010xh} and $C$-parameter \cite{Hoang:2015hka} are among the most precise quoted in the PDG.

One notices, however, that these extractions tend to yield significantly lower values of $\alpha_s$ as compared to the PDG world average, which motivates further investigation. It is generally observed in the literature that lower values of $\alpha_s$ emerge from event-shape predictions that account both for the resummation of logarithms in perturbation theory and a semi-analytic description of non-perturbative power corrections from hadronization. Recently, this discrepancy has motivated studies of models of power corrections in the three-jet region of event-shape distributions \cite{Luisoni:2020efy,Caola:2021kzt,Caola:2022vea} and associated impacts on the determination of $\as$ \cite{Luisoni:2020efy,Nason:2023asn}, indicating  additional systematic uncertainties from such effects. In the current work, we follow a somewhat orthogonal approach that addresses the question whether there exist theoretical uncertainties in the \emph{dijet} factorized prediction for the thrust distribution that have a similarly non-negligible impact on the $\as$ determination. We stress that the purpose of this work is to identify such systematic effects that remain to be better understood or controlled in future fits, and that we do not aim at a competitive $\alpha_s$ extraction here.

While multiple explanations for this tension could exist, a definitive answer will at the very least require a better understanding of the impact of {\bf{(A)}} non-perturbative (NP) physics effects and {\bf{(B)}} perturbative scale uncertainties on the error estimation of the extracted results.  For example, regarding ${\bf{(A)}}$, the SCET thrust and $C$-parameter fit results mentioned above were performed in a two-dimensional plane in $\alpha_s$ and $\Omega_{1}$, where the latter is a universal parameter encoding the leading NP shift in the tail regions of the differential distributions \cite{Webber:1994cp,Dokshitzer:1995zt,Korchemsky:1998ev,Korchemsky:1999kt,Lee:2006nr}:
\be
\label{eq:leadingshift}
\frac{d\sigma}{de} (e) \underset{\text{NP}}{\longrightarrow} 
\frac{d\sigma}{de} \Big(e-c_{e} \frac{\Omega_1}{Q}\Big)\,,
\ee
where $e$ denotes a generic event shape whose dominant power corrections come from the soft sector, and where $c_e$ is a calculable, observable-dependent scaling coefficient given by $c_\tau = 2$ for thrust and $c_C = 3 \pi$ for $C$-parameter.  While \eqref{eq:leadingshift} is only valid in a region of $e$ large enough to justify an operator product expansion of the soft shape function that describes the NP physics, its implication for fitting analyses is evident: shifts in theory distributions due to differing values of $\alpha_s$ can partly be compensated by shifts from differing values of $\Omega_1$.  As a result, extractions are often given in terms of ellipses in the combined $\alpha_s-\Omega_1$ plane, whose sizes correspond to a given statistical confidence level and whose slopes encode the correlation between the two fit parameters $\alpha_s$ and $\Omega_1$.  Furthermore, often embedded in the definition of $\Omega_1$ is an implicit scheme for simultaneously cancelling the leading soft infrared renormalon appearing in the perturbative dijet soft function and certain hadronization model parameter(s) \cite{Hoang:2007vb}.  This renormalon cancellation scheme amounts to a choice, and as recently argued in \cite{Bachu:2020nqn}, many reasonable choices exist.  We stress that when we make the same choices as in prior SCET-based thrust analyses \cite{Abbate:2010xh,Hoang:2015hka}, we obtain results entirely compatible with these works. But here we will also show that multiple well-defined scheme choices may well lead to a significant spread of the fit values, implying a potential systematic theory uncertainty that should be accounted for, while also motivating further strategies to reduce or avoid such systematic effects.

As part of our investigations we will also find that $\alpha_s$ extractions that are constrained to a more restricted domain of the observable---in particular the dominantly dijet region of differential distributions---typically result in better-quality fits than those performed up to or even beyond the multi-jet threshold.  This represents a complementary perspective to the one that was recently presented in \cite{Nason:2023asn}, which advocated $\alpha_s$ fits in the far-tail regions, where three-jet and multi-jet events dominate, and where recent advances in understanding three-jet NP power corrections have focused. While the latter developments are important to improve the theoretical understanding of event-shape distributions, we note that these anyalses are currently model-dependent, and therefore not on the same footing as the corresponding dijet studies. In particular, the universality of the three-jet NP power corrections is currently not well understood.

In this work we study the thrust distribution \mbox{$T=1-\tau$} \cite{Brandt:1964sa,Farhi:1977sg}, a canonical $e^+e^-$ event shape defined as
\begin{equation}
\label{eq:tauadef}
\tau = \frac{1}{Q} \, \underset{i}{\sum} \;
|{\bf p}_{\perp}^{i}|\; e^{-|\eta_i|}\,,
\end{equation}
with $Q$ the center-of-mass (c.o.m.) energy of the collision, and ${\bf p}_{\perp}^{i}$ and $\eta_i$ respectively denoting the transverse momentum and rapidity of the $i$th final-state particle measured with respect to the thrust axis. Thrust is amongst the best studied $e^+e^-$ event shapes, a fact in part due to the ability to calculate its spectra at N$^3$LL$^{\prime}  + \mathcal{O}(\alpha_s^3)$ resummed and matched accuracy \cite{Becher:2008cf,Abbate:2010xh}. As already mentioned, thrust has been utilized for prior effective-field-theory-based extractions of the strong coupling, and this provides a controlled environment for us to study the impacts of certain assumptions embedded in the theoretical framework, and to compare with prior analyses.  Indeed, in addition to revealing the novel physics points mentioned above, our study provides the first independent cross-check of the results in \cite{Abbate:2010xh,Hoang:2015hka}, despite a number of minor systematic differences in our analysis that will be described in detail in Sec.~\ref{sec:COMPARE} below.

The remainder of the paper develops as follows: in Sec.~\ref{sec:THEORY} we will review the dijet SCET factorization theorem employed to predict the thrust distribution, the various perturbative ingredients required therein, and the matching to the fixed-order QCD prediction up to $\mathcal{O}(\alpha_s^2)$ accuracy. We also describe a new high-statistics calculation of the three-loop $\mathcal{O}(\alpha_s^3)$ remainder function using the public \texttt{EERAD3} code \cite{Gehrmann-DeRidder:2014hxk}, and highlight some systematic concerns about it that lead us to not include the matching to this order in our final extraction code.  In Sec.~\ref{sec:NONPERT} we  review the further factorization of the soft function into a non-perturbative shape function and a fully perturbative component, as well as the formalism for achieving the leading infrared renormalon cancellation between the two.  We will then introduce the three different renormalon cancellation schemes we intend to study, and in Sec.~\ref{sec:UNCERTAINTIES} we review two alternative sets of profile functions as well as our method of varying them to estimate the overall perturbative uncertainty.  Finally, in Sec.~\ref{sec:EXTRACT} we present our method of extracting $\lbrace \alpha_s, \Omega_1 \rbrace$, show our results for these quantities in different schemes, discuss their implications, and also touch on the uncertainties associated with the three-loop soft constant $c_{\tilde S}^3$, which is the only N$^3$LL$'$ ingredient that is currently not known exactly. We conclude in Sec.~\ref{sec:CONCLUDE}. App.~\ref{sec:EXPANSIONS} and \ref{sec:INGREDIENTS} respectively collect details on the fixed-order expansion of the resummed cross section and the order-by-order expansion of the renormalon cancellation formulae, while App.~\ref{sec:MORER0} provides more details of an analysis of one of the renormalon schemes introduced in Sec.~\ref{sec:NONPERT} that does not play a central role in the comparison of schemes in the main body of this paper nor in its conclusions.

%%%%%%%%%%%%%%%%%%%%%%%%%%%%%%%%%%%%%%%%%%%%%%%%%%%%%%%%%%
\section{Perturbative Treatment}
\label{sec:THEORY}

The bulk of the theory we implement is thoroughly presented in prior literature, and so for brevity we only outline the core features of our calculation in what follows, leaving many background details to those references (see e.g.~\cite{Abbate:2010xh, Almeida:2014uva, Bell:2018gce}).  

We predict the singular part of the integrated (or cumulative) thrust distribution
\begin{equation}
\label{eq:integrated}
\sigma_c(\tau) = \int_0^\tau d\tau' \frac{d\sigma}{d\tau'}
\end{equation}
with methods from SCET, for which factorization theorems for dijet $e^+e^-$ event shapes are well established, and match to fixed-order QCD to account for non-singular contributions. Hence the overall perturbative cross section consists of two parts, 
\begin{equation}
\label{eq:sigmasum}
    \sigma_c^{PT}(\tau) = \sigma_{c,\text{sing}}(\tau) + \sigma_{c,\text{ns}}(\tau)\,, 
\end{equation}
with the non-singular matching contribution $\sigma_{c,\text{ns}}(\tau)$ implicitly defined in \eqref{eq:difference} below. We will now address both of these contributions in turn.

%%%%%%%%%%%%%%%%%
\subsection{Singular Contribution}

The singular and resummed contribution $\sigma_{c,\text{sing}}(\tau)$, normalized by the Born cross section $\sigma_0$, reads\footnote{We use the form derived in~\cite{Bell:2018gce}, which depends at a given order of perturbation theory on the factorization scales $\mu_{H,J,S}$, but is otherwise explicitly independent of the renormalization scale $\mu$ at \emph{every} order of perturbation theory.}
\begin{align}
&\frac{\sigma_{c,\text{sing}}(\tau)}{\sigma_0} = e^{\wt K(\mu_H,\mu_J,\mu_S;Q) + K_\gamma(\mu_H,\mu_J,\mu_S)}  \biggl(\frac{1}{\tau}\biggr)^{\Omega(\mu_J,\mu_S)} \nn \\
&\times H(Q^2,\mu_H)  
\wt J\Bigl(\partial_\Omega + \ln\frac{\mu_J^{2}}{Q^{2}\tau},\mu_J\Bigr)^2 \;\wt S\Bigl(\partial_\Omega + \ln\frac{\mu_S}{Q\tau},\mu_S\Bigr) \nn  \\
&\times \frac{e^{\gamma_E \Omega}}{\Gamma(1-\Omega)}\,,
\label{eq:cumulant}
\end{align}
where the evolution kernels $\tilde{K}$, $K_\gamma$, and $\Omega$ are given by
\begin{align}
\label{eq:Kkernels}
\wt K(\mu_H,\mu_J,\mu_S;Q) &\equiv -\kappa_H \wt K_\Gamma(\mu,\mu_H;Q) \\
&\!\!\!\! - 2j_J\kappa_J \wt K_\Gamma(\mu,\mu_J;Q) - \kappa_S \wt K_\Gamma(\mu,\mu_S;Q)\,, \nn \\
\nonumber
K_\gamma(\mu_H,\mu_J,\mu_S) &\equiv K_{\gamma_H}(\mu,\mu_H) + 2K_{\gamma_J}(\mu,\mu_J) \nn \\
&\quad + K_{\gamma_S}(\mu,\mu_S)\,, \nn \\
\Omega(\mu_J,\mu_S) &\equiv -2\kappa_J \,\eta_\Gamma(\mu,\mu_J) - \kappa_S \,\eta_\Gamma(\mu,\mu_S)\,. \nn
\end{align}
The hard function $H$ arises from the matching of the effective theory to QCD and it describes virtual corrections to $e^+e^- \rightarrow q \bar{q}$ scattering, while collinear radiation along the jet directions is encoded in the jet functions $\wt{J}$ (with the tilde denoting an evaluation in Laplace space). Finally, background low-energetic radiation that communicates between the two jets is described by the (Laplace-space) soft function $\wt{S}$.  All of these functions  are evaluated in \eqref{eq:cumulant} at an associated `natural' scale $\mu_{H,J,S}$ at which their logarithmic corrections are minimized and their perturbative series well-behaved. In SCET the resummation is achieved via renormalization group (RG) evolution, which generates the kernels $\wt{K}$, $K_\gamma$, $\Omega$ in \eqref{eq:Kkernels} defined in terms of cusp ($\Gcusp$) and non-cusp ($\gamma_F$) anomalous dimensions,
\begin{align}
\nonumber
\wt K_\Gamma(\mu,\mu_F;Q) &\equiv \int_{\mu_F}^\mu \frac{d\mu'}{\mu'} \Gcusp[\as(\mu')] \ln\frac{\mu'}{Q}\,, \\
\nonumber
 K_{\gamma_F}(\mu,\mu_F) &\equiv \int_{\mu_F}^\mu \frac{d\mu'}{\mu'} \gamma_F[\as(\mu')], \\
 \label{eq:KkernelsB}
 \eta_\Gamma(\mu,\mu_F) &\equiv \int_{\mu_F}^{\mu}\frac{d\mu'}{\mu'} \Gcusp[\as(\mu')].
\end{align} 
The corresponding values for the parameters $\kappa_F$ and $j_F$ are given for the thrust distribution by
\begin{align}
\nonumber
j_H &= 1, \,\,\,\,\, &&j_J = 2, \,\,\,\,\, &&j_S=1,\\
\kappa_H &=4,\,\,\,\,\,\,&&\kappa_J = -2,\,\,\,\,\,&&\kappa_S= 4.
\end{align}   
Given the perturbative expansions of the cusp and non-cusp anomalous dimensions, one can solve \eqref{eq:KkernelsB} order by order, achieving approximate analytic expressions for the evolution kernels that resum the logarithmic corrections to the thrust distribution to a given accuracy.\footnote{The systematic uncertainties of this approximation have been studied e.g.~in~\cite{Billis:2019evv,Ebert:2021aoo}.}
The order to which each individual component of \eqref{eq:cumulant} must be calculated can be read off e.g.~from Table 6 of~\cite{Bell:2018gce}. 
As our goal consists in resumming the thrust distribution to N$^3$LL$^{\prime}$ accuracy, we need the non-cusp anomalous dimensions $\gamma_F$ and the fixed-order hard, jet, and soft functions $H, \wt J, \wt S$ up to $\mathcal{O}(\alpha_s^3)$ corrections, while $\mathcal{O}(\alpha_s^4)$ ingredients are required for the cusp anomalous dimension $\Gcusp$ and the QCD $\beta$-function.  

Of these, only the three-loop matching correction $c_{\tilde{S}}^3$ to the Laplace-space soft function $\tilde{S}$ is currently unknown, although an extraction of this quantity with the {\tt{EERAD3}} event-generator results from \cite{Monni:2011gb} was performed in \cite{Bruser:2018rad}, which found
\begin{equation}
\label{eq:cS3:EERAD3}
c_{\tilde{S}}^3 \big \vert_{\tt EERAD3} ~= -19988 \pm 5440 \,,
\end{equation}
where we have added the uncertainties quoted in \cite{Bruser:2018rad} linearly.  However, in Sec.~\ref{sec:OTHER} below we will also examine scenarios with a Pad\'e approximated value for $c_{\tilde{S}}^3$ that has been used in earlier studies of the thrust distribution (in particular in \cite{Abbate:2010xh}) with
\begin{equation}
\label{eq:cS3:Pade}
c_{\tilde{S}}^3 \big \vert_{\rm Pad\acute{e}} ~=~ 691 \pm 1000 \,.
\end{equation}
The goal of that exercise consists in understanding how sensitive $\alpha_s$ determinations are to this unknown N$^3$LL$^{\prime}$ coefficient in different perturbative schemes that we will introduce below. In any case, despite the somewhat large systematic difference between these two estimates of $c_{\tilde{S}}^3$, and the remaining large uncertainty in \eqref{eq:cS3:EERAD3}, we will still label results that depend on these estimates as N$^{3}$LL$^\prime$ in what follows, in accord with previous analyses \cite{Abbate:2010xh,Hoang:2015hka}.

%%%%%%%%%%%%%%%%%
\subsection{Non-singular Contribution}
\label{sec:MATCHING}

In the far-tail region of the distribution, where resummation effects are less important, it is necessary to match the singular SCET cross section from \eqref{eq:cumulant} to the full QCD prediction. Obtaining the additional non-singular component in \eqref{eq:sigmasum} is then a matter of determining remainder coefficients $r_c^i$ by subtracting off the singular SCET prediction from the QCD result, at a given order in perturbation theory,
\begin{align}
\nonumber
&\frac{\sigma^{PT}_c(\tau)}{\sigma_0} -\frac{\sigma_\text{c,sing}(\tau)}{\sigma_0}=
r_c(\tau) \\
\label{eq:difference}
&\simeq \theta(\tau)\,\biggl\{
 \bar{r}_c^1(Q,\tau) + \bar{r}_c^2(Q,\tau)+ \bar{r}_c^3(Q,\tau)
\biggr\}\,,
\end{align} 
where $\bar{r}_c^i(Q,\tau) \equiv \big(\alpha_s(Q)/(2\pi) \big)^i \, r_c^i(\tau)$.
The singular cross section is obtained from \eqref{eq:cumulant} by expanding $F=H,\wt{J}, \wt{S}$ at $\mu_H=\mu_J=\mu_S=Q$ to fixed order in $\alpha_s$, multiplying out these expressions, and then inverse Laplace-transforming the result.  This process is straightforward given the well-known expressions for the fixed-order coefficients $F_n$ defined by
\be
\label{eq:Fexpansion}
 F(L_F,\mu_F)  = \sum_{n=0}^\infty \left(\frac{\as(\mu_F)}{4\pi}\right)^n  F_n(L_F)\,,
\ee
which depend on (logarithmically weighted) anomalous dimensions $\gamma_F^n$, $\Gamma_F^n$, as well as unweighted singular constants $c_F^n$, briefly discussed above for $F=\wt S$.  In App.~\ref{sec:EXPANSIONS} we have collected explicit expressions for these functions up to $\mathcal{O}(\alpha_s^3)$, in both momentum and Laplace space. In terms of these quantities, the singular cross section $\sigma_{\text{c,sing}}(\tau)$ is fully determined, and one only needs to obtain the QCD result $\sigma^{PT}_c(\tau)$ to determine the remainder coefficients. At $\mathcal{O}(\alpha_s)$ this can be done analytically (from, e.g., \cite{Ellis:1996mzs}), while we have implemented the relevant $\mathcal{O}(\alpha_s^2)$  matching from \cite{Bell:2018gce} in this work, which used {\tt{EVENT2}} \cite{Catani:1996vz} to obtain $r_c^2$. This then leaves the $\mathcal{O}(\alpha_s^3)$ remainder coefficient $r_c^3$. 

%------------------------------------------------
\begin{figure}[t!]
\centering
\vspace{0em}
\includegraphics[width=\columnwidth]{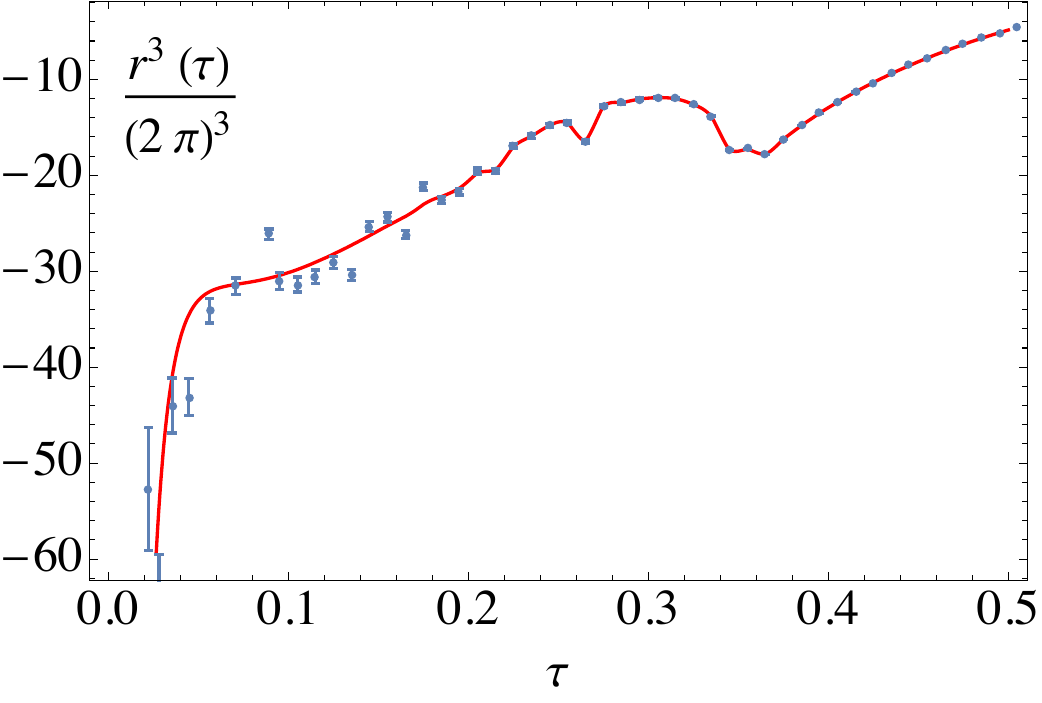}
\vspace{-2em}
\caption{Differential remainder function $r^3(\tau)/(2\pi)^3$ as extracted from {\tt{EERAD3}}. The data points are the {\tt{EERAD3}} data, and the red line is a combined interpolation for large $\tau$ and a fit to a basis of logarithmic functions for small $\tau$ as explained in the text.
}
\vspace{-1em}
\label{fig:r3}
\end{figure}
%------------------------------------------------

While the differential analogue to $r_c^3$, $r^3 = dr_c^3 / d\tau$, is available from the corresponding SCET matching performed in \cite{Abbate:2010xh}, we have opted to perform our own extraction using the {\tt{EERAD3}} generator, which yields a binned approximation to the differential thrust distribution away from $\tau = 0$,
\begin{equation}
    \frac{1}{\sigma_0} \frac{d \sigma_c}{d\tau} \Big\vert_{\tt EERAD3} \simeq \frac{1}{\sigma_0} \frac{d \sigma_c}{d\tau} \Big\vert_{\tau>0} = B(\tau) + r(\tau)\,, 
\end{equation}
where $B(\tau)$ is a singular function that is reproduced by the SCET calculation. Critically, the unknown constant $c_{\tilde{S}}^3$ mentioned above does not appear in this expression, as it only contributes to the coefficient of the $\delta(\tau)$ term.  This can be seen explicitly in \eqref{eq:Ccoeffs} in App.~\ref{sec:EXPANSIONS}, which can also be used to obtain $r^3$ using the differential analogue of \eqref{eq:difference}. The resulting object can then be integrated to obtain the cumulative remainder function $r_c^3$.

Following this procedure, we have computed the remainder function $r^3$ in \texttt{EERAD3} with $1.5\times 10^{10}$ events with the internal infrared cutoff parameter set to $10^{-7}$. The result is shown in \fig{r3}. We note that this number of events is considerably greater than in prior determinations appearing in the literature. With these statistics, we are able to learn about the behavior of the fixed-order prediction of \texttt{EERAD3} in a bit more detail than previously possible, especially in the small $\tau$ region. The remainder $r^3$ is obtained by subtracting all the singular contributions predicted by \eqref{eq:momentumexpansion} down to the single-logarithmic coefficient, i.e.~the logarithms with coefficients $\sigma_{3k}$ with $k\geq 1$. However, we have noticed that the coefficient $\sigma_{31}$ of the single logarithm at three loops is itself not predicted accurately by \texttt{EERAD3}---it is off by nearly a factor of two. Thus, the result in \fig{r3} likely contains an uncancelled singular contribution. We have gone ahead with the exercise of fitting the data in \fig{r3} to a basis of sub-leading (power-suppressed) logarithmic functions as was done in \cite{Abbate:2010xh}, which is represented by the red line for $\tau<0.2$ (the red line for $\tau>0.2$ is simply a direct interpolation of the \texttt{EERAD3} data). In this way we could integrate $r^3$ to obtain a cumulative remainder function $r_c^3$. However, because of the uncancelled singular contribution in the \texttt{EERAD3} data, we do not believe we can obtain a reliable prediction for $r_c^3$ at this time, nor can we use these data to obtain a reliable extraction of the unknown three-loop soft constant $c_{\tilde{S}}^3$ using the procedure described in \cite{Hoang:2008fs}. These observations should not, however, cast doubt on previous extractions of $\as$ using \texttt{EERAD3} calculations of $r^3$, as the `missing logarithmic effect' is less pronounced at larger values of $\tau$ where the $\as$ fits are performed, and the statistical uncertainties for the lower number of events that were used in those works likely encompass the error, anyway.

Nevertheless, as a result of these observations, we choose to match only to the $\mathcal{O}(\alpha_s^2)$ remainder function $r_c^2$ in this work, effectively setting $r_c^3$ to zero. Thus we label our results as N$^3$LL$^\prime +\cO(\as^2)$. We stress, however, that our study of renormalon cancellation scheme and perturbative scale profile choices, and the conclusions drawn therefrom, remain unaffected from this restriction. We have verified this by performing extractions with $r^3$ in \fig{r3} turned on or off. We defer further study of $r^3$ and its integral $r_c^3$ to a future publication. We also note that the more recent $\alpha_s$ determinations in \cite{Kardos:2018kqj,Verbytskyi:2019zhh} used \texttt{CoLoRFulNNLO} \cite{DelDuca:2016ily} to determine the $\mathcal{O}(\alpha_s^3)$ remainder function.

With this set of remainder functions $r^i_c$ at hand, the final perturbative prediction for the  cumulative thrust distribution is given by 
\begin{align}
\label{eq:sigmaPT}
\sigma^{PT}_c(\tau) &= \frac{\sigma_{c,\text{sing}}(\tau; \mu_{H},\mu_{J},\mu_{S})}{\sigma_0} + \frac{\alpha_s(\mu_{ns})}{2\pi} r_c^1(\tau) \nn \\
&+ \left(\frac{\alpha_s(\mu_{ns})}{2\pi}\right)^2 \left[r_c^2(\tau) + \beta_0 r_c^1(\tau) \ln \frac{\mu_{ns}}{Q} \right]\,,  
\end{align}
where we have made the dependence of the singular cross section on the hard, jet and soft scales $\mu_{H,J,S}$ explicit, and we have allowed for an independent scale $\mu_{ns}$ that can be varied to probe the perturbative uncertainty of the non-singular contribution.  In Sec.~\ref{sec:UNCERTAINTIES} we will discuss how we vary these scales to estimate unknown higher-order corrections in all sectors of our calculation.

%%%%%%%%%%%%%%%%%%%%%%%%%%%%%%%%%%%%%%%%%%%%%%%%%%%%%%%%
\section{Non-Perturbative Treatment}
\label{sec:NONPERT}

\subsection{Gapped Shape Function}

 A complete theoretical treatment of the thrust distribution must also take into account non-perturbative effects due to hadronization. These effects are encoded in the dijet soft function in \eqref{eq:cumulant} describing low-energy wide-angle radiation between the two jets.  We use a model for the soft function \cite{Korchemsky:1998ev,Korchemsky:1999kt,Hoang:2007vb},
 \begin{equation}
 \label{eq:softfunction}
     S(k,\mu_S) = \int dk' \,S_{PT}(k-k',\mu_S)\,f_\text{mod}(k'-2\overline\Delta)\,,
 \end{equation}
 where $f_\text{mod}$ is a non-perturbative shape function that modifies the perturbative prediction for the soft function, and $\overline\Delta$ is a gap parameter modeling the minimum soft momentum in a final state due to hadronization. This implies the following formula for the cross section itself:
\begin{align}
\nonumber
\frac{1}{\sigma_0}\,\sigma_c(\tau) &= \int dk\, \sigma_c^{PT}\Bigl(\tau - 
\frac{k}{Q};\mu_H,\mu_J,\mu_S,\mu_{ns}\Bigr) \\
\label{eq:cumulantshape}
&\times f_\text{mod}
\bigl(k - 2\overline\Delta\bigr)
  \,.
\end{align}
In this prediction, the shape function is convolved with both the (resummed) singular and the non-singular parts of the cross section in \eqref{eq:sigmaPT}---as in previous treatments \cite{Abbate:2010xh,Hoang:2015hka,Bell:2018gce}---in order to smooth the transition from the resummation region to the fixed-order region. This is an important point to keep in mind in the following discussion, as any modifications we make to the non-perturbative gap parameter will affect both regions.

For the gapped shape function in \eqref{eq:softfunction}, we adopt the form
\begin{equation}
\label{eq:Smod}
f_{\text{mod}}(k) = 
\frac{1}{\lambda} \left[ \sum_{n=0}^{\infty}\,b_{n} \,f_{n} \left(\frac{k}{\lambda} \right) \right]^{2} \,,
\end{equation}
where the functions $f_n(x)$ form a complete orthonormal basis composed of Legendre polynomials~\cite{Ligeti:2008ac}.  For the present analysis, we set $b_0=1$ and $b_n = 0$ for $n >0$, since we are only interested in the tail region of the  distribution, where the first moment $\lambda$ of the shape function is the only relevant parameter. In the absence of the non-perturbative gap parameter $\overline\Delta$, one then recovers the leading shift from \eqref{eq:leadingshift} when the shape function is convolved with the perturbative cross section. With a non-zero $\overline\Delta$, on the other hand, the relationship between $\overline\Omega_1$ and the shape function is modified to
\be
\label{eq:Omegagap}
2\overline\Omega_1 = 
2\overline\Delta+ \int dk\,k \,f_\text{mod}(k)\,,
 \ee
with $\overline{\Delta} \sim \Lambda_{\rm{QCD}}$, and where it is understood that the $\lambda$ parameter in \eqref{eq:Smod} is given by $\lambda = 2 \left(\overline{\Omega}_1 - \overline{\Delta} \right)$.  Note, however, that our notation in \eqref{eq:Omegagap} has slightly changed from \eqref{eq:leadingshift}, as the presence of the barred notation ($\overline{\Omega}_1, \overline{\Delta}$) indicates objects defined in an $\overline{\text{MS}}$-like renormalization scheme, where the soft function $S_{PT}$ has been calculated.  However, both the perturbative soft function and the gap parameter $\overline{\Delta}$ exhibit renormalon ambiguities \cite{Hoang:2007vb}, which must be cancelled to obtain stable predictions.

To do so, one redefines the gap parameter in \eqref{eq:softfunction} as
\begin{equation}
\overline{\Delta} = \Delta(\mu_\delta,\mu_R) + \delta(\mu_{\delta},\mu_R),
\end{equation}
where $\Delta$ is renormalon free and $\delta$ is chosen to cancel the ambiguity in the perturbative soft function. We will review the procedure for calculating $\delta$ in various schemes in Sec.~\ref{sec:RENORMALON} below. In these schemes, the perturbative series for $\delta$ is calculated starting from the perturbative soft function that is renormalized at a `reference' scale $\mu_\delta$. Moreover, the condition imposed on the Laplace-space soft function is evaluated at an argument proportional to the inverse of a second `subtraction'  scale $\mu_R$. The dependence on these two scales, as well as the details on the chosen subtraction scheme to compute $\delta$, enters the final prediction for the cross section.

Without repeating the details of the derivation, we quote here the final result for our prediction of the renormalon-free cross section \cite{Abbate:2010xh,Hoang:2014wka,Bell:2018gce}:
\begin{align}
\label{eq:ultimate}
\frac{1}{\sigma_0}\,\sigma_c(\tau) &= \int dk\, \sigma_c^{PT}\Bigl(\tau - 
\frac{k}{Q};\mu_H,\mu_J,\mu_S,\mu_{ns}\Bigr) \\
&\times \Bigl[ e^{-2\delta(\mu_\delta,\mu_R)\frac{d}{dk}}f_\text{mod}
\bigl(k - 2\Delta(\mu_\delta,\mu_R)\bigr)
\Bigr]  \,.
\nonumber
\end{align}
The scheme chosen to compute the renormalon subtraction series, and thus the definition of the gap parameter $\Delta$, must be specified in computing the cross section. The scheme includes a definition of the reference scale $\mu_\delta$, and in case it is not equal to $\mu_S$, the terms in brackets should be re-expanded in powers of $\as(\mu_S)$ \cite{Bachu:2020nqn}.

Independent of the chosen renormalon cancellation scheme is the practical approach for evaluating \eqref{eq:ultimate}.  We do so by collecting terms with the same explicit powers of $\as$, such that the cross section necessary for N$^{3}$LL$^\prime$ accuracy can be expanded as 
\begin{equation}
    \frac{\sigma_c(\tau)}{\sigma_0} = \sigma_c^{[0]}(\tau) +\sigma_c^{[1]}(\tau) + \sigma_c^{[2]}(\tau) + \sigma_c^{[3]}(\tau)\,,
\end{equation}
whose components are given order-by-order by 
\begin{equation}
\label{eq:Xsecexpand}
    \sigma_{c}^{[i]}(\tau) = \int dk \, \left[ \sum_{n=0}^i \, \sigma_c^{\text{N$^n$LO}}\Bigl(\tau - \frac{k}{Q}\Bigr) \, f_\text{mod}^{(i-n)}(k-2\Delta)  \right]\,,
\end{equation}
where we have suppressed the scale dependence in both $\sigma_{c}^{\text{N$^n$LO}}$ (the purely perturbative resummed and matched cross section from \eqref{eq:sigmaPT}, with any fixed-order prefactors truncated to N$^n$LO accuracy, following the notation used in Eq.~(4.38) of \cite{Bell:2018gce}) and the shape function $f_\text{mod}$ for brevity. 
The component notation $f_{\text{mod}}^{(i)}$ represents the fact that in practice we expand out the renormalon-corrected shape function as 
\begin{align}
\label{eq:modelexpansion}
e^{-2\delta\frac{d}{d k}} f_\text{mod}(k-2 \Delta)  &= \sum_i f_\text{mod}^{(i)}(k-2\Delta) \,,
\end{align}
where the coefficients $f_{\text{mod}}^{(i)}$ refer to an expansion in $\alpha_s(\mu_S)$, given in \eqref{eq:fmodexpansion}.
Explicit expressions for these coefficients can be found order-by-order in App.~\ref{sec:INGREDIENTS}, where it is clear that these terms \emph{depend} on the particular renormalon cancellation scheme.  It also follows that in the presence of a gapped and renormalon-corrected shape function, the actual shift of the differential distribution will no longer be a constant, as in \eqref{eq:leadingshift}.  Indeed, the scale dependence of both $\Delta$ and $\delta$, which will be discussed in upcoming sections, leads to a $\tau$-dependent `effective' shift, which can be calculated as
\begin{equation}
\label{eq:effectiveshift}
    \zeta_\text{eff}(\tau)\equiv \int dk \, k \left[ \sum_i f_\text{mod}^{(i)}(k-2\Delta)\right]\,.
\end{equation}
The behavior of this effective shift will guide in large part our considerations below on a set of renormalon cancellation schemes for the thrust distribution.

%%%%%%%%%%%%%%%%%%%%%%%%%
\subsection{Renormalon Cancellation Schemes}
\label{sec:RENORMALON}

As mentioned above, both the perturbative soft function $S_{PT}$ and the subtraction term $\delta$ suffer from renormalon ambiguities associated to infrared poles in their (all-order) Borel-series representations. As in prior studies, we adopt a formalism to cancel them against one another, thereby rendering the overall cross section free of the leading soft renormalon. A generalized set of schemes achieving this cancellation was presented in \cite{Bachu:2020nqn}, each defined by imposing a condition on the soft function in Laplace space to render it free of the leading renormalon:
\begin{equation}
\label{eq:schemedefine}
\frac{d^n}{d(\ln \nu)^n} \ln \left[\wt{S}_{PT}(\nu,\mu_{\delta}) \,
e^{-2 \nu \delta(\mu_{\delta},\mu_R)} \right]_{\nu = \xi / \mu_R} = 0 \,,
\end{equation}
where $\nu$ is the Laplace-space variable, $\mu_\delta$ the reference renormalization scale where $\wt S_{PT}$ is evaluated, and $\mu_R$ the renormalon subtraction scale, with the condition on the $n$-th derivative of $\wt S_{PT}$ imposed at the Laplace-space argument $\nu=\xi/\mu_R$. The parameters $n$, $\xi$, along with the choice of $\mu_\delta$, define the renormalon cancellation scheme being used to define the subtraction term $\delta$ and thus the gap parameter $\Delta$. From \eqref{eq:schemedefine} one immediately obtains an expression for the subtraction term $\delta$ in terms of the perturbative soft function,
\begin{equation}
\label{eq:subtractdefine}
\delta(\mu_{\delta},\mu_R) = \frac{\mu_R}{2 \xi} \frac{d^n}{d(\ln \nu)^n} \ln \wt{S}_{PT}(\nu,\mu_{\delta}) \big \vert_{\nu = \xi/\mu_R} \,.
\end{equation}
Here it is clear that we need to control the RG evolution of the subtraction term, both in terms of the reference scale $\mu_{\delta}$ and the subtraction scale $\mu_R$, which is governed by the anomalous dimensions
\begin{align}
\label{eq:RAnomDim}
\gamma_\Delta \left[\alpha_s(\mu_{\delta})\right] &= \frac{d}{d\ln\mu_{\delta}} \Delta(\mu_{\delta},\mu_R) = -\frac{d}{d\ln\mu_{\delta}} \delta(\mu_{\delta},\mu_R) \,,
\nn\\
\gamma_R \left[\alpha_s(\mu_R)\right] &= - \frac{d}{d\mu_R} \Delta (\mu_R,\mu_R) = \frac{d}{d\mu_R} \delta (\mu_R,\mu_R) \,.
\end{align}
The $\mu_{\delta}$-evolution is well-known \cite{Hoang:2008fs}, and is given in terms of the cusp evolution function $\eta_\Gamma$ defined in \eqref{eq:KkernelsB}.  The second equation is the so-called `$R$-evolution' equation \cite{Hoang:2008fs,Hoang:2008yj}, and its anomalous dimension $\gamma_R$ explicitly depends on the subtraction scheme defined in \eqref{eq:subtractdefine}. Note that $\gamma_R$ is computed at $\mu_\delta=\mu_R$. We give the solution for the $\mu_\delta$- and $R$-evolved gap parameter to three-loop order in \eqref{eq:RGapDeltaevolve}.

We now address the freedom in defining renormalon cancellation schemes which, as is evident in \eqref{eq:subtractdefine}, depend on the choice of the derivative rank (the parameter $n$), the overall normalization (the parameter $\xi$), and the reference scale $\mu_{\delta}$.  As long as an appropriate hierarchy of scales is maintained, one may choose these parameters freely.  In \cite{Abbate:2010xh,Hoang:2015hka}, for example, the so-called \emph{R(-Gap) Scheme} was defined by
\begin{align}
\label{eq:Rdefine}
\text{{\bf{R Scheme:}}} \,\,\,\,\, & \lbrace n, \xi, \mu_{\delta}, \mu_{R} \rbrace = \lbrace 1, e^{-\gamma_E}, \mu_S, R \rbrace \,,
\end{align}
where the functional form of the profile $R$ will be given in Sec.~\ref{sec:PROFILES} below. Exact expressions for $n=1$ for both  the subtraction term $\delta$ and the anomalous dimension $\gamma_R$ are given in App.~\ref{sec:INGREDIENTS} up to $\mathcal{O}(\alpha_s^3)$.

\begin{figure}[t!]
\centering
\includegraphics[width=1.0\columnwidth]{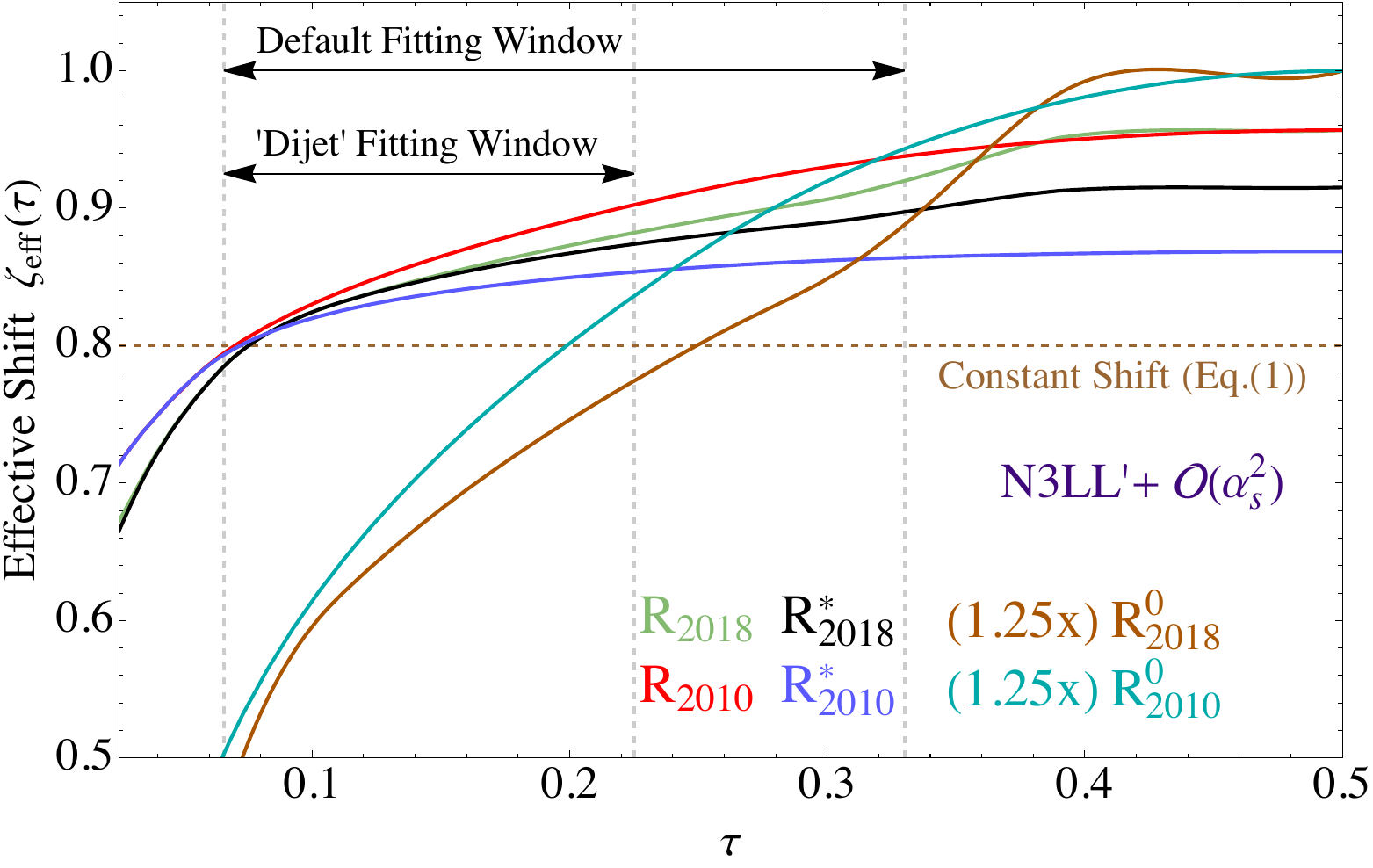}
\vspace{-1.5em}
\caption{The effective shift \eqref{eq:effectiveshift} of the cross section due to different renormalon cancellation and perturbative scale profile schemes, evaluated using `central' profiles at $Q=m_Z$.  The flat, dashed line corresponds to the constant shift of \eqref{eq:leadingshift}, while the vertical gray lines correspond to the two fitting windows discussed below.  All functions are calculated at N$^{3}$LL$^\prime$$ + \mathcal{O}(\alpha_s^2)$ accuracy.
}
\vspace{-0.5em}
\label{fig:effectiveshift}
\end{figure}

In \fig{effectiveshift} we plot the effective shift \eqref{eq:effectiveshift} of the differential distribution that results from various scheme choices across the $\tau$-domain relevant for the $\alpha_s$ fits. Specifically, the red and green curves refer to the R scheme for two different perturbative scale choices (`2010' and `2018') that we will introduce in Sec.~\ref{sec:PROFILES} below.  A notable feature of \fig{effectiveshift} is that the effective shift of the cross section \emph{grows} as $\tau$ increases, even into the region of multi-jet configurations, where the applied formalism for the description of NP corrections, which is based on a dijet factorization theorem, loses its validity. We stress that the growth of $\zeta_\text{eff}$ in \eqref{eq:effectiveshift} is a direct consequence of the $R$ evolution of the gap parameter $\Delta$, predicted by \eqref{eq:RAnomDim}. Its lasting effect in the large $\tau$ region of the cross section, in particular, is a result of the choice made in \eqref{eq:cumulantshape} to convolve the shape function with both the singular and non-singular parts of the cross section. As argued above, this choice was made in order to smooth the transition between the dijet and multi-jet regions. Interestingly, we note that similar effects of a growing non-perturbative shift are also mimicked by models for power corrections to the three-jet region, recently studied in \cite{Caola:2022vea}, although the quantitative effect found in that work is more pronounced. In the context of renormalon cancellation-induced effects in a \emph{dijet} factorization formula, however, the growth of the effective shift $\zeta_\text{eff}$ well into the multi-jet region can appear surprising.

These observations led us to explore the space of renormalon cancellation schemes that may yield an effective shift whose growth is mitigated for large $\tau$ values, while still achieving the cancellation of the leading soft renormalon. As the growth of \eqref{eq:effectiveshift} is related to the RG evolution of the gap parameter $\Delta(\mu_{\delta},\mu_R)$, and therefore the scale profiles of Sec.~\ref{sec:PROFILES}, this motivated us to look for other choices of the reference scale $\mu_\delta$ and/or the subtraction scale $\mu_R$ to achieve this goal.

To that end, we also define the following \emph{R$^\star$~Scheme}:
\begin{align}
\label{eq:Rstar} 
\text{{\bf{R$^\star$ Scheme:}}}\,\,\,\,\, & \lbrace n, \xi, \mu_{\delta}, \mu_{R} \rbrace = \lbrace 1, e^{-\gamma_E}, R^\star, R^\star \rbrace\,,
\end{align}
where we have chosen to identify the reference scale $\mu_{\delta}$ not with the soft scale $\mu_S$, but rather with the subtraction scale $\mu_R$ itself. This eliminates right away any logarithms of $\mu_\delta/\mu_R$ in the renormalon subtraction term $\delta$, as evident from \eqref{eq:deltan0} and \eqref{eq:deltan1}, though there will still be logarithms of $\mu_S/\mu_R$ when re-expanding the subtractions in $\alpha_s(\mu_S)$. In this particular scheme, however, these logarithms do not show up until $\mathcal{O}(\as^3)$, since the one-loop subtraction vanishes exactly, see \eqref{eq:deltan1LO}. 

The scheme choice \eqref{eq:Rstar} gives us some freedom to tweak the behavior of the subtraction scale $R^*$ without inducing too large logarithms with respect to the soft scale $\mu_S$. For instance, we may \emph{freeze} the growth of $R^*$ as a function of $\tau$ with an appropriate choice of the profile functions in Sec.~\ref{sec:PROFILES}. In particular, we will choose a functional form for this new scale, $\mu_\delta=\mu_R=R^\star$, to be a piecewise function:
\begin{equation}
\label{eq:Rstar:profile}
    R^\star(\tau) = \begin{cases}
     R(\tau) & R(\tau) < R_{\rm{max}}\\
     R_{\rm{max}} & R(\tau)\ge R_{\rm{max}} \,.
    \end{cases}
\end{equation}
where $R(\tau)$ is the profile function one might have chosen for the R scheme, and the value $R_{\rm{max}}$ is the value of $R(\tau)$ at the point $\tau=t_1$, which marks the transition between the non-perturbative and resummation regions, as will be discussed in more detail in Sec.~\ref{sec:PROFILES}. Importantly, due to the choice $\mu_\delta=R^\star$ all logarithms in $\mu_{\delta}/R$ appearing in the subtraction term $\delta$ are turned off in the (dominantly non-perturbative) domain with $\tau<t_1$, while for $\tau>t_1$ we allow for small logarithmic contributions in $\mu_{\delta}/R$. In contrast to this, imposing the modified profile in \eqref{eq:Rstar:profile} directly in the R scheme would have induced large logarithmic corrections in $\mu_{\delta}/R$, since the reference scale $\mu_{\delta}$ traces the soft scale $\mu_S$ in this case.

The resulting effective shift in the R$^\star$ scheme is plotted in black and blue in \fig{effectiveshift}, respectively, for the two perturbative scale choices that we mentioned before. In particular, one observes a leveling off in comparison to the standard R scheme, for both scale choices, as  intended. This figure also illustrates that the range of $n=1$ scheme choices we consider corresponds to a variation of power corrections in the multi-jet region of order 10\%.

 %------------------------------------------------
%------------------------------------------------

As a third and entirely independent scheme, we also consider an instance with $n=0$, which we refer to as the \emph{R$^0$ Scheme}:
\begin{align}
\label{eq:R0} 
\text{{\bf{R$^0$ Scheme:}}}\,\,\,\,\, & \lbrace n, \xi,  \mu_{\delta}, \mu_{R} \rbrace = \lbrace 0, 2 \pi, \mu_S, R \rbrace
\,.
\end{align}
We also give explicit expressions for $\delta$ and $\gamma_R$ in the $n=0$ case in App.~\ref{sec:INGREDIENTS}, where one notices that the cancellation ingredients generically appear with one higher logarithmic power than in the $n=1$ case.  The overall effective shift of the R$^0$ Scheme, with varying scale choices, is also shown in Fig.~\ref{fig:effectiveshift} in brown and cyan, where one notices that while both curves increase with $\tau$ with a much steeper slope, their overall magnitude is noticeably reduced with respect to the $n=1$ schemes for all $\tau$ (notice that the curves are multiplied by 1.25 in the figure). 
These effects are jointly associated to the additional power of logarithms appearing in the $n=0$ cancellation terms with respect to $n=1$ counterparts (cf. App.~\ref{sec:MORER0}), and to the $\mathcal{O}(10)$ difference in the normalization factor $\xi$ between R$^{(\star)}$ ($e^{-\gamma_E}$) and R$^0$ ($2\pi$) scheme definitions we have chosen.\footnote{Note that the gap parameter $\Delta$ appearing in $\zeta_\text{eff}$ in \eqref{eq:effectiveshift} evaluated at the reference scale $R_\Delta$ remains the same between the $R^0$ and $R^{(\star)}$ schemes:  $\Delta(R_\Delta,R_\Delta)=0.1$ GeV.}
However, due to less stable perturbative convergence and harder-to-control uncertainties in this scheme, we will not consider it as a candidate for our main analysis below. Regardless, for completeness, we do show some results in this scheme in Sec.~\ref{sec:OTHER} and App.~\ref{sec:MORER0}.

The important point to emphasize is that there are a number of consistent schemes that can be used to cancel the leading soft renormalon of the perturbative soft function $S_{PT}$ and the gap parameter $\Delta$, and in upcoming sections we will study the impact of this choice on the numerical extraction of $\lbrace \alpha_s, \Omega_1 \rbrace$. There are, of course, other considerations one can use to determine if one scheme is preferable or better behaved than another, e.g.~perturbative stability and convergence, and, if so, one could thus argue that the theoretical uncertainty coming from varying schemes can be removed (as we have so chosen for $n=0$). However, in the absence of such arguments, the variation of the subtraction scheme ought to be considered to be its own source of systematic theoretical uncertainty.

%%%%%%%%%%%%%%%%%%%%%%%%%%%%%%%%%%%%%%%%%%%%%%%%%%%%%%%%%%%%%
\section{Estimating Theory Uncertainties}
\label{sec:UNCERTAINTIES}

\subsection{Profile Functions}
\label{sec:PROFILES}

The resummed, matched and NP-corrected cross section in \eqref{eq:ultimate} depends on a set of scales characterizing the physics hierarchies present in our factorization framework, and these should smoothly transition across the full $\tau$ domain we study.  Those in the perturbative cross section $\sigma_c^{PT}$ should of course be chosen to live at values that minimize the logarithms present in the hard, jet and soft functions in the $\tau$ domain most sensitive to resummation effects, i.e. the tail region of the distribution.  For smaller values of $\tau$, towards the peak of the distribution, the full shape function of \eqref{eq:ultimate}  becomes necessary to describe the non-perturbative physics at play, and we force all scales to plateau at some value $\mu_0$ just above $\Lambda_{\rm{QCD}}$.  For larger values of $\tau$, on the other hand, in the far-tail region where matching to fixed-order QCD is required, we merge the scales onto the hard scale $\mu_H$.  All of these mergers are achieved with particular choices of profile functions \cite{Ligeti:2008ac,Abbate:2010xh} and, as mentioned above, demonstrating the impact of this choice on the $\alpha_s$ extractions is one of our central messages in this work.  To that end we  will present the explicit forms for the profiles designed in \cite{Bell:2018gce} (`2018 profiles') and \cite{Abbate:2010xh} (`2010 profiles') in what follows.

\subsection*{2018 Profiles}

In \cite{Bell:2018gce} we designed a set of profile functions for a generalized class of event shapes that encompasses thrust---the angularity distributions. Specifically for thrust the profiles, which were inspired by those presented in \cite{Hoang:2014wka,Hoang:2015hka}, were chosen as:
\begin{align}
\nonumber
\mu_{H} &= e_H Q\,, \\
\nonumber
\mu_{S}(\tau) &= \left[1+ e_{S} \, \theta(t_{3} - \tau) 
\left( 1 - \frac{\tau}{t_{3}} \right)^{2}\right] \mu_{\rm{run}}(\tau)\,,\\
\nonumber
\mu_{J}(\tau)&= \left[1+ e_{J} \, \theta(t_{3} - \tau) 
\left( 1 - \frac{\tau}{t_{3}} \right)^{2}\right]\, 
\sqrt{\mu_H \, \mu_{\rm{run}}(\tau)}
\,,\\
\nonumber
\mu_{ns}(\tau) &=
\begin{cases}
\frac{1}{2}\big(\mu_H+\mu_J(\tau)\big) & n_s = 1 \\
\mu_H & n_s = 0 \\
\frac{1}{2}\big(3\mu_H-\mu_J(\tau)\big) & n_s = -1 \\
\end{cases} \,, \\
\label{eq:allscales2018}
\mu_R(\tau) &= R(\tau) \equiv \mu_S(\tau) \,\,\,\,\,\text{with}\,\,\,\,\, \mu_0 \rightarrow R_0, 
\end{align}
where the function $\mu_{\rm{run}}(\tau)$ ensures that the profiles evolve smoothly over the full $\tau$ domain. Its specific form is given by
\begin{equation}
\label{eq:murun}
\mu_{\rm{run}}(\tau) = 
\begin{cases}
\mu_{0} & \tau \le t_{0} \\
\zeta\Big(\tau; \lbrace t_{0}, \mu_{0}, 0 \rbrace, \lbrace t_{1}, 0, 2r\mu_H  \rbrace \Big) & t_{0} \le \tau \le t_{1}\\
2r\mu_H\tau & t_{1} \le \tau \le t_{2} \\
\zeta\Big(\tau; \lbrace t_{2},  0, 2r\mu_H \rbrace, \lbrace t_{3}, \mu_H, 0 \rbrace \Big) & t_{2} \le \tau \le t_{3} \\
\mu_H & \tau \ge t_{3}
\end{cases}
\end{equation}
where $\zeta$ controls the interpolation between different regions, with
\begin{align}
\label{eq:zeta}
&\zeta\left(\tau; \lbrace t_{0}, y_{0}, r_{0} \rbrace, \lbrace t_{1}, y_{1}, r_{1} \rbrace \right)  \\
&\quad =\begin{cases} 
a + r_0(\tau-t_0) + c(\tau-t_0)^2 & \tau \le \frac{t_{0} + t_{1}}{2} \\
A + r_1(\tau-t_1) + C(\tau-t_1)^2 & \tau \ge \frac{t_{0} + t_{1}}{2} 
\end{cases} \nn
\end{align}
and the various coefficients therein are determined by the continuity requirement of this function and its first derivative,
\begin{align}
\label{eq:zetacoefficients}
a &= y_0 + r_0 t_0\,, &  c&= 2\,\frac{A-a}{(t_0-t_1)^2} + \frac{3r_0+r_1}{2(t_0-t_1)}\,, \\
A &= y_1 + r_1 t_1\,, &  
C &= 2\,\frac{a-A}{(t_0-t_1)^2} + \frac{3r_1 + r_0}{2(t_1-t_0)} \,. \nn
\end{align}
As is clear, the shape of the scale $\mu_{\rm{run}}(\tau)$ in \eqref{eq:murun}, and therefore of all $\tau$-dependent scales in \eqref{eq:allscales2018}, is controlled by the transition points $t_i$, which we choose to be
\begin{align}
\nonumber
t_{0} &=  \frac{n_{0}}{Q} ,\,\,\, &&t_{2} = 
0.295 \,n_2\,,\\
\label{eq:tparameters}
t_{1} &= \frac{n_{1}  }{Q} ,\,\,\,&&t_{3} = 
0.5\,n_3\,. 
\end{align}
The particular forms in \eqref{eq:tparameters} were designed somewhat empirically with particular purposes: $t_{0,1}$ demarcate the boundary between the fully non-perturbative and resummation regions; they roughly track the peak of the differential distribution, and thus scale as the leading power correction $1/Q$. Meanwhile $t_2$ approximates the crossover where singular and non-singular contributions to the cross section are of equal magnitude, which indicates that resummation effects become less important and should hence be turned off. Finally, we set $t_3$ to be just below the maximal thrust value of the spherically symmetric configuration, $\tau^{\text{sph}}=1/2$, so that our predictions reduce to their fixed-order values slightly below this kinematic endpoint.

\subsection*{2010 Profiles}

In their earlier study \cite{Abbate:2010xh}, the authors designed the following set of profile functions:
\begin{align}
\nonumber
\mu_{H} &= e_H Q\,, \\
\nonumber
\mu_{S}(\tau) &=
\begin{cases}
\mu_0 + \frac{b}{2t_1}\tau^2, & \tau \le t_1, \\
b \tau + d, & t_1 \le \tau \le t_2, \\
\mu_H - \frac{b}{1-2t_2}\left(\frac{1}{2}-\tau\right)^2, & \tau \ge t_2
\end{cases},\\
\nonumber
\mu_{J}(\tau)&= \left[1+ e_{J} \,
\left(\frac{1}{2} - \tau \right)^{2}\right]\, \sqrt{\mu_H \, \mu_S(\tau)},\\
\nonumber
\mu_{ns}(\tau) &=
\begin{cases}
\mu_H & n_s = 1 \\
\mu_J(\tau) & n_s = 0 \\
\frac{1}{2}\big(\mu_J(\tau)+\mu_S(\tau)\big) & n_s = -1 \\
\end{cases} \,, \\
\label{eq:allscales2010}
\mu_R(\tau) = R(\tau) &\equiv 
\begin{cases}
R_0 + \mu_1 \tau + \mu_2 \tau^2, & \tau \le t_1 \\
\mu_S(\tau), & \tau \ge t_1
\end{cases}, 
\end{align}
with parameters $t_i$ that are akin to those introduced in \eqref{eq:tparameters}, whereas the parameters $b,d$ are fixed by the continuity requirement of the soft profile and its first derivatives at $\tau = t_i$---see \cite{Abbate:2010xh} for complete details. We also note that these profiles have been somewhat superseded by the 2015 thrust analysis of \cite{Hoang:2015hka}, in which the change of profile functions was found to have only a minor impact on the final $\{\as,\Omega_1\}$ extractions. We nevertheless decided to compare against the 2010 profiles here, since we will see in the following that this impact can be greater in different renormalon cancellation schemes.

Besides their overall functional form, the profiles in \eqref{eq:allscales2018} and \eqref{eq:allscales2010} depend on multiple parameters that allow us to vary these scales in order to estimate unknown perturbative corrections in all sectors of our calculation. The parameters $e_{H,J,S}$ control the overall magnitude and width of the hard, jet, and soft profile bands,  $R_0 \lesssim \mu_0$ ensures that no large logarithms between the soft and renormalon scales arise while ensuring a non-zero subtraction in the peak region \cite{Abbate:2010xh}, the adjustable parameters $n_i$ allow a variation about our estimates for the transition points between the different regions, the parameter $r$ represents a variable slope in the transition regions, and $n_s$ picks out different values of the non-singular scale $\mu_{ns}$ entering the matching contributions to $\sigma^{PT}_c$ in \eqref{eq:sigmaPT}, whose variation probes the size of missing higher-order terms in these fixed-order predictions. Varying between the choices for $\mu_{ns}$ in \eqref{eq:allscales2018} and \eqref{eq:allscales2010} that track the scales $\mu_J$ and/or $\mu_S$ also probes the missing effect of resumming sub-leading logarithms present in the non-singular contribution. The choice of whether $\mu_H$ or $\mu_J$ is the default $\mu_{ns}$ scale, and thus whether the range of variations probe higher or lower values of $\mu_{ns}$, represents the primary difference between the two sets of profile functions. Both are valid and free choices based on considerations of improving perturbative convergence and obtaining reasonable theory uncertainty estimates. As we will see below, however, this choice can have a measurable effect on the (far-)tail region of the distribution and therefore the $\alpha_s$ extractions that include these bins.

%-------------------------------------------------------------------------
\begin{table}[tp!]
\renewcommand{\arraystretch}{1.5}
\begin{center}
\begin{tabular}{|c|c|c|}
\hline
& 2018 Profiles of \cite{Bell:2018gce} & 2010 Profiles of \cite{Abbate:2010xh} \\
\hline
\hline 
$e_H$ & $0.5\leftrightarrow 2$ & $0.5\leftrightarrow 2$  \\ 
\hline 
$e_J$ & $-0.75\leftrightarrow 0.75$ & $-1\leftrightarrow 1$ \\ 
\hline 
$e_S$ & 0 & N.A.  \\ 
\hline
$n_0$ & $1\leftrightarrow 2$ GeV & N.A. \\
 \hline
$n_1$ & $8.5\leftrightarrow 11.5$ GeV & $2\leftrightarrow 8$ GeV \\ 
\hline
$n_2$ & $0.9\leftrightarrow 1.1$ & $0.678\leftrightarrow 1.017$ \\ 
\hline
$n_3$ & $0.8\leftrightarrow 0.9$ & N.A. \\ 
\hline
$\mu_0$ & $0.8\leftrightarrow 1.2$ GeV & $1.5\leftrightarrow 2.5$ GeV \\ 
\hline
$R_0$ & $\mu_0 - 0.4$ GeV & $0.85 \,\mu_0$  \\ 
\hline
$r$ & $0.75 \leftrightarrow 1.33$ & N.A. \\
 \hline
$\delta c_{\tilde S}^3$ & $-1\leftrightarrow1$ &  $-1\leftrightarrow1$ \\
 \hline
$\delta r^2$ & $-1\leftrightarrow1$ & $-1\leftrightarrow1$ \\
 \hline
$n_s$ & $\{-1,0,1\}$ & $\{-1,0,1\}$  \\ 
\hline
\end{tabular}
\end{center}
\vspace{-0.5em}
\caption{
Parameter ranges for both sets of profile functions we consider in this work. In both scenarios, all parameters are chosen randomly within the ranges shown, and the `central' values are the centers of the given ranges, except for the parameters $e_H$ and $r$, whose central values are given by $e_H=r=1$. Note that we have refined this variation for the R$^0$ scheme---see Section \ref{sec:OTHER} for details.
}
\label{tab:scan}
\vspace{-3mm}
\end{table}
%-------------------------------------------------------------------------

%------------------------------------------------
\begin{figure*}[tp!]
\centering
\vspace{-4pt}
\includegraphics[width=0.95\columnwidth]{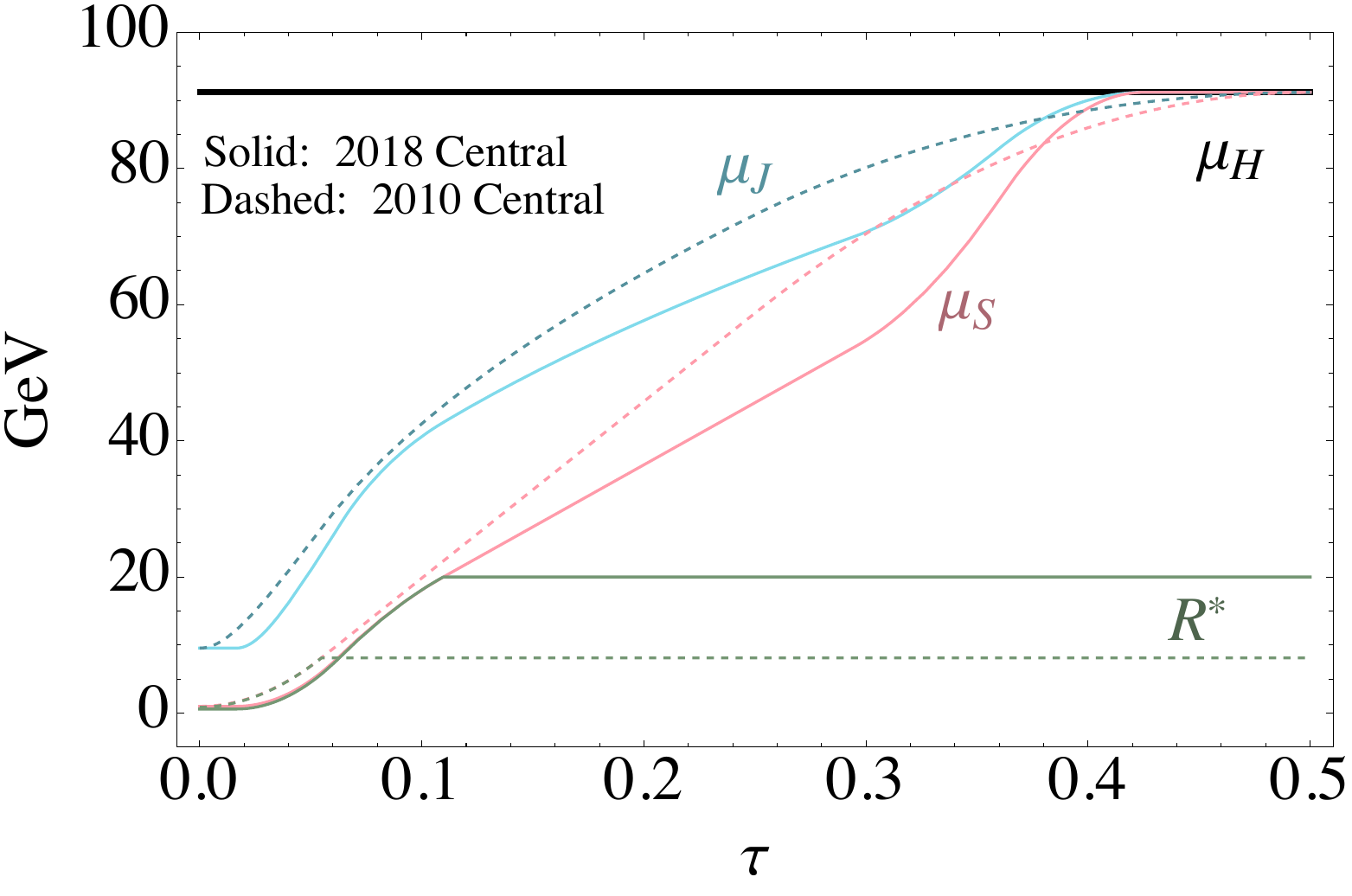}
\includegraphics[width=0.95\columnwidth]{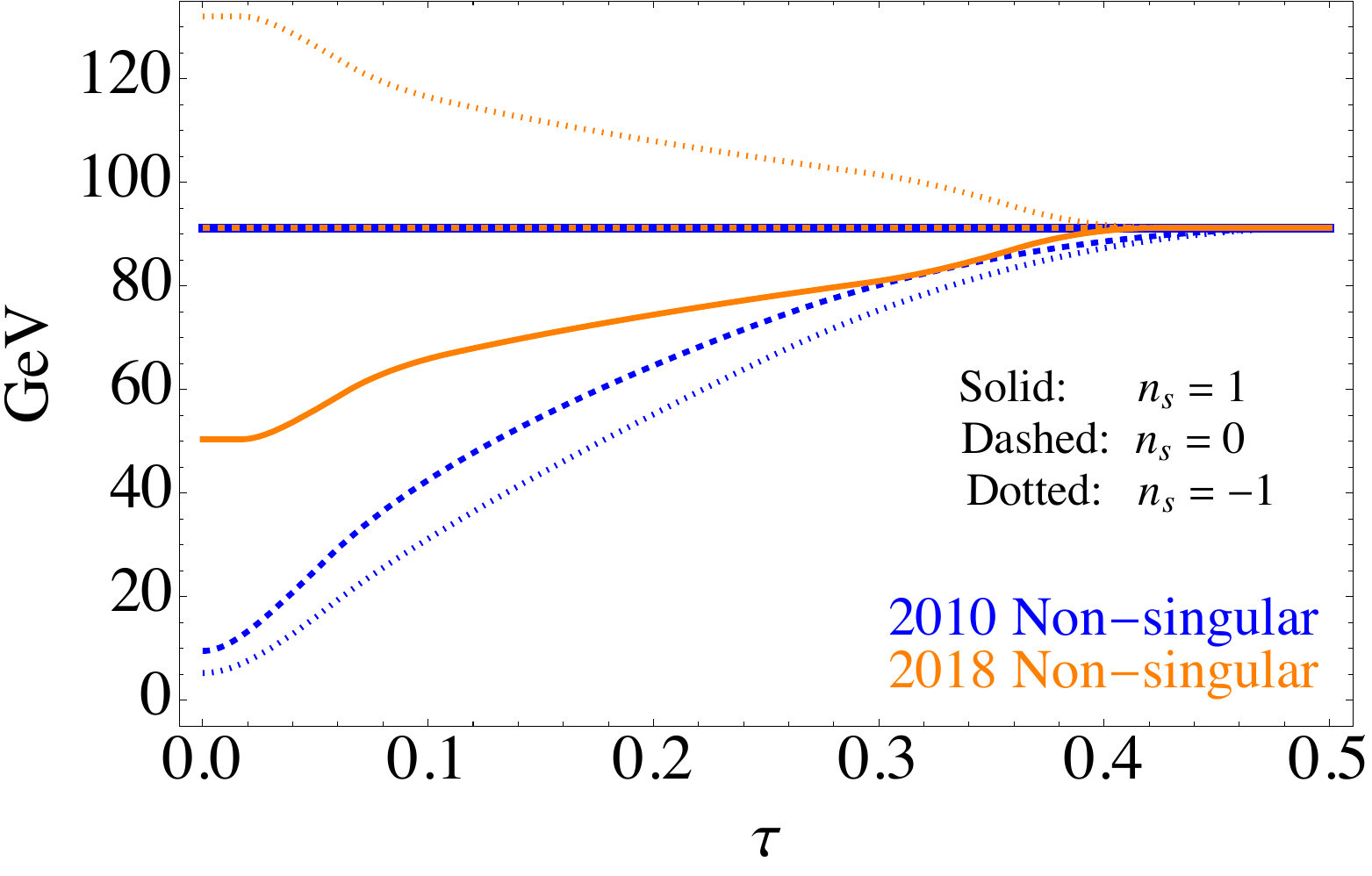}
\
\includegraphics[width=0.95\columnwidth]{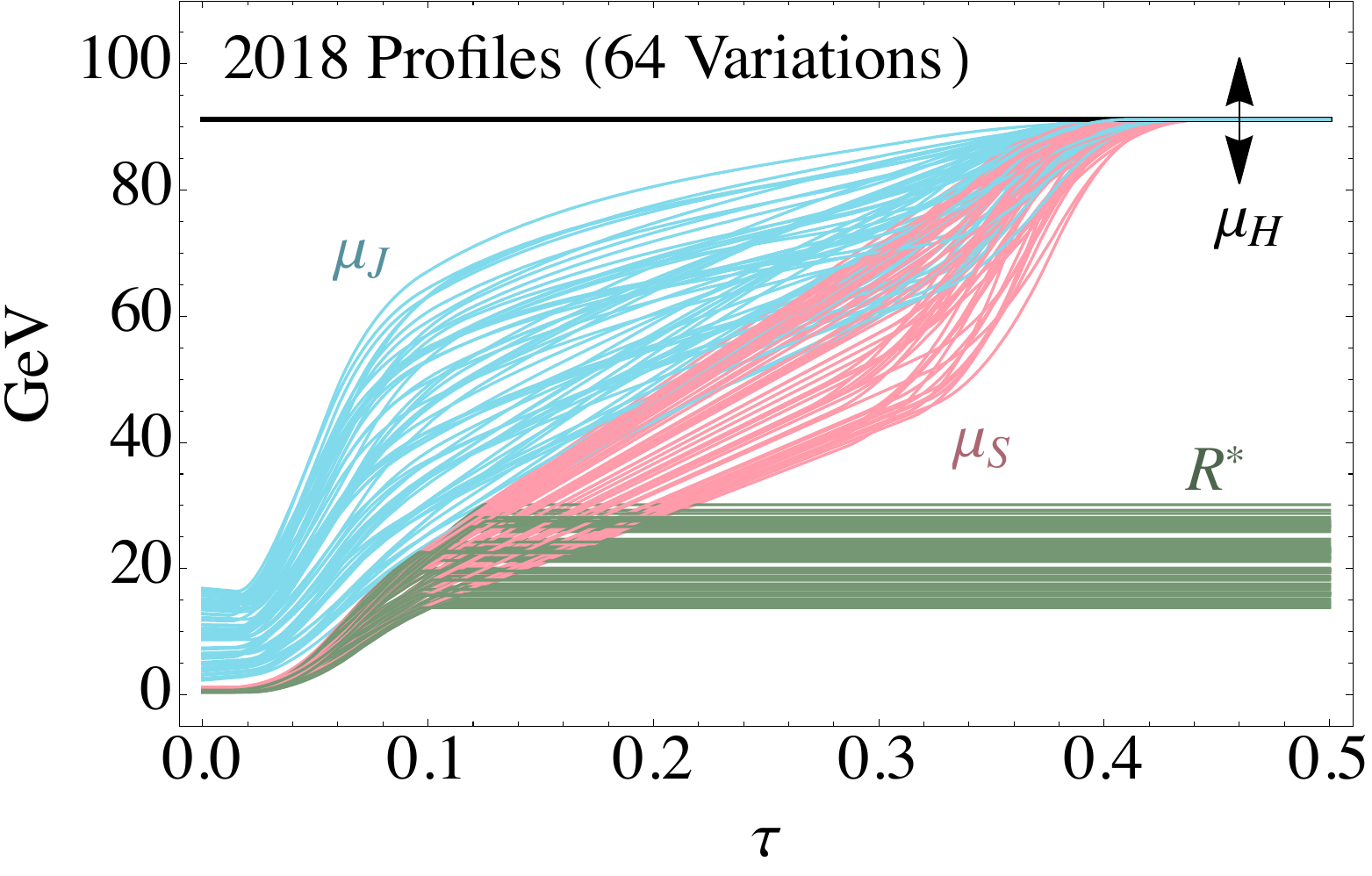}
\includegraphics[width=0.95\columnwidth]{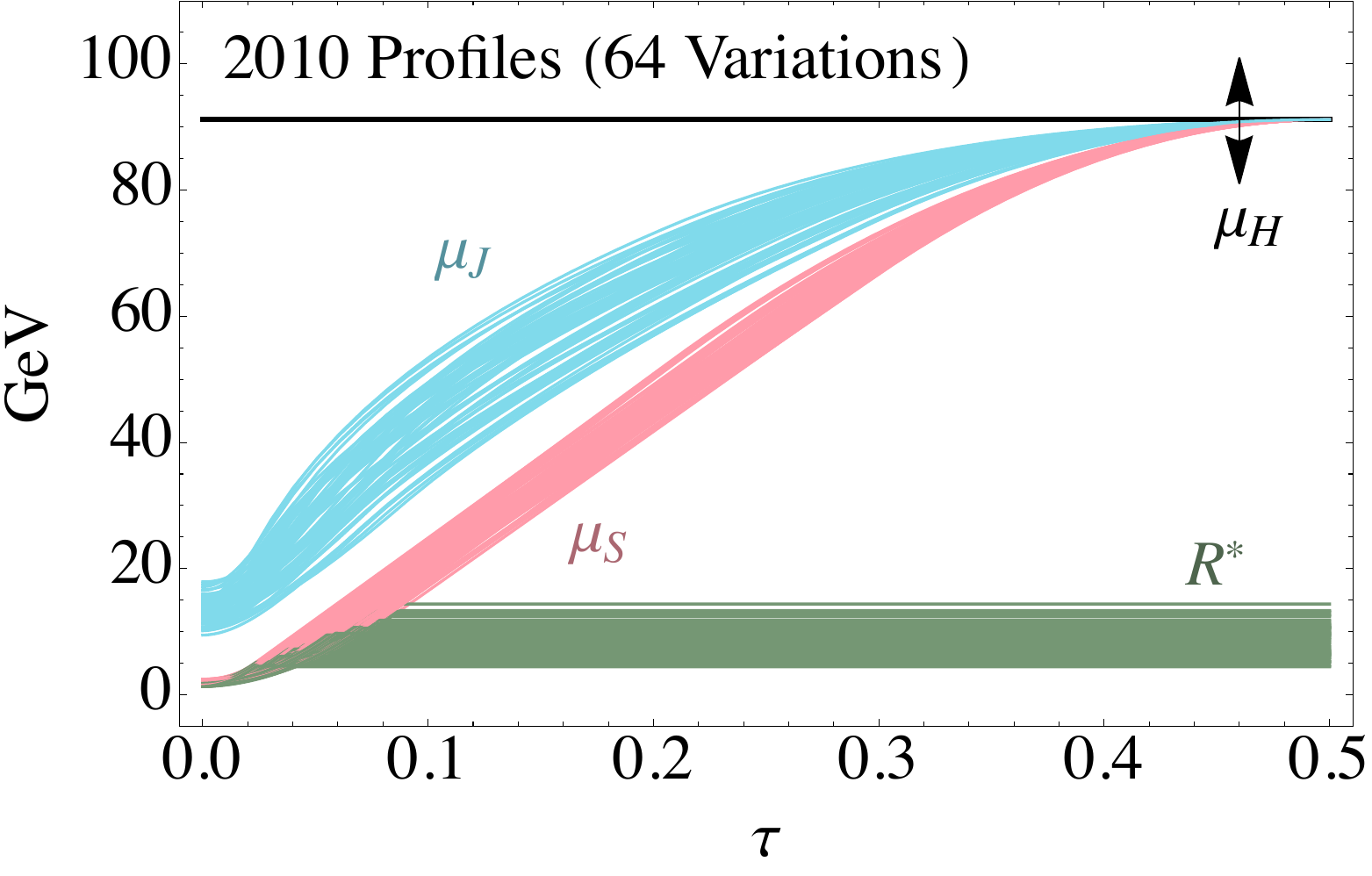} 
\vspace{-0.5em}
\caption{The 2018 and 2010 profiles we implement. The top left figure compares the central values for $\mu_{H,J,S,R}$, while the top right figure shows the non-singular scale $\mu_{ns}$.  Bottom plots show the variations of the 2018 and 2010 profiles from a random scan of parameters in the ranges shown in Tab.~\ref{tab:scan} (as $\mu_H$ controls the overall scale of all profiles, its variation is indicated by the black arrows). In the bottom plots $\mu_S<\mu_J$ is realized for any given set of scales, despite the fact that the overall bands overlap. The R$^\star$ scales from \eqref{eq:Rstar} are shown in all figures, and all plots correspond to $Q=m_Z$.} 
\vspace{-0.5em}
\label{fig:ThrustProfiles}
\end{figure*}
%------------------------------------------------

%------------------------------------------------
\begin{figure*}[tp!]
\centering
\begin{tabular}{cc}
\includegraphics[width=.925\columnwidth]{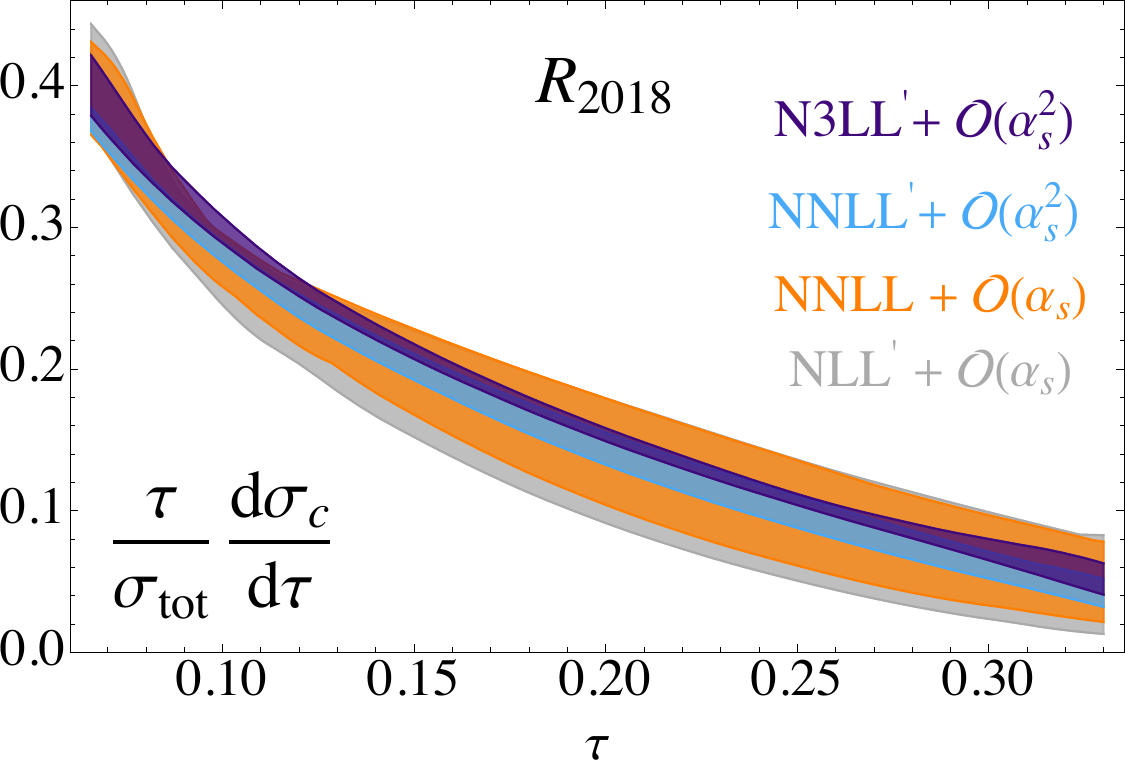} \quad
\includegraphics[width=.925\columnwidth]{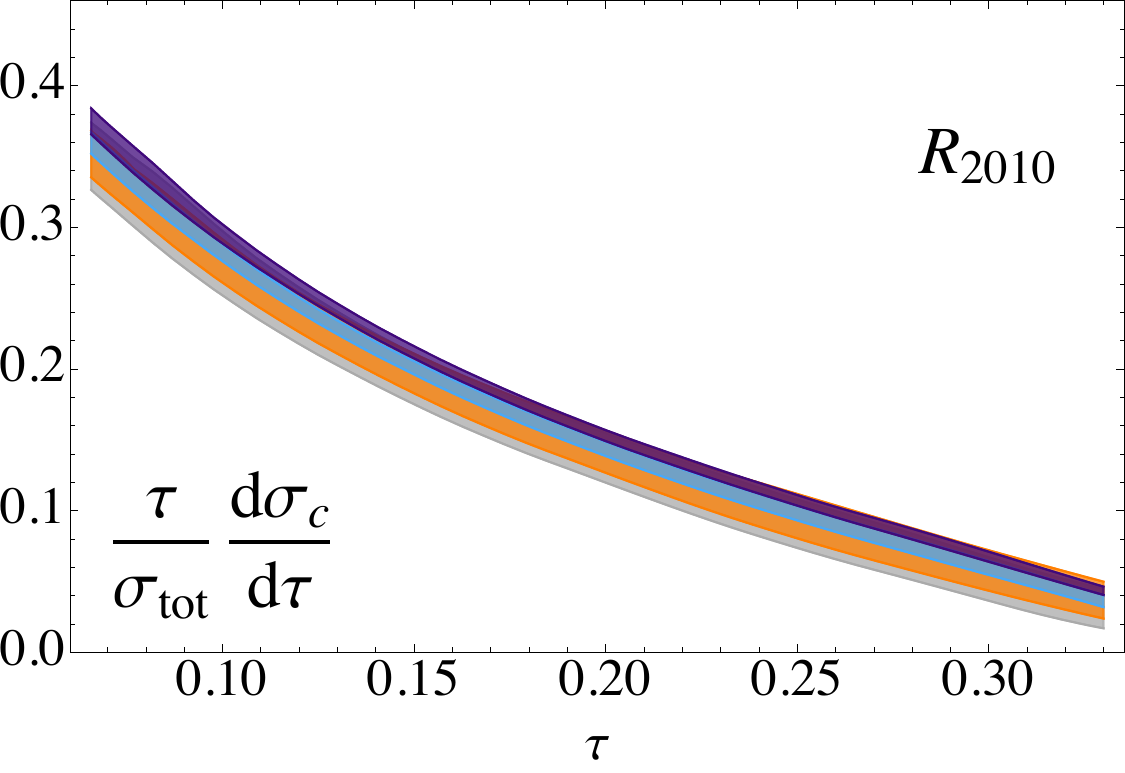} 
\\
\includegraphics[width=.925\columnwidth]{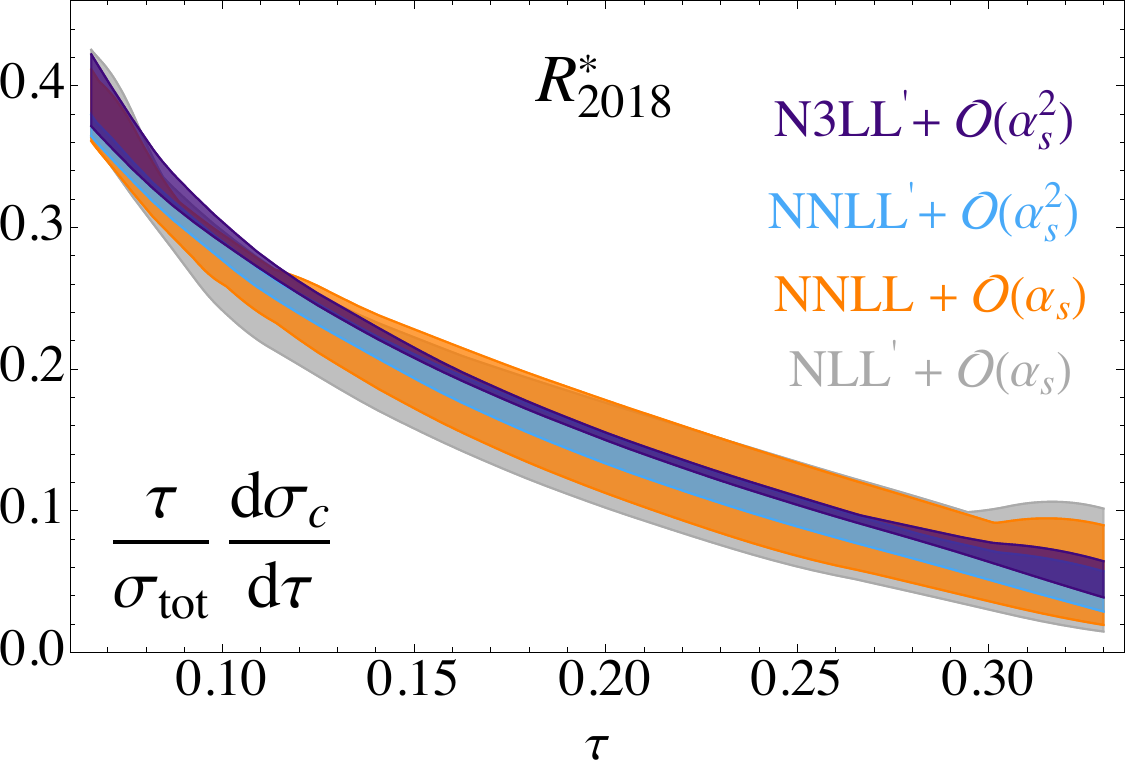} \quad
\includegraphics[width=.925\columnwidth]{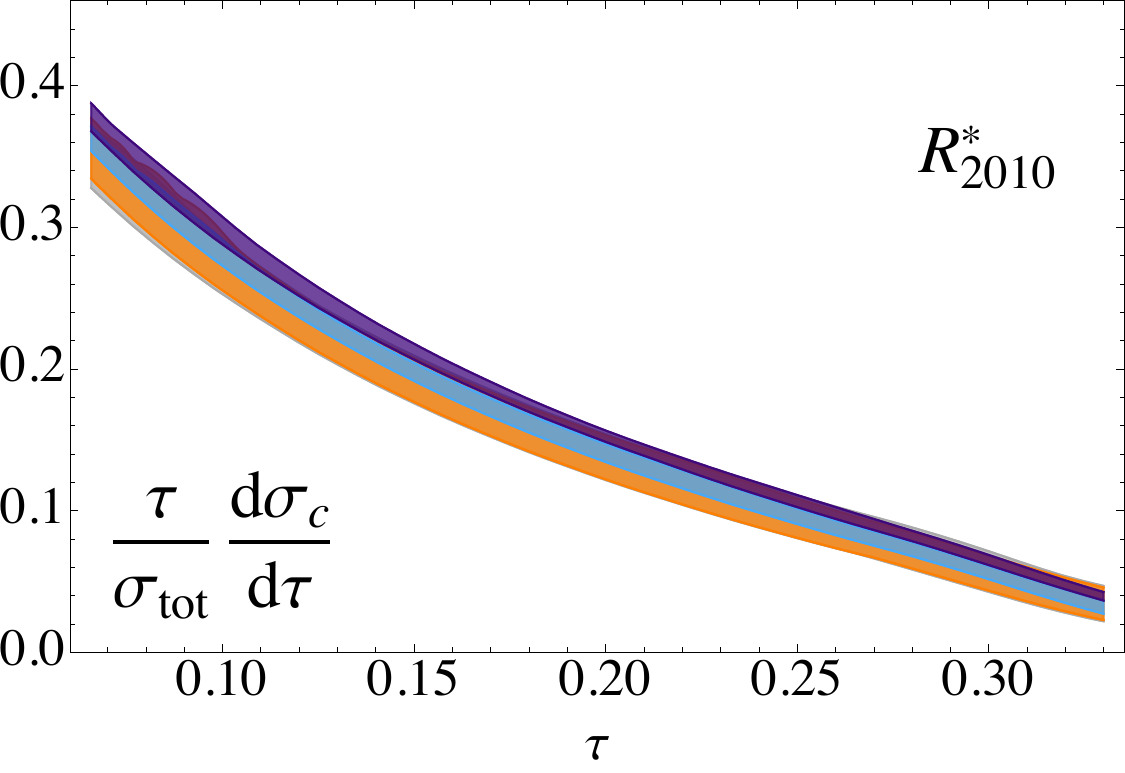}
\end{tabular}
\vspace{-0.5em}
\caption{{\bf{Top Row:}} Normalized thrust distribution across the  $\tau$ domain relevant for the $\alpha_s$ extractions in the R scheme at different resummed and matched accuracies as indicated by the colors. The plots are generated with 64 variations of the embedded profile parameters, with 2018 (2010) profiles shown in the left (right) column. {\bf{Bottom Row:}} The same for the R$^\star$ scheme.}
\vspace{-0.5em}
\label{fig:Rconvergence&dataPlot}
\end{figure*}
%------------------------------------------------

Apart from these two sets of profile functions, we recall that our purpose consists in comparing schemes that use different prescriptions to subtract the leading soft renormalon, as described in Section \ref{sec:RENORMALON}. Specifically, we choose $\mu_R(\tau) = R(\tau)$ as shown in \eqref{eq:allscales2018} and \eqref{eq:allscales2010} in the R and R$^0$ schemes, whereas we impose a cutoff on the subtraction scale $\mu_R$ in the R$^*$ scheme via the modified profile $\mu_R(\tau) = R^*(\tau)$ given in \eqref{eq:Rstar:profile} (where the function $R(\tau)$ that enters \eqref{eq:Rstar:profile} is the same function that is used in the R and R$^0$ schemes). Where to impose the cutoff is a free choice, and we have chosen, for both sets of profile functions, to define
\begin{equation}
\label{eq:Rmax}
    R_\text{\rm{max}} = R(t_1)\,,
\end{equation}
i.e. we freeze the growth of $\mu_R$ to the value it takes at $t_1$, which marks the transition from the non-perturbative (peak) region of the distribution to the resummed (tail) region. We freely admit this is an arbitrary choice---it represents an attempt to limit the growth of the effective shift $\zeta_\text{eff}$ as illustrated in Fig.~\ref{fig:effectiveshift} to a mild but measurable extent and to study its impact on the $\as$ determination. This will be a primary focus of our studies in the remainder of this paper. There are certainly other choices of $R_\text{max}$ and/or the shape of the $\mu_R$ profile one could choose to study.

In order to estimate the total perturbative uncertainty, we adopt the procedure used in \cite{Abbate:2010xh} and \cite{Bell:2018gce}, where random values of all parameters that enter the profile functions are scanned over a predetermined range with each instantiation yielding a different profile (including values of $t_1$ and thus $R_\text{\rm{max}}$ per \eqref{eq:Rmax}).  The overall envelope of these profiles is then taken as the total theory uncertainty. The ranges we scan over for both sets of profile functions are shown in Tab.~\ref{tab:scan}.

The central values for the 2018 and 2010 $\mu_{H,J,S,R}$ profiles are shown in the top left panel of Fig.~\ref{fig:ThrustProfiles}, while the top right panel gives the central values of the non-singular scale $\mu_{ns}$ for each choice of $n_s \in \lbrace -1,0,1 \rbrace$ in both scenarios. The results of 64 random scans of profile parameters are then illustrated in the bottom two panels of Fig.~\ref{fig:ThrustProfiles} for both sets of profile functions.\footnote{Our $\{\as,\Omega_1\}$ fits in \sec{EXTRACT} will actually scan over $\cO(500-1000)$ random profile parameters within the same ranges.}  Note that in all panels we have plotted the renormalon scale with the modified R$^\star$ prescription of \eqref{eq:Rstar}, as otherwise it would be visibly indistinguishable from $\mu_S$ across the bulk of the domain. The hard scale parameter $e_H$ furthermore controls the overall scale of all profiles, and for illustration purposes the curves have been normalised appropriately, whereas its variation is indicated by the black arrows. All plots in \fig{ThrustProfiles} are shown for $Q=m_Z$.

From \fig{ThrustProfiles} one notices that the choices embedded in the two sets of profile functions in \eqref{eq:allscales2018} and \eqref{eq:allscales2010}, which are both legitimate and robust frameworks, lead to a qualitatively different behavior of the various scales throughout the bulk of the $\tau$ domain.  This statement is especially true when considering the non-singular scale $\mu_{ns}$ in the top right panel of Fig.~\ref{fig:ThrustProfiles}.  In \sec{RESULTS} we will study the impact this has on the $\alpha_s$ extractions.

%%%%%%%%%%%%%%%%%%%%%%%%%
\subsection{Predictions for Differential Distributions}
\label{sec:PREDICTIONS}

In addition to the uncertainties from unknown higher-order corrections, which we estimate via the procedure described above, we must also account for systematic errors associated to the numerical extraction of some of the theory parameters from event generators.  In our setup there are two such parameters of concern, the three-loop soft matching coefficient $c_{\tilde{S}}^3$, which we take from \cite{Bruser:2018rad}, and the $\mathcal{O}(\alpha_s^2)$ remainder functions $r^2$, which we have extracted from {\tt{EVENT2}}. Following \cite{Bell:2018gce}, we assign an error function to the central values found for these objects $X \in \lbrace c_{\tilde{S}}^3, r^2\rbrace$ as
\begin{align}
\label{eq:systerror}
    X &= X^{\rm{central}} + 
    \begin{cases}
        \delta X \;\Delta X^{\rm{upper}} & \left( \delta X > 0\right)\\
        \delta X \;\Delta X^{\rm{lower}} & \left( \delta X < 0\right)
    \end{cases}\,,
\end{align}
where $\delta X$ is varied between $\pm 1$ as presented in Tab.~\ref{tab:scan}, and $\Delta X$ represents the associated uncertainty on the extracted parameters, which is given in \eqref{eq:cS3:EERAD3} for $c_{\tilde{S}}^3$, whereas we follow the strategy described in \cite{Bell:2018gce} to assign an uncertainty to the remainder function $r^2$. This procedure allows us to account for systematic uncertainties associated to these quantities, despite the fact that these errors are not the primary focus of our study.

\begin{figure}
\centering
 \includegraphics[width=0.95\columnwidth]{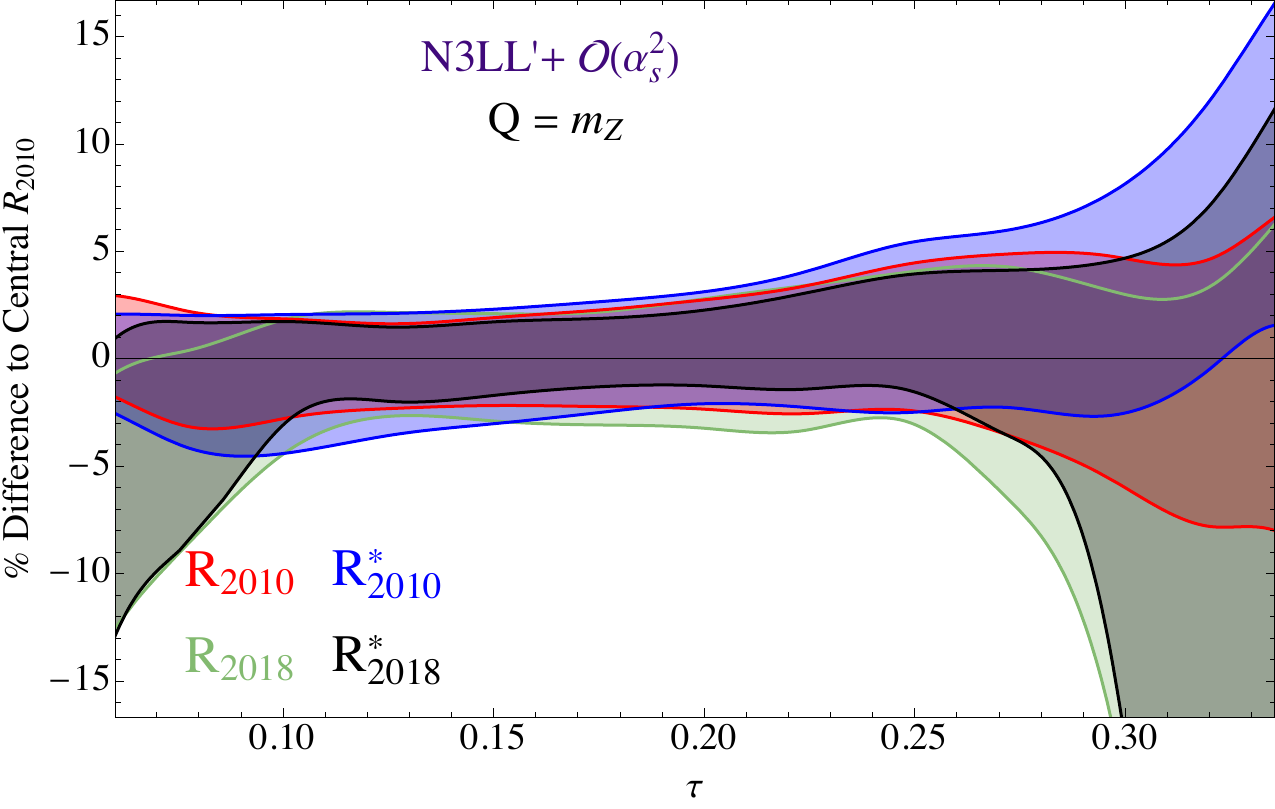} 
\vspace{-0.5em}
\caption{Comparison of the differential distributions shown in Fig.~\ref{fig:Rconvergence&dataPlot} for four renormalon-cancellation and profile-variation schemes, normalized to the central profile of the R$_{2010}$ scheme. 
} 
\vspace{-0.5em}
\label{fig:renormcomps}
\end{figure}

Given the theory inputs described above, we can use \eqref{eq:ultimate} to predict cumulative and differential thrust distributions with well-defined theory uncertainties.\footnote{We stress that all our predictions are based on the cumulative cross section in \eqref{eq:ultimate}. In order to determine uncertainty bands for the differential distributions, we have calculated the derivative of \emph{all} cumulative distribution curves coming from individual profile variations and then maximized/minimized these across the $\tau$ domain, as opposed to simply taking the derivative of the max/min cumulative curves.} We do so in Fig.~\ref{fig:Rconvergence&dataPlot} for the R and R$^\star$ schemes in the top and bottom row, respectively, whereas we relegate a discussion of the R$^0$ scheme to Sec~\ref{sec:OTHER} and App.~\ref{sec:MORER0}.
Here the left (right) column of panels corresponds to predictions made with the 2018 (2010) profile choices, and for convenience we indicated the scale choice in each renormalon cancellation scheme with a subscript. Each panel shows the prediction for the differential cross section, normalized by the total cross section and multiplied by a factor of $\tau$, across the $\tau$ domain that will be relevant for the $\alpha_s$ fits in Sec.~\ref{sec:EXTRACT}. We use the $\mathcal{O}(\alpha_s^3)$ fixed-order hadronic expression from \cite{Chetyrkin:1996ela} for the total cross section normalization. 
The various colors in Fig.~\ref{fig:Rconvergence&dataPlot}  indicate different resummed and matched accuracies, ranging from NLL$^\prime + \mathcal{O}(\alpha_s)$ in gray to our best N$^{3}$LL$^\prime$$  + \mathcal{O}(\alpha_s^2)$ prediction in purple.  All panels, including Fig. \ref{fig:renormcomps}, are produced with $Q = m_Z$, $\alpha_s(m_Z) = 0.11$, $\Omega_1(R_\Delta,R_\Delta) = 0.4$ GeV, and $R_\Delta =  1.5$~GeV.

Broadly speaking, we observe excellent convergence between different resummed and matched accuracies, using both 2018 and 2010 profile scales and considering different renormalon cancellation schemes.  This convergence is evident even towards large $\tau$ values where, at least for the 2018 profiles, we observe a slight widening of the uncertainty bands around the $t_2$ transition point, which controls the $\tau$-dependent scales in these schemes via \eqref{eq:murun}, and which we also vary according to Tab.~\ref{tab:scan}.  We also observe from the bottom panels of Fig.~\ref{fig:Rconvergence&dataPlot} that logarithms of $\mu_S/\mu_{\delta}$, which are non-zero above $t_1$ in the R$^\star$ scheme (but absent in the R scheme), do not qualitatively impact the perturbative convergence of the differential distributions---these logarithms, which are of $\mathcal{O}(1)$ across the relevant domain,  therefore seem safe. Finally, we have checked that all schemes in Fig.~\ref{fig:Rconvergence&dataPlot} exhibit excellent qualitative agreement when compared against data in the relevant fitting regime.   

Independent of the renormalon cancellation scheme, we also notice throughout Fig.~\ref{fig:Rconvergence&dataPlot} that the error bands associated to 2010 profile variations are generally smaller than those associated to 2018 profile variations; the latter can thus be considered more conservative than the former.  Of course, this behavior could have been anti\-cipated from the profile variations shown in Fig.~\ref{fig:ThrustProfiles}, where the width of the 2018 profiles is already wider than its 2010 counterpart.  This feature is largely due to the desire to predict a global set of observables (angularities) in \cite{Bell:2018gce}, where the 2018 profiles were determined, all with well-behaved convergence and uncertainties, as opposed to thrust alone.  Doing so required a somewhat more conservative set of variations. We have, however,  checked that narrowing the 2018 soft, jet, and renormalon scale variations in Fig.~\ref{fig:ThrustProfiles} to widths heuristically similar to the 2010 profiles does not alter our conclusions below in Sec.~\ref{sec:EXTRACT}. We have therefore chosen to leave the 2018 variations in the range used in \cite{Bell:2018gce}. 

In Fig.~\ref{fig:renormcomps} we examine the differences between the differential distributions shown in Fig.~\ref{fig:Rconvergence&dataPlot} more closely. Specifically,
we display $\tau/\sigma_{\rm{tot}}\, d\sigma_c/d\tau$ for the four schemes in in Fig.~\ref{fig:Rconvergence&dataPlot}, normalized to the central profiles of the R$_{2010}$ scheme. All predictions are made at N$^3$LL$^\prime+\mathcal{O}(\alpha_s^2)$ accuracy with $Q=m_Z$. As can be seen, scheme variations can lead to multi-percent effects on the differential distributions, and these effects are especially pronounced towards the far tails of the distributions.  Later, in Sec.~\ref{sec:EXTRACT}, we will explore the possibility of performing fits within a restricted $\tau$ domain isolated more towards the purely dijet region (i.e. $\tau\leq0.225$), where the differences between the various schemes is less pronounced, rather than the default fitting window that is shown in  Fig.~\ref{fig:renormcomps}. One should also keep in mind that the results in \fig{renormcomps} are computed with fixed values of $\alpha_s(m_Z)$ and $\Omega_1(R_\Delta,R_\Delta)$ as reported above, and therefore do not necessarily translate into overlapping best-fit values for $\{\as,\Omega_1\}$ for a given theory profile---this merely provides a guide to a potentially more stable fit window that we will analyse below.

%------------------------------------------------
%------------------------------------------------

%%%%%%%%%%%%%%%%%%%%%%%%%%%%%%%%%%%%%%%%%%%%%%%%%%%%%%%
\section{Global $\{\alpha_s, \Omega_1\}$ Extractions}
\label{sec:EXTRACT}

We will now fit the differential distributions that we designed in the previous sections to the global thrust data in order to extract the strong coupling $\alpha_s$ and the dominant NP parameter $\Omega_1$. To do so, we first collect the data and describe the method we use for this extraction, before we present our main observations regarding the impact of varying renormalon schemes and perturbative profile scale choices on the fit results. We also perform a more detailed comparison of our theory framework to the one that was used in prior SCET-based thrust analyses, and briefly explore other sources of systematic uncertainties that may affect the extraction. Throughout this section, we report values for the fit parameters at standard scales, i.e.~$\alpha_s (m_Z)$ and $\Omega_1(R_\Delta, R_\Delta)$, where $R_\Delta = 1.5$~GeV is an arbitrary (perturbative) reference scale, although we will suppress this scale dependence in the following for brevity.

%%%%%%%%%%%%%%%%%%%%%%
\subsection{Experimental Data}
\label{sec:DATA}

As is already evident from \eqref{eq:leadingshift} and the related discussion in the introduction, the fit parameters $\alpha_s$ and $\Omega_1$ will turn out to be fairly correlated, since the effects of varying one can to some extent be compensated by the other. In order to lift this degeneracy, it will be important to include data at varying c.o.m.~energies $Q$. We will therefore perform a global fit to the available thrust data that includes 52 different datasets with c.o.m.~energies spread over $Q \in \lbrace 35, 207 \rbrace$ GeV. Specifically, we include data from {\tt{ALEPH}} \cite{ALEPH:2003obs} at 91.2, 133, 161, 172, 183, 189, 200 and 206 GeV, {\tt{DELPHI}} \cite{DELPHI:1999vbd,DELPHI:2000uri,DELPHI:2003yqh} at 45, 66, 76, 91.2, 133, 161, 172, 183, 189, 192, 196, 200, 202, 205 and 207 GeV, {\tt{JADE}} \cite{MovillaFernandez:1997fr} at 35 and 44 GeV, {\tt{L3}} \cite{L3:2004cdh}  at 41.4, 55.3, 65.4, 75.7, 82.3, 85.1, 91.2, 130.1, 136.1, 161.3, 172.3, 182.8, 188.6, 194.4, 200 and 206.2 GeV, {\tt{OPAL}} \cite{OPAL:1997asf,OPAL:1999ldr,OPAL:2004wof} at 91, 133, 161, 172, 177, 183, 189 and 197 GeV, {\tt{SLD}} \cite{SLD:1994idb} at 91.2 GeV and {\tt{TASSO}} \cite{TASSO:1990cdg} at 35 and 44 GeV. 
We note that this dataset largely mimics the one that was used in \cite{Abbate:2010xh}, in an effort to minimize systematic differences between the two extractions that are associated to the experimental input.

In the following our default fits will include 488 bins whose centers fall within the $6/Q \le \tau \le 0.33$ domain. This fit window corresponds to the default choice that was also used in \cite{Abbate:2010xh}. Roughly speaking, it corresponds to a domain which starts at the onset of the resummation-sensitive tail region of the distribution, going up to the kinematic endpoint of a three-parton final-state, beyond which further corrections to those captured in the dijet factorization theorem \eqref{eq:cumulant} become important.  

%%%%%%%%%%%%%%%
\subsection{Extraction Method}
\label{sec:METHOD}

%------------------------------------------------
\begin{figure}
\centering
\includegraphics[width=0.95\columnwidth]{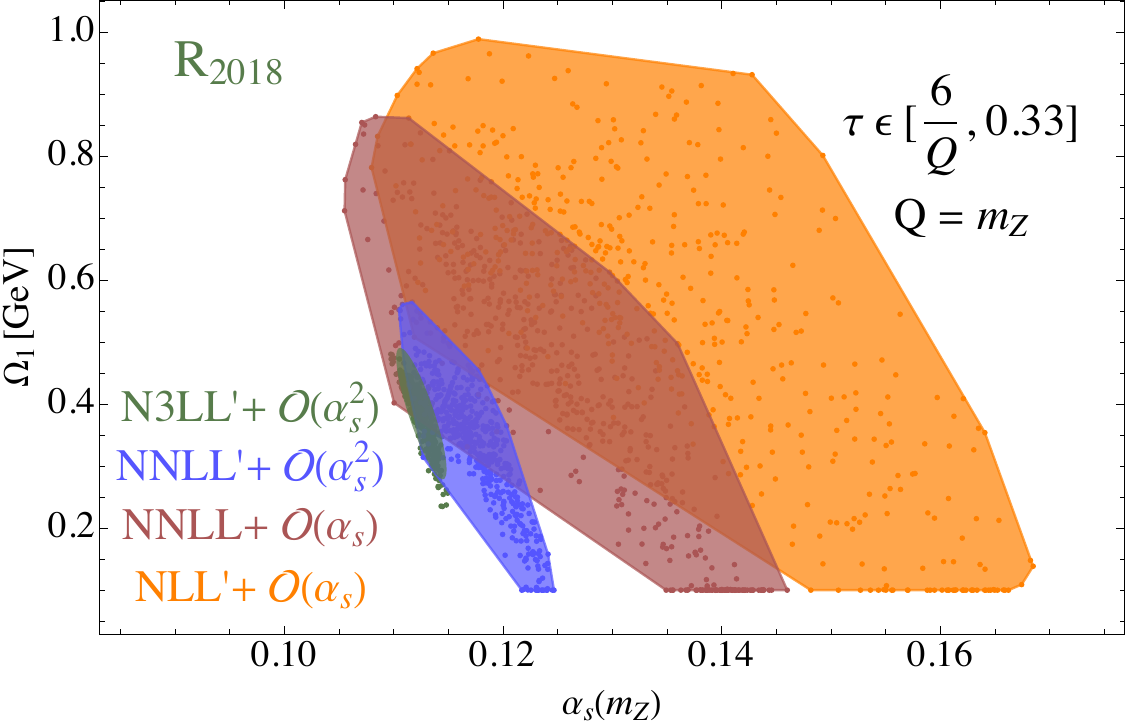} \\[1em]
\includegraphics[width=0.95\columnwidth]{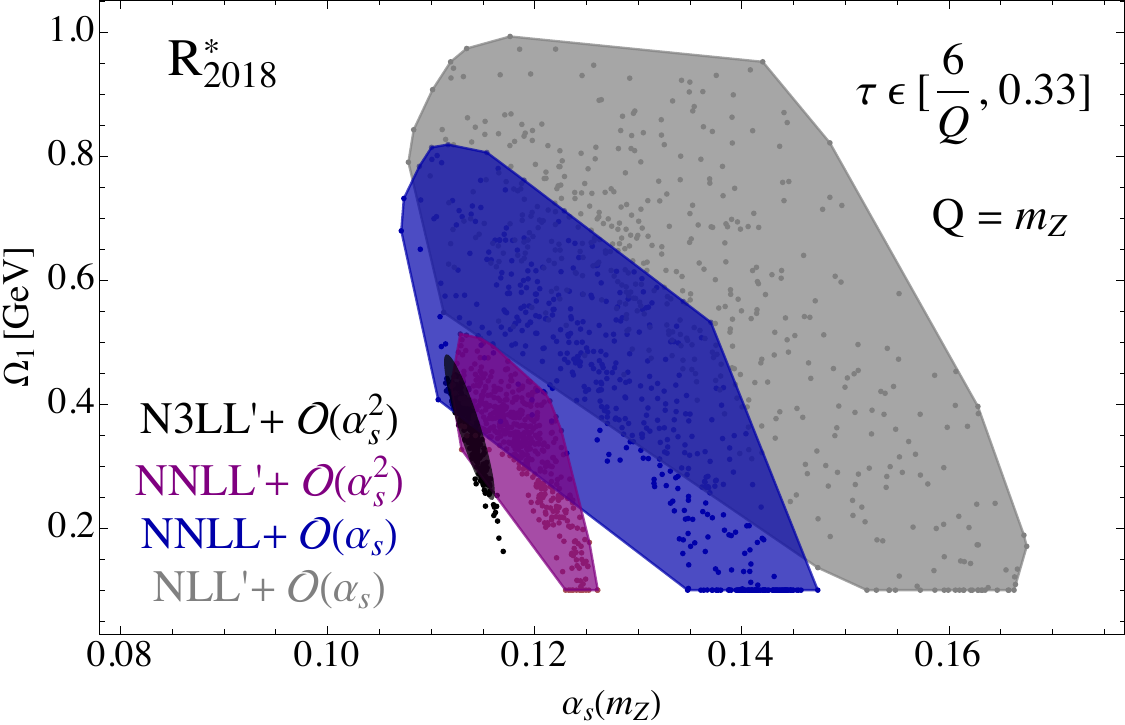} \\
\vspace{-0.5em}
\caption{Extractions of $\lbrace \alpha_s, \Omega_1 \rbrace$ in the R$_{2018}$ scheme (top panel) and the R$^*_{2018}$ scheme (bottom panel) for different resummed and matched accuracies as indicated by the colors. Here $Q=m_Z$, and we note that we have restricted $\Omega_1 \ge \Delta(R_\Delta,R_\Delta) = 0.1$ GeV in these scans.} 
\vspace{-0.5em}
\label{fig:mZconverge}
\end{figure}
%------------------------------------------------

%------------------------------------------------
\begin{figure*}[t!]
\centering
\begin{tabular}{cc}
\includegraphics[width=.9555\columnwidth]{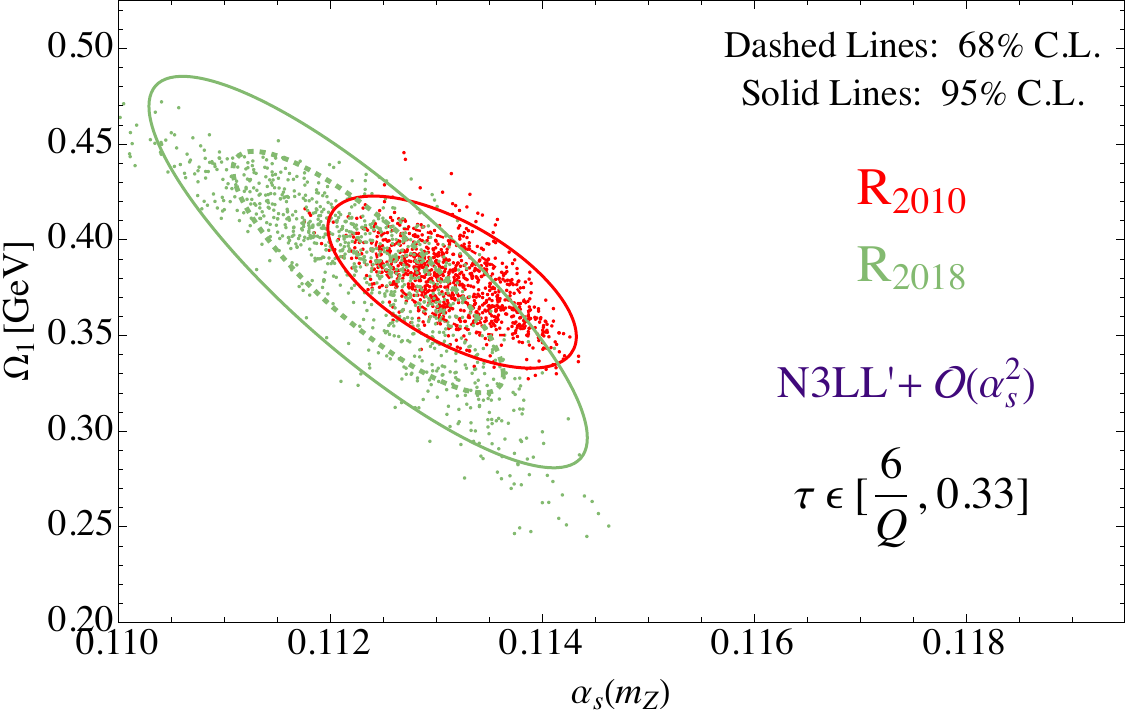} \quad
\includegraphics[width=.9555\columnwidth]{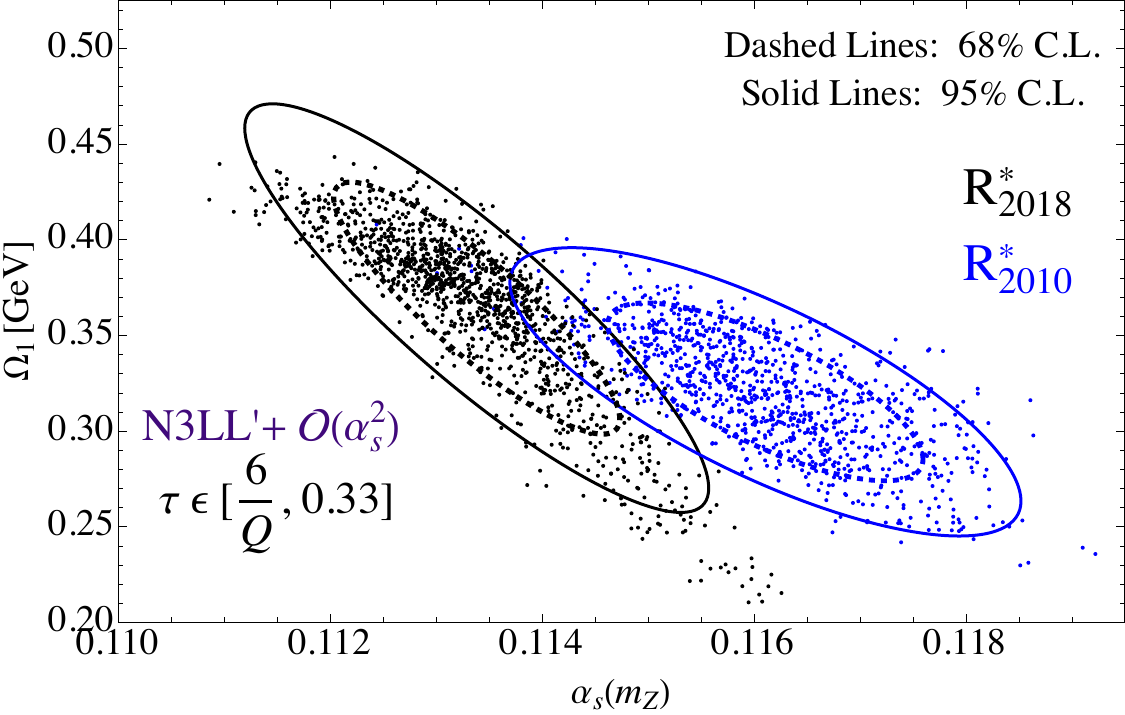}
\\[1em]
\includegraphics[width=.9555\columnwidth]{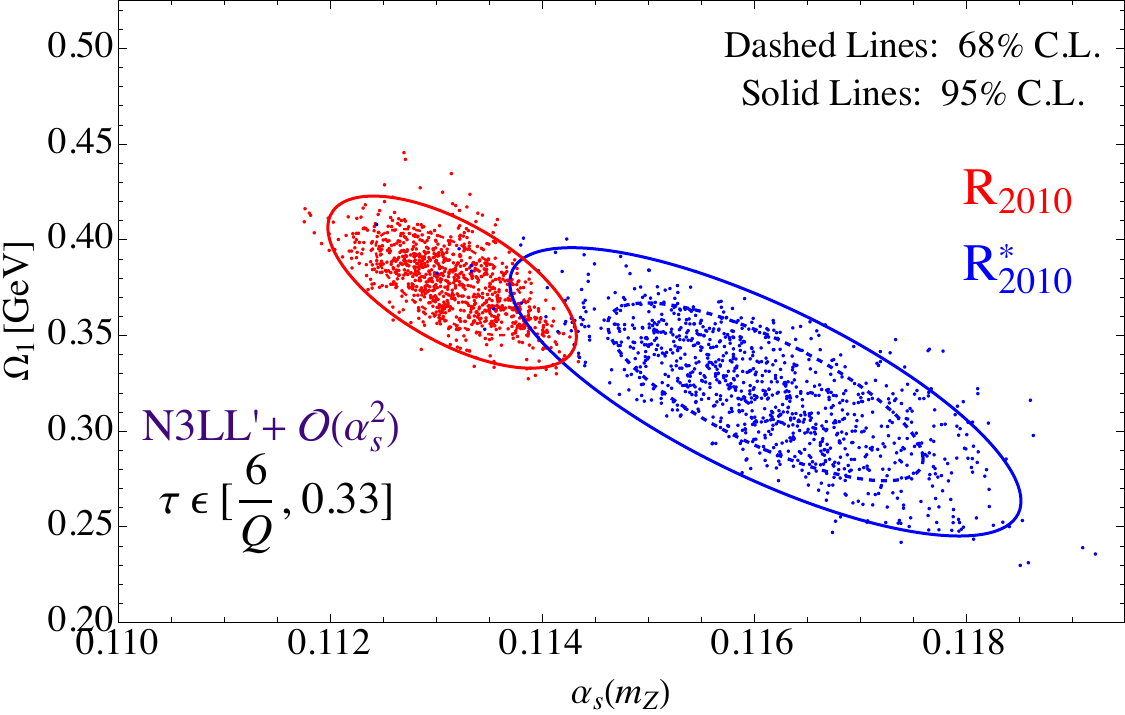} \quad
\includegraphics[width=.9555\columnwidth]{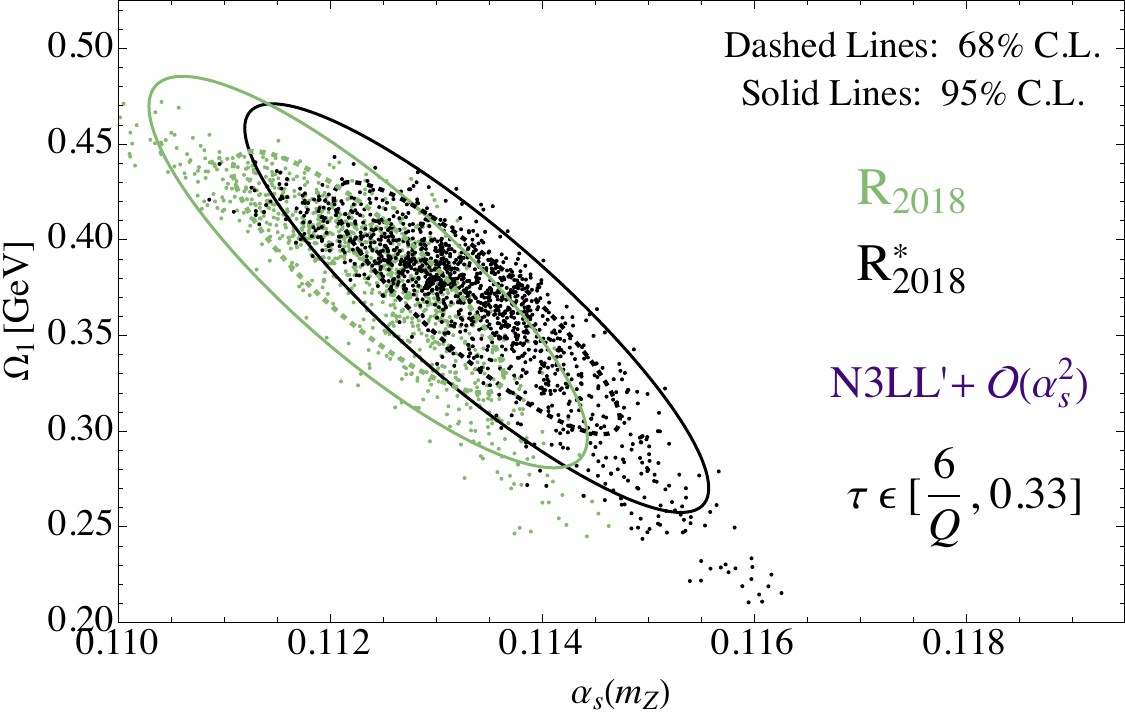} 
\\
\end{tabular}
\vspace{-0.5em}
\caption{Extractions of $\lbrace \alpha_s, \Omega_1 \rbrace$ using 
N$^3$LL$^\prime + \mathcal{O}(\alpha_s^2)$ theory predictions fitted to global thrust data in our default fitting window, $6/Q \le \tau \le 0.33$. Upper panels compare the systematic impact of varying 2010 vs.~2018 profile scales within a given renormalon cancellation scheme, while bottom panels vary renormalon schemes within a given set of profile functions.  Dashed (solid) ellipses correspond to 68\% (95\%) C.L..  Note that $\Omega_1$ is formally defined differently in the R and R$^*$ schemes. 
}
\vspace{-0.5em}
\label{fig:Results1}
\end{figure*}
%------------------------------------------------

Given the framework outlined above, we are now in the position to compare various scheme-dependent theory predictions to the available thrust data, with the aim of extracting values for the strong coupling constant $\alpha_s$ and the leading NP shift parameter $\Omega_1$ in each setup. To assess the quality of these fits quantitatively, we will perform a $\chi^2$ analysis at the binned level:
\begin{equation}
\label{eq:chi2}
\chi^2 \equiv \sum_{i,j} \,\Delta_i \,V_{ij}^{-1} \Delta_j \,,
\end{equation}
where we have defined $\Delta_i$ as the difference between the theoretical prediction and the data in the $i$-th bin,
\begin{equation}
\label{eq:chi2Delta}
\Delta_i \equiv 
\frac{1}{\sigma}\frac{d\sigma}{d\tau} (\tau_{i}) \Bigr\rvert^{\text{exp}} - \frac{1}{\sigma}\frac{d\sigma}{d\tau} (\tau_{i}) \Bigr\rvert^{\text{th}} \,,
\end{equation}
and where the theory predictions for a bin between $\left[\tau_1, \tau_2\right]$ (with $\tau_2 > \tau_1$) are calculated from the difference of the cumulative thrust distribution evaluated at the endpoints of the bin, but with profile scales evaluated at the bin's center $\overline{\tau} \equiv (\tau_1+\tau_2)/2$: 
\begin{equation}
\label{eq:midpoint}
 \frac{1}{\sigma}\frac{d\sigma}{d\tau} (\tau_{i}) \Big\vert^{\text{th}}_{\rm{MP}} \equiv   \frac{1}{\sigma_{tot}} \frac{\sigma_c\big(\tau_2,\mu_a(\overline{\tau})\big) - \sigma_c\big(\tau_1,\mu_a(\overline{\tau})\big)}{\tau_2-\tau_1}\,,
\end{equation}
as advocated in \cite{Abbate:2010xh}. Here $\mu_a$ refers to any of the $\tau$-dependent scales present in our framework.  
We deem this the \emph{midpoint} (MP) binning procedure.

Then, for a given dataset, $V_{ij}$ incorporates the statistical and systematic errors quoted by the experimental collaborations. The statistical errors of each bin, $e^{\rm{stat}}_i$, are considered to be independent and contribute to the diagonal entries of $V_{ij}$. However, the correlated systematic uncertainties between different bins yield non-zero off-diagonal entries. As no information on these correlations is available in the literature, we estimate them using the Minimal Overlap Model (MOM), as advocated by the LEP QCD working group \cite{OPAL:2004wof,ALEPH:2003obs}, and which was also employed in the analyses of \cite{Abbate:2010xh,Hoang:2015hka}. According to the MOM prescription, the off-diagonal elements of $V_{ij}$ can be estimated as the minimum of the two systematic errors of the individual bins $i$ and $j$, such that 
\begin{equation}
    V_{ij} \big{|}_{\text{MOM}} = (e^{\rm{stat}}_i)^2 \delta_{ij}+ \min(e_i^{\rm{sys}},e_j^{\rm{sys}})^2 \,.
\end{equation}  
For a given value of the c.o.m.~energy $Q$ and profile function parameters, we may then find the values $\lbrace\alpha_s,\Omega_1 \rbrace$ that minimize $\chi^2$.

To estimate the theoretical uncertainty on the extracted $\{\alpha_s, \Omega_1\}$ values, we repeat the same procedure for multiple random draws of the profile function parameters within the ranges given in Tab.~\ref{tab:scan}.  In particular, for scans evaluated at N$^3$LL$^\prime + \mathcal{O}(\alpha_s^2)$ accuracy  we perform $\mathcal{O}(1000)$ scans, while we perform $\mathcal{O}(500)$ scans at lower perturbative accuracies.  Upon collecting all of the minimized $\lbrace\alpha_s,\Omega_1 \rbrace$ pairs, we then define the theoretical error estimate as the 68\%/95\% confidence-level (C.L.) ellipse for the two parameters, i.e.~by the ellipse centered at the coordinates $\lbrace \mu_{\alpha}, \mu_{\Omega} \rbrace$, with $\mu_X$ the mean value from all theory draws, and with the standard covariance matrix for the dataset multiplied by the 68th/95th percentile value for $\chi^2$ distributions with two degrees of freedom ($\sim$2.279/5.991), serving as a weight matrix. We then recognize the overall error ellipse defined by these two inputs (center and weight matrix) as an object that can be parameterized as 
\begin{equation}
\label{eq:Ktheory}
    K_{\rm{theory}} =  \begin{pmatrix}
    \sigma_{\alpha}^2 &  \rho_{\alpha \Omega}\, \sigma_{\alpha} \sigma_{\Omega}\\
    \rho_{\alpha \Omega} \, \sigma_{\alpha} \sigma_{\Omega} & \sigma_{\Omega}^2 
      \end{pmatrix} \,,
\end{equation}
where we interpret $\sigma_X$ as weighted variances for the two parameters $\alpha_s$ and $\Omega_1$, and with off-diagonal entries involving the weighted correlation coefficients $\rho_{\alpha \Omega}$ of the two parameters.  We report these parameters for our global scans in the following sections.  To conclude, we inter\-pret \eqref{eq:Ktheory} as the area within which, upon randomly drawing new sets of theory parameters and executing the $\chi^2$ minimization procedure mentioned above, roughly 68\%/95\% of the extracted $\{\alpha_s, \Omega_1\}$ values will fall.

One could also consider incorporating a further experimental uncertainty as described in \cite{Hoang:2015hka}, by constructing a $\chi^2$ distribution as a function of $\alpha_s$ and $\Omega_1$ around the minimum $\chi^2$ obtained using the central profiles in Tab.~\ref{tab:scan}.  However, this uncertainty is expected \cite{Hoang:2015hka} to be much smaller in comparison to the perturbative theory errors parameterized in \eqref{eq:Ktheory} and, given that our motivation consists in highlighting  systematic theory uncertainties, we will only quote the errors embedded in \eqref{eq:Ktheory}.

%%%%%%%%%
\subsection{Results and Discussion}
\label{sec:RESULTS}

%---------------------------------------------------------------------------
\begin{table*}[t]
\centering
{\renewcommand{\arraystretch}{2.}
\begin{tabular}{|c|c|c|c|c|c|}
\hline
\textbf{Profiles} & \textbf{Parameters} & \boldmath{$R$} (Default) & \boldmath{$R^\star$} (Default) & \boldmath{$R$} (Dijet) & \boldmath{$R^\star$} (Dijet)
\\
\hline \hline
\multirow{3}{*}{2018 Profiles} & $\lbrace \alpha_s, \Omega_1 \rbrace$ & $\lbrace 0.1124, 0.383 \rbrace$ & $\lbrace 0.1134, 0.364 \rbrace$ & $\lbrace 0.1126, 0.379 \rbrace$ & $\lbrace 0.1135, 0.353 \rbrace$
\\
\cline{2-6} & $\lbrace \sigma_\alpha, \sigma_\Omega \rbrace$ & $\lbrace 0.0021, 0.102 \rbrace$ & $\lbrace 0.0022, 0.107 \rbrace$ & $\lbrace 0.0016, 0.098 \rbrace$ & $\lbrace 0.0021, 0.118 \rbrace$
\\
\cline{2-6} & $\rho_{\alpha\Omega}$ & $-0.847$ & $-0.881$ & $-0.872$ & $-0.908$
\\
\hline
\multirow{3}{*}{2010 Profiles} & $\lbrace \alpha_s, \Omega_1 \rbrace$ & $\lbrace 0.1132, 0.378 \rbrace$ & $\lbrace 0.1161, 0.320 \rbrace$ & $\lbrace 0.1133, 0.373 \rbrace$ & $\lbrace 0.1161, 0.308 \rbrace$
\\
\cline{2-6} & $\lbrace \sigma_\alpha, \sigma_\Omega \rbrace$ & $\lbrace 0.0012, 0.045 \rbrace$ & $\lbrace 0.0024, 0.075 \rbrace$ & $\lbrace 0.0010, 0.049 \rbrace$ & $\lbrace 0.0023, 0.072 \rbrace$
\\
\cline{2-6} & $\rho_{\alpha\Omega}$ & $-0.633$ & $-0.758$ & $-0.703$ & $-0.731$
\\
\hline
\end{tabular}}
\vspace{0.5em}
\caption{Central values for the 95\% C.L.~$\{\alpha_s, \Omega_1\}$ ellipses, and their associated weighted variances $\sigma_{\alpha,\Omega}$ and correlation coefficient $\rho_{\alpha\Omega}$, in different renormalon-cancellation and profile-variation schemes. Values for $\Omega_1$ are given in units of GeV. The columns `Default' and `Dijet' indicate different fitting windows---see text for details (the numbers in the `Default' columns correspond to the ellipses shown in Fig.~\ref{fig:Results1}). 
}
\vspace{-0.5em}
\label{tab:BestFit}
\end{table*}
%--------------------------------------------------------------------------

We first study the convergence of the $\lbrace \alpha_s, \Omega_1 \rbrace$ extractions when progressively higher perturbative orders are used in the theoretical predictions. Specifically, we display fit results for NLL$^\prime + \mathcal{O}(\alpha_s)$ up to N$^3$LL$^\prime + \mathcal{O}(\alpha_s^2)$ resummed and matched accuracies for the R$_{2018}$ scheme (top panel) and the R$^\star_{2018}$ scheme (bottom panel) for $Q=m_Z$ in Fig.~\ref{fig:mZconverge}.\footnote{Note that we have restricted $\Omega_1  \ge \Delta(R_\Delta,R_\Delta)$ in our extraction code, hence some lower-order scan results are artificially aligned at this boundary, but we have verified that this does not affect any of the N$^3$LL$^\prime + \mathcal{O}(\alpha_s^2)$ fits shown in the following. Also note that the polygonal, lower-order results represent the bounding region for all fits in this plot, not just those lying within a certain confidence level.} As expected from Fig.~\ref{fig:Rconvergence&dataPlot}, the convergence to the highest perturbative accuracy is excellent in each scheme. In particular, the trend is towards lower values of $\alpha_s$ and $\Omega_1$ for these $n=1$ schemes, when the perturbative accuracy is increased. This observation is consistent with prior analyses \cite{Abbate:2010xh,Hoang:2015hka}.  We find a similar behaviour for the R$_{2010}$ and R$^\star_{2010}$ schemes. However, the pattern turns out to be rather different in the R$^0$ schemes, as will be described in App.~\ref{sec:MORER0}. Because of the instabilities found in these schemes, we do not consider the R$^0$ schemes any further in this section.

In Fig.~\ref{fig:Results1} we show the result of global $\lbrace \alpha_s, \Omega_1 \rbrace$ extractions at the highest perturbative accuracy in four different renormalon-cancellation and profile-variation schemes. For illustration purposes, we only display two of these schemes in each panel. Although our framework does not exactly correspond to the one used in \cite{Abbate:2010xh,Hoang:2015hka} (a detailed comparison will be given in Sec.~\ref{sec:COMPARE} below), the results in red for the R$_{2010}$ scheme are certainly closest to these numbers, and they may hence serve as a proxy for these results. 

By comparing the uncertainty ellipses in all four schemes, the one in the R$_{2010}$ scheme clearly sticks out to be significantly smaller than the others. From the upper left panel, we furthermore observe that the impact of switching profiles is only moderate in the R scheme, which is in line with the observations made in \cite{Hoang:2015hka}. This effect is, however, more pronounced in the R$^*$ scheme, as can be seen in the upper right panel. More importantly, the lower panels compare different renormalon-cancellation schemes for the same set of profile functions (2010/2018 in the bottom left/right panel). It is worth emphasizing that switching between the R and R$^*$ schemes changes the formal definition of the $\Omega_1$ parameter, and one should therefore be careful in interpreting the vertical axis in these plots. For what concerns the horizontal axis, on the other hand, $\as$ is of course independent of the chosen renormalon scheme, and is defined by the usual $\overline{\text{MS}}$ prescription. The definition of $\Omega_1$ \emph{does}, however, affect the $\alpha_s$ extraction in an indirect way, since the fitting routine determines the best fit values of the correlated $\lbrace \alpha_s, \Omega_1 \rbrace$ pair.

From the lower panels in Fig.~\ref{fig:Results1} we observe that, relative to the R scheme, the error ellipses are noticeably shifted to larger $\alpha_s$ values in the R$^*$ scheme, with the effect being more significant for 2010 profiles mainly because of its lower value of $t_1$. On a more quantitative level, the results from Fig.~\ref{fig:Results1}, which were obtained using the default fitting window with $6/Q \le \tau \le 0.33$, are summarized in the columns labelled `Default' in Tab.~\ref{tab:BestFit}. Here $\sigma_{\alpha}$ and $\sigma_{\Omega}$ are the weighted variances of $\alpha_s$ and $\Omega_1$, respectively, whereas $\rho_{\alpha\Omega}$ quantifies their correlation as described above. 

%------------------------------------------------
\begin{figure*}[t!]
\centering
\includegraphics[width=.95\columnwidth]{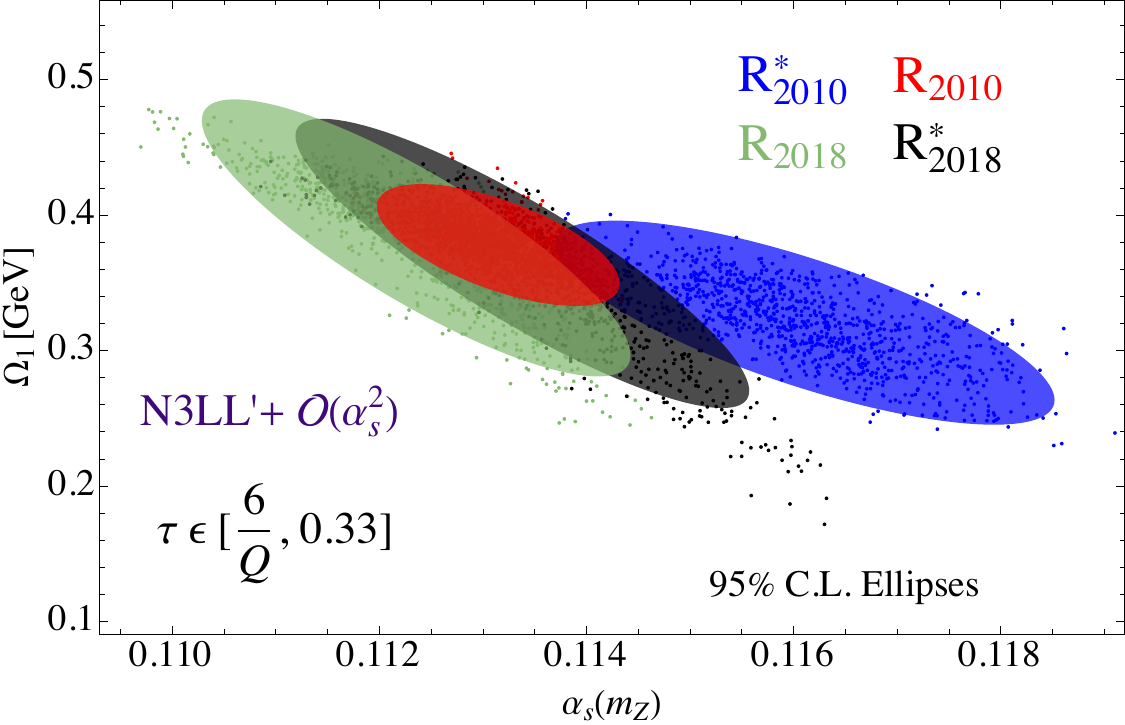}
\includegraphics[width=.95\columnwidth]{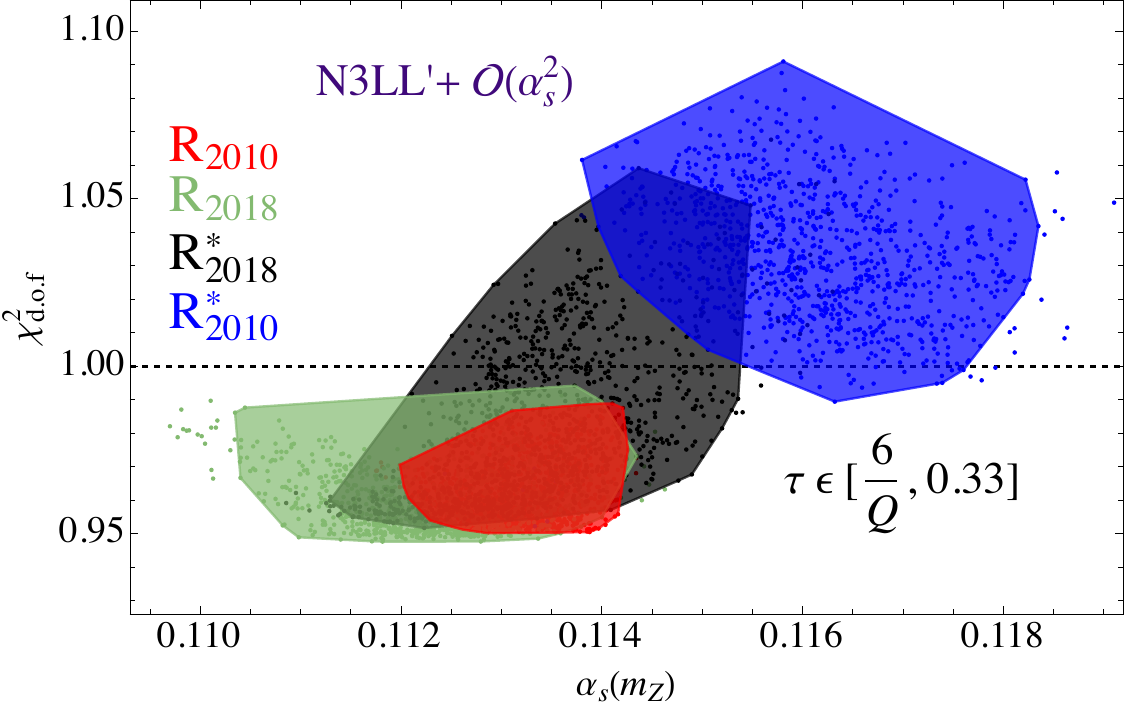}
\vspace{-0.5em}
\caption{Result of a global fit to thrust data using 
N$^3$LL$^\prime + \mathcal{O}(\alpha_s^2)$ theory predictions and a fitting window $6/Q \le \tau \le 0.33$. The plots show the 
$\alpha_s - \Omega_1$ plane ({\bf{Left}}) and $\alpha_s - \chi^2_{\rm{dof}}$ plane ({\bf{Right}}) for four different renormalon-cancellation and profile-variation schemes. The ellipses in the left panel correspond to 95\% C.L..
}
\vspace{-0.5em}
\label{fig:Results2}
\end{figure*}
%------------------------------------------------

In Fig.~\ref{fig:Results2} we combine all of the 95\% C.L.~ellipses from Fig.~\ref{fig:Results1} into the left panel, and show the corresponding $\alpha_s$-$\chi^2_{\rm{dof}}$ plane in the right panel. We re-emphasize that the vertical axis should not be misinterpreted in the left plot, since the definition of the $\Omega_1$ parameter depends on the chosen renormalon scheme. From the $\chi^2_{\rm{dof}}$ distribution one observes that all schemes provide good fits to the data, with the R$^*_{2010}$ scheme, which yields the largest $\alpha_s$ values, being slightly less preferred than the others. Moreover, the left plot shows that, upon considering all four schemes, the extracted 95\% C.L.~values of $\alpha_s$ and $\Omega_1$, while largely consistent with one another, span a range that extends well beyond that of any one scheme alone.  This represents the \emph{first core observation} of our work, namely that changing between different, well-defined renormalon-cancellation schemes and/or profile scale variations can lead to noticeably different extractions of $\lbrace \alpha_s, \Omega_1\rbrace$.  This may be viewed as an indication of additional systematic theory uncertainties. On the other hand, a theoretically motivated reason to stick to a single scheme could, of course, remove this uncertainty. That being said, the schemes we include in \fig{Results2} represent only a subset of possible schemes exhibiting reasonably good perturbative convergence and quality of fit---two criteria that could be used to prefer a particular scheme.

The NP corrections embedded in our framework have been derived  from a dijet soft function that enters the factorization theorem  \eqref{eq:cumulant}. It has recently been argued \cite{Luisoni:2020efy,Caola:2021kzt,Caola:2022vea} that other sources of NP corrections become relevant in the far-tail region of the distribution that may not be negligible over the entire domain $6/Q \le \tau \le 0.33$ used in the previous fits. Related to this, we observe in Fig.~\ref{fig:renormcomps} that the scheme dependence we consider has its most prominent effects in the far-tail region as well. We therefore consider an alternative fit window with $6/Q \le \tau \le 0.225$ in the remainder of this section that concentrates more on purely dijet events. The total number of bins in this setup is then reduced from 488 to 371. As an indication, the perturbative variations of the $Q=m_Z$ thrust distributions in the different schemes shown in \fig{renormcomps} can be reduced from $\lesssim 12\%$ at the upper boundary of our default fit window ($\tau = 0.33$) to $\lesssim 5\%$ at the upper boundary of the reduced fit window ($\tau = 0.225$). We  examine next if this has a noticeable imprint on the $\lbrace \alpha_s, \Omega_1 \rbrace$ extractions.

%------------------------------------------------
\begin{figure*}[t!]
\centering
\includegraphics[width=1.0\columnwidth]{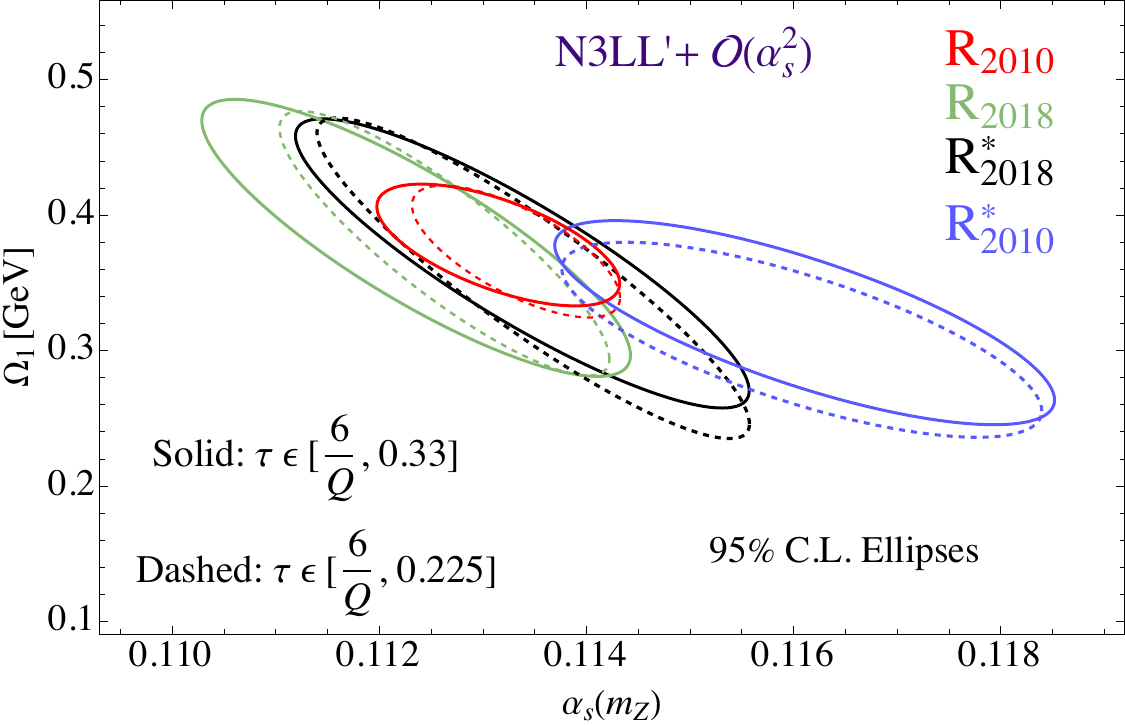}
\includegraphics[width=1.0\columnwidth]{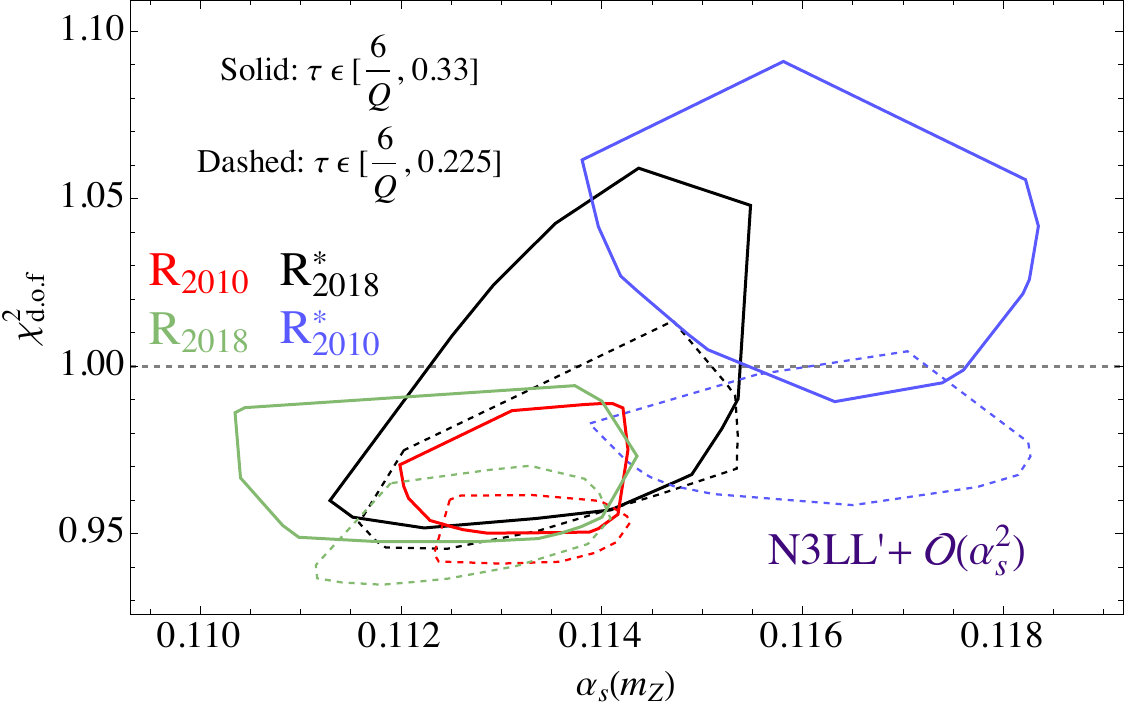}\\
\vspace{-0.5em}
\caption{The same as in Fig~\ref{fig:Results2} for the default fit window with $6/Q \le \tau \le 0.33$ (solid contours) and a reduced dijet fit window with $6/Q \le \tau \le 0.225$ (dashed contours).}
\vspace{-0.5em}
\label{fig:Results3}
\end{figure*}
%------------------------------------------------

In the two panels of Fig.~\ref{fig:Results3} we compare  results that were obtained using this `dijet fit window' (dashed contours) to the ones with the default fit window (solid contours) that were already shown in Fig.~\ref{fig:Results2}. In the left plot one sees that this change has only a mild effect on the  $\lbrace \alpha_s, \Omega_1 \rbrace$ extractions (as previously noted in \cite{Abbate:2010xh}), with the most prominent effect being a shift of the R$_{2018}$ ellipse to slightly larger $\alpha_s$ values. As a result, the overall spread of the fit results among the four considered schemes is just slightly reduced in this setup. These observations are also given numerically in the last two columns of Tab.~\ref{tab:BestFit}. In the right panel of Fig.~\ref{fig:Results3}, on the other hand, one observes that the more prominent effect of narrowing the fit window is a universal trend towards lower $\chi^2_{\rm{dof}}$ values among all considered schemes, despite the fact that the number of bins used in these analyses has been reduced significantly. This improvement is especially noticeable for the R$^\star_{2010}$ scheme (in blue), where the overwhelming number of fits drops below the $\chi^2_{\rm{dof}} = 1$ contour, and which yields $\alpha_s$ values that are more compatible with the PDG world average than the other schemes.

To summarize, we find that fits that are based on a more central dijet-type $\tau$ domain seem to yield higher-quality results for $\lbrace \alpha_s, \Omega_1 \rbrace$ extractions than those including data from the far-tail region, where multi-jet events start to dominate.  This represents the \emph{second core observation} of our analysis, which suggests that precision fits on a more limited dijet window may provide an alternative to the strategy proposed e.g.~in \cite{Nason:2023asn}, in the absence of a model-independent understanding of NP corrections associated with tri- and multi-jet events. From Fig. \ref{fig:Results3}, on the other hand, we only see slight evidence  for any substantial qualitative improvement in the agreement between different schemes when considering more central $\tau$ fits. In other words, over the range of schemes we consider, using a narrower, higher-quality fit window does not by itself remove the potential systematic uncertainty on $\{\as,\Omega_1\}$ coming from this scheme dependence.

%%%%%%%%%
\subsection{Comparison to Prior Results}
\label{sec:COMPARE}

While there exist a number of dedicated thrust-based $\alpha_s$ extractions in the literature (see e.g.~\cite{Davison:2009wzs,Abbate:2010xh,Gehrmann:2012sc,Hoang:2015hka}), our framework is particularly close to the one used in \cite{Abbate:2010xh,Hoang:2015hka}, and we will therefore perform a more detailed comparison to these analyses in this section. First of all, both setups have in common that they use methods from SCET to resum large logarithmic corrections in the dijet limit, and the SCET predictions are matched to fixed-order QCD calculations to account for non-singular corrections in the far-tail region of the distributions. Even more importantly, both setups use a gapped dijet shape function to account for NP corrections in a scheme that is free of the leading soft renormalon. Higher-order corrections are furthermore estimated using a scan over profile scale parameters, and the resulting theory predictions are  fitted to global thrust data for c.o.m~energies $Q \in \lbrace 35, 207 \rbrace$ GeV. Apart from these similarities, there exist, however, a number of differences between the two frameworks that we will now explain in detail, but which do not make a significant impact on our conclusions:
\begin{itemize}
\item 
There are two differences concerning the perturbative treatment described in \sec{THEORY}. First, we only match the N$^3$LL$^\prime$ resummed prediction to $\mathcal{O}(\alpha_s^2)$ fixed-order calculations, due to our findings in \sec{MATCHING} regarding instabilities of \texttt{EERAD3} in the small $\tau$-region. In contrast to this, \cite{Abbate:2010xh,Hoang:2015hka} do include the $\mathcal{O}(\alpha_s^3)$ matching using \texttt{EERAD3}, though with lower statistics and thus larger errors that overshadow these instabilities. In any case they do not measurably affect the region used for the fits. The second point concerns the three-loop soft constant $c_{\tilde{S}}^3$, for which \cite{Abbate:2010xh,Hoang:2015hka} used the Pad\'e approximant in \eqref{eq:cS3:Pade}, whereas we implemented the estimate in \eqref{eq:cS3:EERAD3} that only became available in 2018 \cite{Bruser:2018rad}. As we will show in the following section, switching to the Pad\'e value brings our predictions into even better quantitative agreement with the ones from \cite{Abbate:2010xh,Hoang:2015hka}.
\item 
Our NP treatment in the R schemes, on the other hand, closely resembles the one used in \cite{Abbate:2010xh,Hoang:2015hka} for what concerns both the renormalon scheme definition and the profile functions used.  The differences between the two frameworks are absolutely minor in this respect, and they concern e.g.~only slightly different input values for the gap parameter $\Delta(R_\Delta,R_\Delta)=0.05$~GeV at a reference scale $R_\Delta=2$~GeV used in \cite{Abbate:2010xh,Hoang:2015hka}, compared to our value $\Delta(R_\Delta,R_\Delta)=0.1$~GeV at $R_\Delta=1.5$~GeV. Note that the gap parameter anyway gets absorbed into $\Omega_1$, which we extract from data, according to \eqref{eq:Omegagap}. 
\item
The analyses in \cite{Abbate:2010xh,Hoang:2015hka} furthermore include a number of small effects (bottom and hadron masses, QED effects), which we did not consider, mainly because we were not aiming at a competitive $\alpha_s$ extraction in this work. In \cite{Abbate:2010xh,Hoang:2015hka} the authors showed that these have only a minor impact on the fits, and given that they are unrelated to the main concern of our paper---renormalon schemes and profile variations---we do not expect that these will change any of our findings in a significant way.
\item 
In addition we use a slightly different method for calculating binned distributions. Whereas \cite{Abbate:2010xh,Hoang:2015hka} integrate directly the resummed differential cross section, we start from the cumulative distribution \eqref{eq:integrated}, and calculate the bins according to the midpoint prescription in \eqref{eq:midpoint}. As explained in \cite{Abbate:2010xh}, this avoids the presence of any spurious contributions, and the difference to the method from \cite{Abbate:2010xh,Hoang:2015hka} is expected to be at the sub-percent level in the considered fit windows.
\item
Finally, there are some minor differences in the applied fitting procedure. While the two approaches use highly similar datasets, the authors of \cite{Abbate:2010xh,Hoang:2015hka} calculate the uncertainty ellipses by calculating the best-fit values for $\lbrace \alpha_s, \Omega_1 \rbrace$ obtained from 500 profile variations, drawing the minimum ellipse surrounding the minimum convex polygon encapsulating all of these points, and centered at the average of max and min $\alpha_s/\Omega_1$ values obtained in each direction. Our method for obtaining $K_{\rm{theory}}$ (and the associated error ellipse) is instead outlined in Sec.~\ref{sec:METHOD} above. Moreover, the authors of \cite{Abbate:2010xh,Hoang:2015hka} quote an additional experimental uncertainty in their final results, which is however much smaller than the theory error we also implement.
\end{itemize}
Despite these differences, it is a non-trivial result that our final numbers in the R schemes shown in Fig.~\ref{fig:Results2} and summarized in Tab.~\ref{tab:BestFit} are very similar to the ones obtained in \cite{Abbate:2010xh,Hoang:2015hka}. To make this statement more quantitative, we quote the final numbers obtained in \cite{Abbate:2010xh},
\begin{align}
\label{eq:final:Abbate2010}
    \alpha_s(m_Z) &= 0.1135 \pm 0.0011\,,
    \nonumber\\
    \Omega_1(R_\Delta, R_\Delta) &= (0.323 \pm 0.051) \text{~GeV}\,,
\end{align}
which should be compared to our R$_{2010}$ numbers, whereas the results of the thrust analysis in \cite{Hoang:2015hka},
\begin{align}
    \alpha_s(m_Z) &= 0.1128 \pm 0.0012\,,
    \nonumber\\
    \Omega_1(R_\Delta, R_\Delta) &= (0.322 \pm 0.068) \text{~GeV}\,,
\end{align}
were derived in a setup that is closer to our R$_{2018}$ scheme. Note that the NP parameter is evaluated here at a slightly larger reference scale $R_\Delta = 2$~GeV than for the numbers quoted above
(the evolution to $1.5$~GeV is only a minor effect that reduces the value of $\Omega_1$ by $\sim 4\%$).
Roughly speaking, these numbers translate into an error ellipse that is similar in size to the red one in Fig.~\ref{fig:Results2}, but slightly shifted downwards. We will, in fact, identify one effect that drives the ellipse into this direction in the following section. 

In view of this agreement, we may thus state that our analysis, which is based on a completely independent set of codes, for the first time confirms the results of \cite{Abbate:2010xh,Hoang:2015hka}, regarding the degree to which their extracted value of $\as$ sits lower than the PDG world average.  We consider this important cross-check to be the \emph{third core conclusion} of our work.

%%%%%%%%%%%%%%%%%%%%%%
\subsection{On the Three-Loop Soft Constant $c_{\tilde{S}}^3$}
\label{sec:OTHER}

As mentioned in Sec.~\ref{sec:THEORY}, there exist two approximations for the three-loop soft matching coefficient $c_{\tilde{S}}^3$ in the literature, namely the {\tt{EERAD3}} extraction from \cite{Bruser:2018rad} we have reported in \eqref{eq:cS3:EERAD3} and which is used in our analysis, and the Pad\'e approximant given in \eqref{eq:cS3:Pade} that was implemented in prior $\alpha_s$ fits \cite{Abbate:2010xh,Hoang:2015hka}.  A priori the impact of varying this constant should be small, given that it represents a three-loop effect.  In this section we point out at least two ways that this statement should be qualified.

\begin{figure}[tp!]
\centering
\includegraphics[width=0.955\columnwidth]{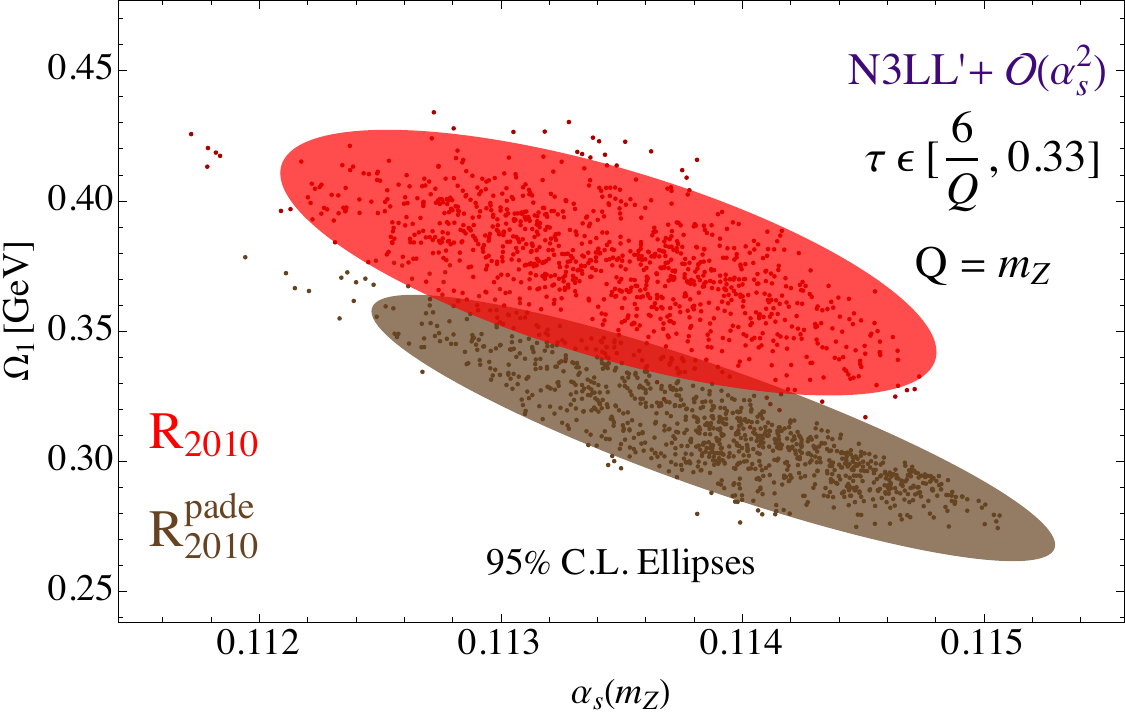}
\vspace{-0.5em}
\caption{Comparison of two $\lbrace \alpha_s, \Omega_1 \rbrace$ extractions that either use the {\tt EERAD3} value $c_{\tilde{S}}^3 = -19988 \pm 5440$ (red) or the Pad\'e approximant $c_{\tilde{S}}^3 = 691 \pm 1000$ (brown).
} 
\vspace{-0.5em}
\label{fig:systematics}
\end{figure}

First, we study the role of $c_{\tilde{S}}^3$ in $\lbrace \alpha_s, \Omega_1 \rbrace$ fits, and provide a more direct comparison between our results presented above and those of \cite{Abbate:2010xh,Hoang:2015hka}. To this end, we compare in Fig.~\ref{fig:systematics} the 95\% C.L.~ellipses for $\lbrace \alpha_s, \Omega_1 \rbrace$ obtained when N$^3$LL$^\prime + \mathcal{O}(\alpha_s^2)$ theory predictions are fitted to $Q=m_Z$ datasets, in the R$_{2010}$ scheme. In particular, the red ellipse corresponds to our default choice of $c_{\tilde{S}}^3$ given in \eqref{eq:cS3:EERAD3}, including its error, whereas the brown ellipse in Fig.~\ref{fig:systematics} uses instead the Pad\'e approximant in \eqref{eq:cS3:Pade}, including again its quoted error. The errors quoted in \eqref{eq:cS3:EERAD3}-\eqref{eq:cS3:Pade} are by themselves actually very minor contributions to the total error ellipses in \fig{systematics}. The differences in the two central values, however, is much larger. As a result, one notices a significant downward shift of our default $95\%$ C.L.~ellipse, which in fact brings our numbers  into even better agreement with the results from \cite{Abbate:2010xh}, as can be verified by comparing to the numbers quoted in \eqref{eq:final:Abbate2010}. As the two values of $c_{\tilde{S}}^3$ differ by more than $3\sigma$, on the other hand, the two ellipses barely overlap. We thus consider the variation of $c_{\tilde{S}}^3$ as another systematic theory uncertainty that may be larger than previously expected (especially due to the instablities in \texttt{EERAD3} at small $\tau$ described in \sec{MATCHING}), although its impact on the determination of $\alpha_s$ itself may only be limited as the plot in Fig.~\ref{fig:systematics} suggests. 

%------------------------------------------------
%------------------------------------------------

%------------------------------------------------
\begin{figure*}[tp!]
\centering
\vspace{-1.2em}
\includegraphics[width=.925\columnwidth]{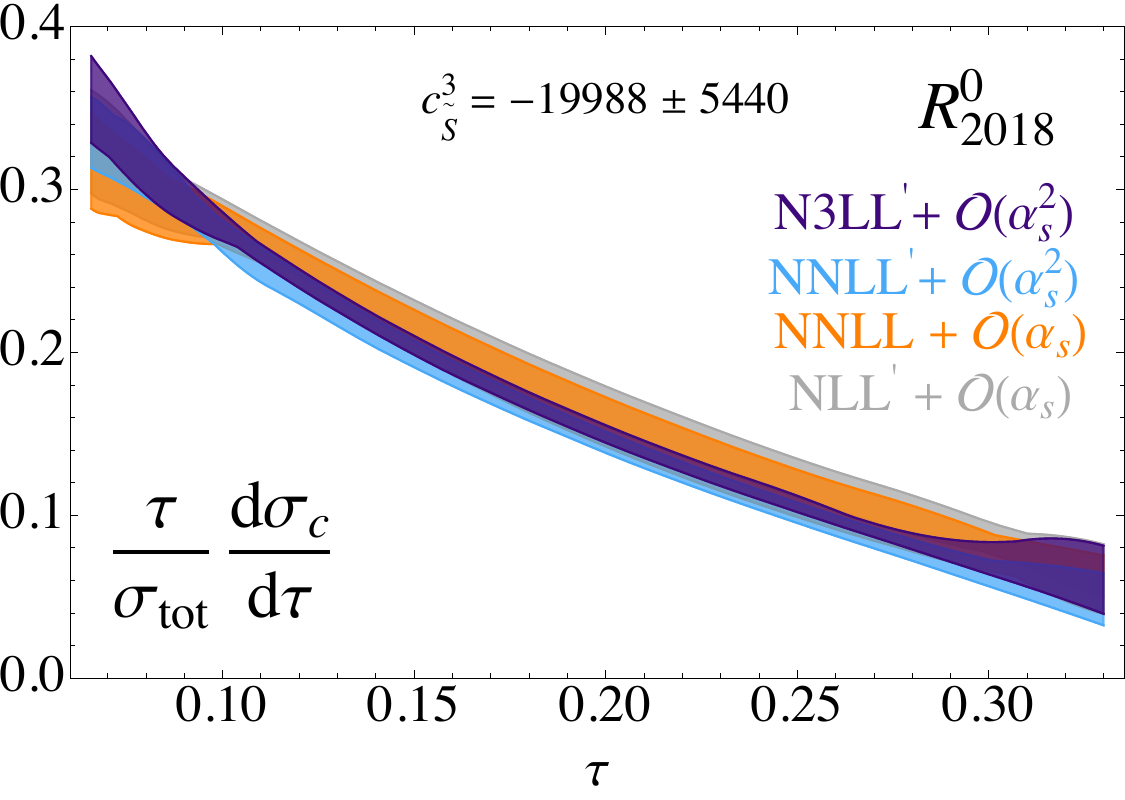} \quad
\includegraphics[width=.925\columnwidth]{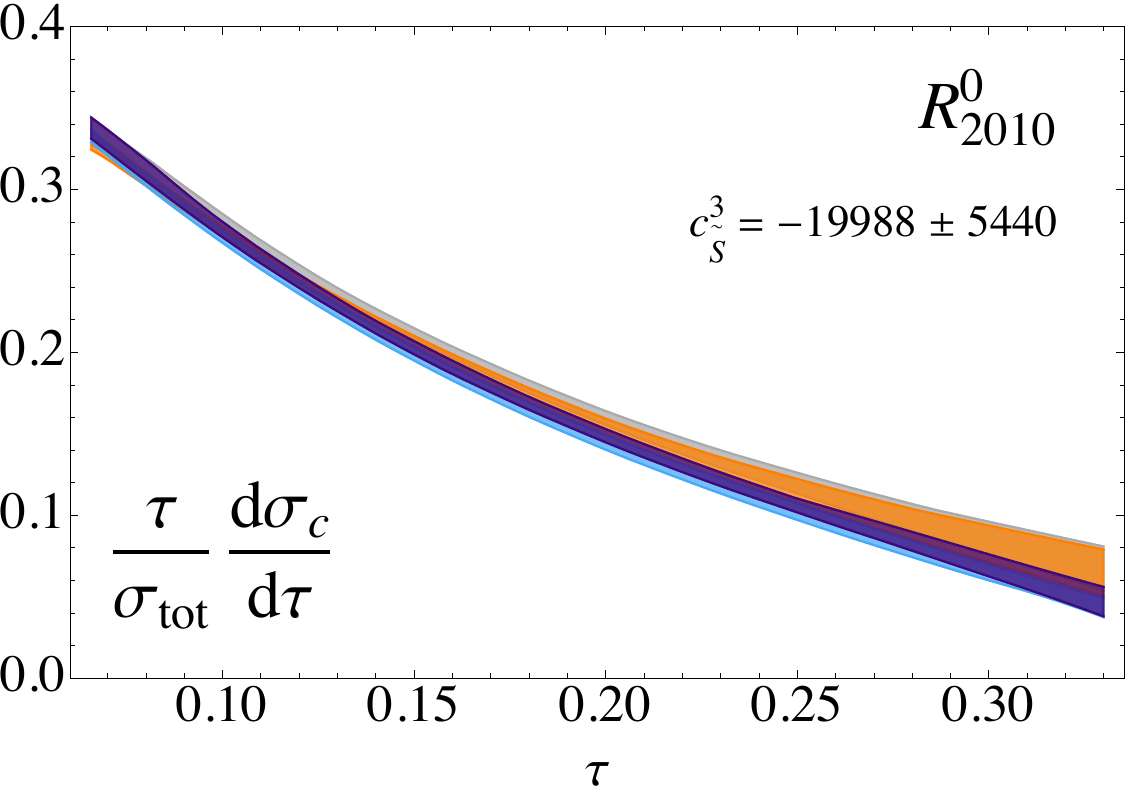} 
\\
\includegraphics[width=.925\columnwidth]{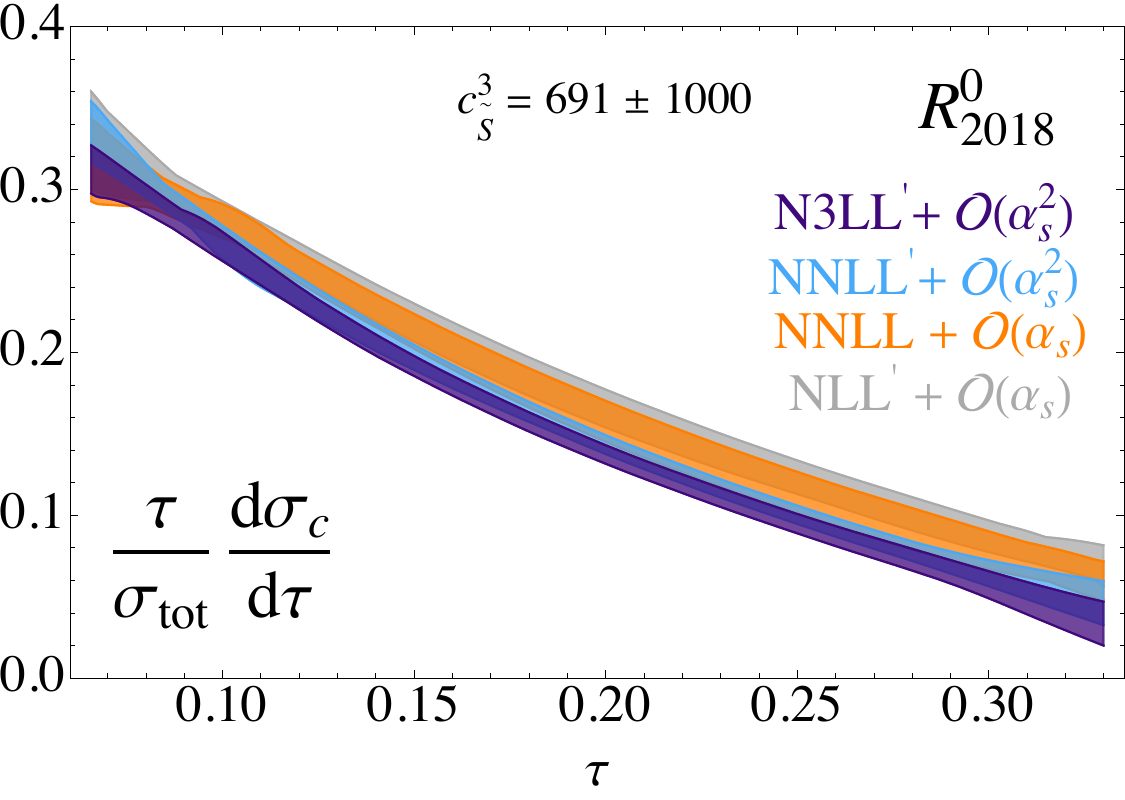}\quad 
\includegraphics[width=.925\columnwidth]{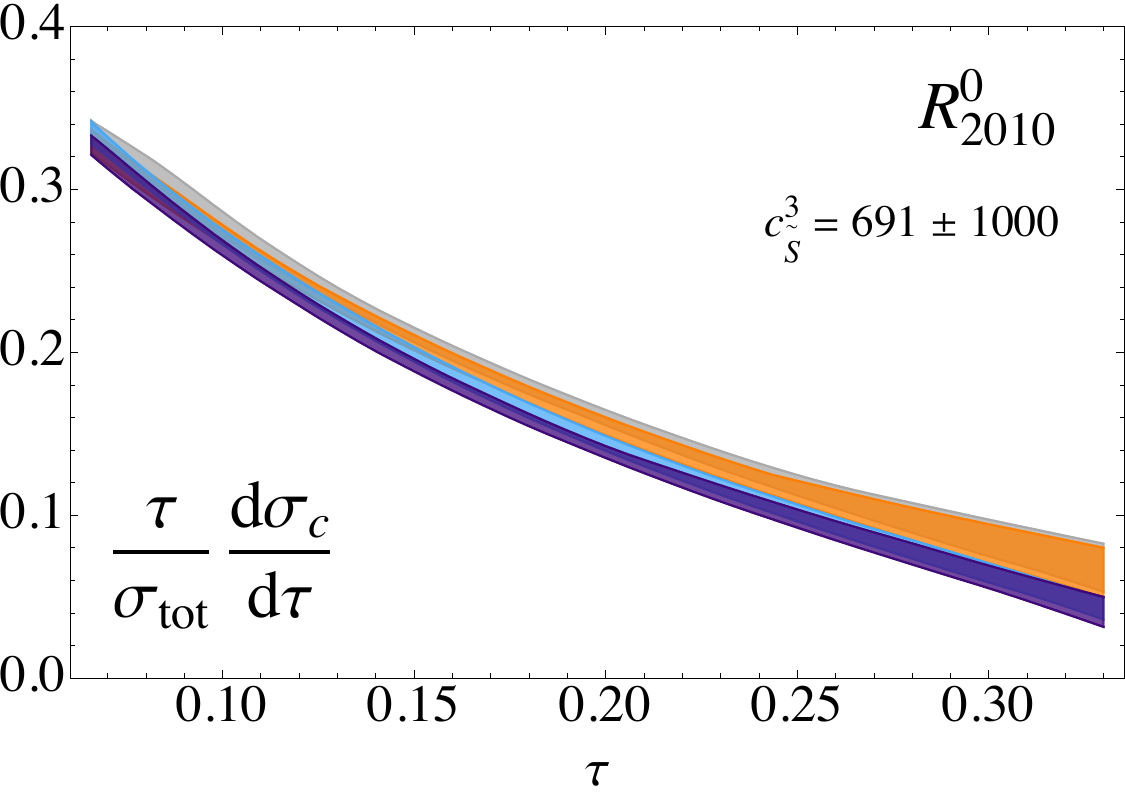}
\vspace{-1em}
\caption{The same as in Fig.~\ref{fig:Rconvergence&dataPlot} for the R$^0$ scheme. The upper plots use the {\tt{EERAD3}} value for $c_{\tilde{S}}^3$ from \eqref{eq:cS3:EERAD3}, and the lower plots the Pad\'e approximant given in \eqref{eq:cS3:Pade}. Left (right) plots refer to 2018 (2010) profile functions.
} 
\vspace{-1em}
\label{fig:R0convergence}
\end{figure*}
%------------------------------------------------

In addition there is a second reason why a better determination of the three-loop soft constant may be warranted, which is related to the R$^0$ renormalon scheme that we introduced in Sec.~\ref{sec:RENORMALON}, but which we largely disregarded in this section because of stability issues. To illustrate these, we show in Fig.~\ref{fig:R0convergence} the analogous plots to the ones in Fig.~\ref{fig:Rconvergence&dataPlot}, but for the R$^0$ schemes. In particular, the upper panels show the result for 2018 profiles (left) and 2010 profiles (right), when the default value of $c_{\tilde{S}}^3$ from \eqref{eq:cS3:EERAD3} is used. Away from the central $\tau$ domain, one clearly observes that the theoretical predictions are not improved when increasingly higher  perturbative orders are included. This effect is particularly pronounced for the 2018 profile scans, and it is true despite the fact that we have already tuned the ranges for some of the parameters in Tab.~\ref{tab:scan}, and only allowed for $0.25 \le e_H \le 1.25$ and 
$-0.5 \le e_J \le 0.5$ ($-0.75 \le e_J \le 0.75$) variations in the R$^0_{2018}$ (R$^0_{2010}$) scenarios.
We recall that the R$^0$ scheme is special, since it is sensitive to one higher power of logarithms in its subtraction terms and, critical to the present discussion, the three-loop soft  constant $c_{\tilde{S}}^3$, which is not yet exactly known---cf. App.~\ref{sec:INGREDIENTS}.  We therefore expect that a concrete determination of this constant, and perhaps a more refined set of profile variations, could eventually stabilize these curves as well.  Regardless, to probe the impact of different $c_{\tilde{S}}^3$ values on the distributions we have included in the lower panels of Fig.~\ref{fig:R0convergence} the corresponding distributions when the Pad\'e approximant in \eqref{eq:cS3:Pade} is used.   Here one notices that the convergence is somewhat improved in comparison to the corresponding upper plots. Indeed, we have also checked more generally that the breadth of the purple N$^{3}$LL$^\prime$ $+ \mathcal{O}(\alpha_s^2)$ bands in the R$^{(\star)}$ schemes of Fig. \ref{fig:Rconvergence&dataPlot} is also artificially enhanced due to the present uncertainty on $c_{\tilde{S}}^3$ given in \eqref{eq:cS3:EERAD3}.  Hence we can readily conclude that $c_{\tilde{S}}^3$ has a visually noticeable impact on the quality of perturbative convergence across all the considered schemes.

To summarize, the three-loop soft matching coefficient $c_{\tilde{S}}^3$ is the only N$^3$LL$^\prime$ ingredient that is currently not known precisely. Despite being a higher-order coefficient, the spread of the two existing estimates for this coefficient has a noticeable impact on both the $\lbrace \alpha_s, \Omega_1 \rbrace$ extraction and the perturbative stabilty of the distributions in the R$^0$ scheme, whose renormalon cancellation terms explicitly depend on $c_{\tilde{S}}^3$. This issue also grows in importance in view of our observations in Sec.~\ref{sec:THEORY} about the accuracy of \texttt{EERAD3} calculations for small $\tau$ values, which the estimate in \eqref{eq:cS3:EERAD3} has relied upon.  A resolution could come from an independent numerical extraction, or, ideally, a complete analytic calculation of the three-loop thrust soft function, of which partial results have already been published \cite{Baranowski:2022khd,Chen:2020dpk}. These observations about the importance of a more precise $c_{\tilde{S}}^3$ determination represent the \emph{fourth core conclusion} of our work.

%%%%%%%%%%%%%%%%%%%%%%%%%%%%%%%%%%%%%%%%%%%%%%%%%%%%%%%%
\section{Conclusion}
\label{sec:CONCLUDE}

We have revisited extractions of the strong \mbox{coupling $\alpha_s$} from a global fit to thrust data motivated by the current tension between SCET-based extractions in~\cite{Abbate:2010xh,Hoang:2015hka} and the PDG world average. To this end, we have used N$^{3}$LL$^\prime$ resummed theory predictions that are matched to $\mathcal{O}(\alpha_s^2)$ fixed-order calculations, and we have implemented a gapped, renormalon-free, dijet shape function to account for non-perturbative (NP) corrections. In the tail region of the distributions, where the fits are performed, the dominant  NP effects then manifest as a shift of the distributions that is driven by the first moment of the shape function $\Omega_1$. The fits are therefore formulated as two-parameter extractions in the $\alpha_s-\Omega_1$ plane.

Whereas recent analyses with similar motivations  have focused on NP effects from three-jet configurations \mbox{\cite{Luisoni:2020efy,Caola:2021kzt,Caola:2022vea}} that are most relevant in the far-tail region of the distributions, our study addressed systematic effects in the dijet factorization theorem itself. Specifically, we examined the impact from changing between different renormalon cancellation schemes and profiled scale variations. Concerning the former, we observed that the specific implementation of a renormalon-free gap parameter that models the minimal energy of a hadronic final state leads to an effective $\tau$-dependent shift as shown in Fig.~\ref{fig:effectiveshift}. We then selected and studied alternative renormalon schemes, following the approach in \cite{Bachu:2020nqn}, that tame the growth of this shift for larger $\tau$ values, and which amounts to a $\mathcal{O}(10\%)$ variation of the NP corrections in the considered fit windows. As for scale variations, we implemented two different sets of profile functions that are very close to the ones used previously in~\cite{Abbate:2010xh} and~\cite{Hoang:2015hka}, respectively. While the choice of profile functions did not make a major difference in those extractions, we found that it can be magnified if varied together with the renormalon scheme. In total this defines six different combinations of renormalon-cancellation and profile-variation schemes.

As our predictions in the R$^0$ schemes, which are more sensitive to higher-order terms in the subtractions, turned out to be less stable, we focused on the remaining four schemes for the $\lbrace \alpha_s,\Omega_1\rbrace$ extractions. In order to lift the degeneracy between the fit parameters, we performed a global fit to thrust data with c.o.m.~energies $Q \in \lbrace 35, 207\rbrace$ GeV. Our results are shown in Fig.~\ref{fig:Results2} and Tab.~\ref{tab:BestFit}. In particular, we found that they are consistent with previous extractions when the same renormalon-cancellation and profile-scale choices are implemented, which can be viewed as an independent confirmation of the analyses in \cite{Abbate:2010xh,Hoang:2015hka}. The scheme dependence of our fit results in the $\alpha_s-\Omega_1$ plane, on the other hand, turned out to be unexpectedly large, possible hinting at additional systematic theory uncertainties. Whereas we do not advance any one particular scheme to be favored over others on theoretical grounds, we remark that some of our $\alpha_s$ extractions are more consistent with the PDG average. This provides significant motivation for further scrutinizing the observed scheme dependence of the dijet predictions in the future.

Since the scheme dependence is more pronounced in the far-tail region of our predictions, we also performed fits that concentrate more on a central dijet-like $\tau$ domain. While these fits turned out to be of higher quality (see Fig.~\ref{fig:Results3}), the pattern in the $\alpha_s-\Omega_1$ plane did not change significantly in this case. Finally, we explored the impact of the three-loop soft matching coefficient $c_{\tilde{S}}^3$ on both the $\lbrace \alpha_s,\Omega_1\rbrace$ extractions and the perturbative stability of the R$^0$ schemes, and we argued that an improved determination of this quantity would help to further improve the theoretical predictions.

As our primary goal consisted in bringing the scheme dependence of the SCET-based $\alpha_s$ extractions to attention, we did not include a number of sub-leading effects that will not change the main conclusions of our analysis, but which may nevertheless be included in future precision fits to thrust data. These include, first and foremost, the $\mathcal{O}(\alpha_s^3)$ remainder function $r_c^3$ and the aforementioned soft three-loop constant $c_{\tilde S}^3$, which we could not extract using \texttt{EERAD3} because of instabilities in the small $\tau$ region. Furthermore, one may account for QED effects or non-zero bottom and hadron masses that were included in the analysis of \cite{Abbate:2010xh,Hoang:2015hka}. Finally,  depending on the chosen fit window, it could be helpful to incorporate resummation effects at the Sudakov shoulder $\tau=1/3$ \cite{Bhattacharya:2022dtm}.  

Our findings also motivate renewed attention to sub-leading power contributions and the associated resummation of logarithmic effects. As discussed in Sec.~\ref{sec:PROFILES}, the main difference between the considered sets of profile functions is sourced by the non-singular scale $\mu_{ns}$, which was chosen to track the other dynamical scales in order to partially account for missing resummation effects that appear in the non-singular contribution. Hence a dedicated resummation analysis of next-to-leading power corrections could eliminate the need to correlate $\mu_{ns}$ to the other scales, and in this way possibly alleviate the systematic uncertainty due to scale-profile choices that our study indicates. Resummations at sub-leading power to the thrust distribution have been studied e.g.~in \cite{Moult:2018jjd,Moult:2019uhz,Beneke:2022obx}.

As a final point, we mention that, regardless of the rich physics discussed in this analysis, a definitive extraction of $\alpha_s$ from $e^+e^-$ event shapes will have to lift its degeneracy with the NP shift parameter $\Omega_1$. While we have used thrust data from different c.o.m energies $Q$ for this purpose in the current analysis, an additional promising method consists in extending the class of observables to $e^+e^-$ angularity distributions $\tau_a$ \cite{Berger:2003iw}. These observables generalize the thrust variable, and their dominant NP effect is controlled by the same parameter $\Omega_1$ (see \eqref{eq:leadingshift}) with a  coefficient 
\cite{Berger:2003pk,Berger:2004xf,Lee:2006nr,Becher:2013iya}
\begin{equation}
\label{eq:Ccoefficient}
c_{\tau_a} = \frac{2}{1-a}\,\,\,\,\,\text{with}\,\,\,\,\,a<1
\end{equation}
that depends on the angularity $a$ ($a=0$ corresponds to thrust). This allows for the possibility of performing $\lbrace \alpha_s,\Omega_1\rbrace$ extractions globally in \emph{both} $Q$ and the new theory parameter $a$, offering a dual probe in the decorrelation effort. Since angularities have been resummed and matched to N$^2$LL$^\prime + \mathcal{O}(\alpha_s^2)$ accuracy \cite{Bell:2018gce}, a precision fit to available data \cite{L3:2011kil} is well overdue.  

%%%%%%%%%%%%%%%%%%%%%%%%%%%%%%%%%%%%%%%%%%%%%%%%%%%%%%%%
\section*{Acknowledgements}

We thank Andre Hoang, Vicent Mateu and Iain Stewart for helpful comments and discussions, and in particular for pointing us to the different renormalon schemes introduced in \cite{Bachu:2020nqn}. We thank Thomas Gehrmann for help with \texttt{EERAD3}, and Boram Yoon for advice and assistance with our large numerical computations. The work of C.L., Y.M.~and B.Y.~was supported by the U.S.~Department of Energy, Office of Science, through the Office of Nuclear Physics, an Early Career Research Award (C.L.) and the LDRD Program at Los Alamos National Laboratory. The research of G.B.~was supported by the Deutsche Forschungsgemeinschaft (DFG, German Research Foundation) under grant 396021762 - TRR 257. J.T.~gratefully acknowledges funding from the European  Union’s Horizon 2020 research and innovation programme under the Marie Sk\l{}odowska-Curie grant agreement No. 101022203, and guest support from the T-2 group at LANL. This research used resources provided by the Los Alamos National Laboratory Institutional Computing Program. LANL is operated by Triad National Security, LLC, for the National Nuclear Security Administration of the U.S.~Department of Energy under Contract No.~89233218CNA000001. J.T. also thanks the HPC and HEP teams at the DAMTP for their assistance with local cluster computing. J.T. and C.L. are grateful to the organizers and participants of the INT workshops `New physics searches at the precision frontier' and `Probing QCD at High Energy and Density with Jets' for their hospitality and discussions during the completion of this work.

%%%%%%%%%%%%%%%%%%%%%%%%%%%%%%%%%%%%%%%%%%%%%%%%%%%%%%%%
\appendix

\begin{widetext}

%%%%%%%%%%%%%%%%%%%%%%%%%%%%%%%%%%%%%%%%%%%%%%%%%%%%%%%%%%
\section{Perturbative Expansions}
\label{sec:EXPANSIONS}

Here we expand the prediction \eqref{eq:cumulant} for the resummed $\tau$ cross section in fixed order to $\cO(\as^3)$, which we will use to extract the non-singular remainder functions from numerical computations. It is expedient to begin by computing the fixed-order expansion of the Laplace-transformed cross section,
\begin{equation}
\wt \sigma(\nu) = \frac{1}{\sigma_0}\int_0^\infty d\tau e^{\nu\tau} \frac{d\sigma}{d\tau}\,,
\end{equation}
which obeys the factorized form,
\begin{equation}
\label{eq:Laplacecs}
\wt\sigma(\nu) = H(L_H,\mu)\tilde J^2(\tilde L_J,\mu)\wt S(\tilde L_S,\mu)\,,
\end{equation}
where $L_H = \ln(\mu/Q)$, $L_J = \ln(\mu^j e^{\gamma_E}\nu/Q^j)$ (where $j = 2$), $L_S= \ln(\mu e^{\gamma_E}\nu/Q)$, and $\mu$ is generic factorization scale. 
Each function $H,\wt J,\wt S$ obeys a perturbative expansion predicted by their RG evolution,
\begin{equation}
\label{eq:Fexp}
F(L_F,\mu) = F(0,Q_F) e^{K_F(\mu,\mu_F)}\,,
\end{equation}
where $Q_H = Q, Q_J = Qe^{\gamma_E}/\nu^{1/j}$, and $Q_S=Qe^{\gamma_E}/\nu$, and $K_F$ is given by
\begin{equation}
K_F(\mu,Q_F) = -j_F \kappa_F K_\Gamma(\mu,Q_F) + K_{\gamma_F}(\mu,Q_F)\,,
\end{equation}
with $\kappa_H = 4,\kappa_J = -2$ and $\kappa_S = 4$.  $K_\Gamma,K_\gamma$ have the fixed-order expansions \cite{Almeida:2014uva}
\begin{align}
K_\Gamma(\mu,Q_F) &= \frac{\as(\mu)}{4\pi} \frac{\Gamma_0}{2} \ln^2\frac{\mu}{Q_F} \\
&\quad + \Bigl(\frac{\as(\mu)}{4\pi}\Bigr)^2 \Bigl( \frac{1}{3}\Gamma_0\beta_0\ln^3\frac{\mu}{Q_F} + \frac{\Gamma_1}{2}\ln^2\frac{\mu}{Q_F}\Bigr) \nn \\
&\quad + \Bigl(\frac{\as(\mu)}{4\pi}\Bigr)^3 \biggl[ \frac{1}{3}\Gamma_0\beta_0^2\ln^4\frac{\mu}{Q_F} + \frac{1}{3}(\Gamma_0\beta_1 + 2\Gamma_1\beta_0)\ln^3\frac{\mu}{Q_F} + \frac{\Gamma_2}{2}\ln^2\frac{\mu}{Q_F}\biggr] \nn \,,
\end{align}
and
\begin{align}
K_{\gamma_F}(\mu,Q_F) &= \frac{\as(\mu)}{4\pi} \gamma_F^0 \ln\frac{\mu}{Q_F} \\
&\quad + \Bigl(\frac{\as(\mu)}{4\pi}\Bigr)^2 \Bigl( \gamma_F^0\beta_0\ln^2\frac{\mu}{Q_F} + \gamma_F^1\ln\frac{\mu}{Q_F}\Bigr) \nn \\
&\quad + \Bigl(\frac{\as(\mu)}{4\pi}\Bigr)^3 \biggl[ \frac{4}{3}\gamma_F^0\beta_0^2\ln^3\frac{\mu}{Q_F} + \gamma_F^0\beta_1 + 2\gamma_F^1\beta_0)\ln^2\frac{\mu}{Q_F} + \gamma_F^2\ln\frac{\mu}{Q_F}\biggr] \nn \,,
\end{align}
where $\Gamma_n,\gamma_F^n$ are the coefficients in the perturbative expansions of the anomalous dimensions,
\begin{equation}
\Gamma_\text{cusp}[\as(\mu)] = \sum_{n=0}^\infty \Bigl(\frac{\as(\mu)}{4\pi}\Bigr)^{n+1} \Gamma_n \,,\quad \gamma_F[\as(\mu)] = \sum_{n=0}^\infty \Bigl(\frac{\as(\mu)}{4\pi}\Bigr)^{n+1} \gamma_F^n\,.
\end{equation}
In \eqref{eq:Fexp}, the functions $F$ evaluated at their ``natural'' scale $Q_F$ have no explicit logs in their expansions:
\begin{align}
\label{eq:Fcoeffs}
F(0,Q_F) = 1 + \sum_{n=1}^\infty \Bigl(\frac{\as(Q_F)}{4\pi}\Bigr)^n c_F^n \,.
\end{align}
Combining all the pieces in \eqref{eq:Laplacecs}, we can express the fixed-order Laplace-space cross section in the form
\begin{equation}
\label{eq:sigmaFOexp}
\wt\sigma(\nu) = \wt C(L_\nu,Q) e^{K(L_\nu,Q)}\,,
\end{equation}
with $L_\nu = \ln(e^{\gamma_E}\nu)$, and the `constant' in front has the form
\begin{equation}
\wt C(L,Q) = 1 + \frac{\as(Q)}{4\pi}C_{10} + \Bigl(\frac{\as(Q)}{4\pi}\Bigr)^2 (C_{20}+C_{21} L) + \Bigl(\frac{\as(Q)}{4\pi}\Bigr)^3 (C_{30} + C_{31}L + C_{32}L^2) + \cdots\,,
\end{equation}
where the extra log terms come from expanding out $\as(Q_{J,S})$ appearing in $F(Q_{J,S})$ in \eqref{eq:Fcoeffs} in powers of $\as(Q)$. These coefficients are given up to $\cO(\as^3)$ by
\begin{align}
\label{eq:Ccoeffs}
C_{10} &= c_H^1 + 2c_{\tilde J}^1 + c_{\tilde S}^1 \,,\\
C_{20}&= c_H^2 + 2c_{\tilde J}^2 + (c_{\tilde J}^1)^2 + c_{\tilde S}^2 + c_H^1 (2c_{\tilde J}^1 + c_{\tilde S}^1) + 2c_{\tilde J}^1 c_{\tilde S}^1 \,,\quad C_{21} = 2\beta_0\Bigl(\frac{2c_{\tilde J}^1}{j} + c_{\tilde S}^1\Bigr) \,,\nn \\
C_{30}&= c_H^3 + 2(c_{\tilde J}^3 + c_{\tilde J}^2c_{\tilde J}^1) + c_{\tilde S}^3 + c_H^2(2c_{\tilde J}^1 + c_{\tilde S}^1) + c_H^1 [2c_{\tilde J}^2 + (c_{\tilde J}^1)^2 + c_{\tilde S}^2 + 2c_{\tilde J}^1c_{\tilde S}^1] + 2c_{\tilde J}^1 c_{\tilde S}^2 + [2c_{\tilde J}^2+ (c_{\tilde J}^1)^2]c_{\tilde S}^1\,, \nn \\
C_{31} &= 2\beta_1\Bigl(\frac{2c_{\tilde J}^1}{j} + c_{\tilde S}^1\Bigr) + 4\beta_0\Bigl(\frac{2c_{\tilde J}^2+(c_{\tilde J}^1)^2}{j} + c_{\tilde S}^2\Bigr) + 2\beta_0 \Bigl[ \frac{2c_{\tilde J}^1}{j}  (c_H^1 + c_{\tilde S}^1) + (c_H^1 + 2 c_{\tilde J}^1) c_{\tilde S}^1\Bigr] \,,\quad C_{32} = 4\beta_0^2\Bigl(\frac{2c_{\tilde J}^1}{j^2} + c_{\tilde S}^1\Bigr) \,. \nn
\end{align}
Meanwhile, the expansion of the exponent $K$ in \eqref{eq:sigmaFOexp} is given by
\begin{align}
K(L,Q) &= \quad \Bigl(\frac{\as(Q)}{4\pi}\Bigr) \bigl( K_{12} L^2 + K_{11} L\bigr) \\ 
&\quad + \Bigl(\frac{\as(Q)}{4\pi}\Bigr)^2 \bigl( K_{23} L^3 + K_{22} L^2 + K_{21} L) \nn \\
&\quad + \Bigl(\frac{\as(Q)}{4\pi}\Bigr)^3 \bigl ( K_{34} L^4 + K_{33} L^3 + K_{32} L^2 + K_{31} L\bigr) + \cdots \nn\,,
\end{align}
with coefficients given up to $\cO(\as^3)$ by
\begin{align}
\label{eq:Kcoeffs}
K_{12}&= -\frac{\Gamma_0}{2}\Bigl(\frac{2\kappa_J}{j}+\kappa_S\Bigr) \,,\qquad K_{11} = \frac{2\gamma_J^0}{j} + \gamma_S^0\,, \\
K_{23}&= -\frac{1}{3}\Gamma_0 \beta_0\Bigl(\frac{2\kappa_J}{j^2}+\kappa_S\Bigr)\,, \quad K_{22} = -\frac{\Gamma_1}{2}\Bigl(\frac{2\kappa_J}{j}+\kappa_S\Bigr) + \Bigl(\frac{2\gamma_J^0}{j^2} + \gamma_S^0\Bigr)\beta_0\,,\quad K_{21} = \frac{2\gamma_J^1}{j} + \gamma_S^1\,,  \nn \\
K_{34}&= -\frac{1}{3}\Gamma_0\beta_0^2\Bigl(\frac{2\kappa_J}{j^3}+\kappa_S\Bigr)\,, \quad K_{33} = -\frac{1}{3}(\Gamma_0\beta_1 + 2\Gamma_1\beta_0)\Bigl(\frac{2\kappa_J}{j^2}+\kappa_S\Bigr) + \frac{4}{3}\beta_0^2 \Bigl(\frac{2\gamma_J^0}{j^3} + \gamma_S^0\Bigr) \,, \nn \\
K_{32} &= -\frac{\Gamma_2}{2}\Bigl(\frac{2\kappa_J}{j}+\kappa_S\Bigr) + \Bigl(\frac{2\gamma_J^0}{j^2}+\gamma_S^0\Bigr)\beta_1 + 2\beta_0\Bigl(\frac{2\gamma_J^1}{j^2}+\gamma_S^1\Bigr)\,,\quad K_{31} = \frac{2\gamma_J^2}{j} + \gamma_S^2 \,, \nn
\end{align}
recalling $\gamma_S^0 = 0$ (though we have kept it to make the patterns more obvious).

Plugging the expansions \eqref{eq:Ccoeffs}-\eqref{eq:Kcoeffs} into \eqref{eq:sigmaFOexp}, we obtain the fixed-order expansion of the Laplace-space cross section,
\begin{equation}
\label{eq:Laplaceexpansion}
\wt\sigma(\nu) = 1 + \sum_{n=1}^\infty \Bigl(\frac{\as(Q)}{4\pi}\Bigr)^n\sum_{k=0}^{2n} \wt\sigma_{nk} L_\nu^k \,,
\end{equation}
where up to $\cO(\as^3)$,
\begin{align}
\label{eq:Laplacecoeffs}
\wt{\sigma}_{12} &= K_{12}, \qquad \wt{\sigma}_{11} = K_{11}, \qquad   \wt{\sigma}_{10} = C_{10}, \\[10pt]
 \nonumber
 \wt{\sigma}_{24} &= \frac{K_{12}^2}{2} ,\,\,\,\,\, \wt{\sigma}_{23} = K_{11} K_{12} + K_{23} , \,\,\,\,\,  \wt{\sigma}_{22} = \frac{K_{11}^2}{2} + C_{10} K_{12} + K_{22}, \,\,\,\,\, \wt{\sigma}_{21} = C_{21} + C_{10} K_{11} + K_{21}, \,\,\,\,\, \wt{\sigma}_{20} = C_{20},\\[10pt]
\nonumber
\wt{\sigma}_{36} &= \frac{K_{12}^3}{6},\qquad \wt{\sigma}_{35} = K_{12} \left(K_{23} + \frac{K_{11}K_{12}}{2} \right),\qquad \wt{\sigma}_{34} = K_{12} \left(K_{22} + \frac{K_{11}^2}{2} + \frac{C_{10}K_{12}}{2}\right) + K_{11}K_{23}+K_{34} \,,\\
\wt{\sigma}_{33} &= C_{10}( K_{11}K_{12} + K_{23}) + C_{21}K_{12} + K_{12}K_{21} + K_{11}K_{22} + K_{33} + \frac{K_{11}^3}{6}, \nn \\
\wt{\sigma}_{32} &= C_{32} + C_{21} K_{11} + C_{20} K_{12} + K_{11} K_{21} +C_{10} \left(K_{22} + \frac{K_{11}^2}{2}\right) + K_{32} \,, \nn \\
\wt{\sigma}_{31} &= C_{31} + C_{20} K_{11} + C_{10} K_{21} + K_{31}, \qquad \wt{\sigma}_{30} = C_{30}\,. \nn
\end{align}
\\
It is also useful to know how to transform \eqref{eq:Laplaceexpansion} back to momentum space. A quick way to do this is from \eqref{eq:cumulant}, which implies that the fixed-order $\tau$ cross section is given simply by
\begin{equation}
\frac{1}{\sigma_0}{\sigma_c(\tau)} = \wt\sigma(\nu;L_\nu\to \partial_\Omega + L_\tau)\frac{e^{\gamma_E\Omega}}{\Gamma(1-\Omega)}\Bigr\rvert_{\Omega\to 0}\,,
\end{equation}
where each log $L_\nu$ in \eqref{eq:Laplaceexpansion} is replaced by the differential operator shown, with $L_\tau = \ln(1/\tau)$.  Then the cross section is expanded in the form,
\begin{equation}
\label{eq:momentumexpansion}
\frac{1}{\sigma_0}{\sigma_c(\tau)} = 1 + \sum_{n=1}^\infty \Bigl(\frac{\as(Q)}{4\pi}\Bigr)^n \sum_{k=0}^{2n}\sigma_{nk}L_\tau^k \,,
\end{equation}
where the coefficients $\sigma_{nk}$ are given in terms of the Laplace-space coefficients $\wt \sigma_{nk}$ in \eqref{eq:Laplacecoeffs} by
\begin{align}
\label{eq:finalmomentumspace}
\sigma_{12} &= \wt{\sigma}_{12}, \,\,\,\,\, \sigma_{11} = \wt{\sigma}_{11}, \,\,\,\,\, \sigma_{10} = \wt{\sigma}_{10} - \frac{\pi^2}{6} \wt{\sigma}_{12} , \\[1em]
\nonumber
\sigma_{24} &= \wt{\sigma}_{24}, \,\,\,\,\, \sigma_{23} = \wt{\sigma}_{23}, \,\,\,\,\,\, \sigma_{22} = \wt{\sigma}_{22} - \pi^2 \wt{\sigma}_{24} , \,\,\,\,\, \sigma_{21} = \wt{\sigma}_{21} - \frac{\pi^2}{2}\wt{\sigma}_{23}  - 8 \zeta_3 \wt{\sigma}_{24} , \,\,\,\,\, \sigma_{20} = \wt{\sigma}_{20} - \frac{\pi^2}{6} \wt{\sigma}_{22} - 2 \zeta_3\wt{\sigma}_{23}  + \frac{\pi^4}{60}\wt{\sigma}_{24}  \\[1em]
\nonumber
\sigma_{36} &= \wt{\sigma}_{36}, \,\,\,\,\, \sigma_{35} = \wt{\sigma}_{35}, \,\,\,\,\, \sigma_{34} = \wt{\sigma}_{34} - \frac{5 \pi^2}{2}\wt{\sigma}_{36} , \,\,\,\,\, \sigma_{33} = \wt{\sigma}_{33} - \frac{5 \pi^2}{3}\wt{\sigma}_{35}  -40 \zeta_3 \wt{\sigma}_{36} , \,\,\,\\
\nonumber
 \sigma_{32} &= \wt{\sigma}_{32} -\pi^2 \wt{\sigma}_{34}  - 20 \zeta_3\wt{\sigma}_{35}  + \frac{\pi^4}{4}\wt{\sigma}_{36} , \qquad \sigma_{31} = \wt{\sigma}_{31} - \frac{\pi^2}{2} \wt{\sigma}_{33} - 8 \zeta_3\wt{\sigma}_{34}  + \frac{\pi^4}{12}\wt{\sigma}_{35}  + \left(20 \pi^2 \zeta_3 - 144 \zeta_5 \right) \wt{\sigma}_{36} ,\\
\sigma_{30} &= \wt{\sigma}_{30} - \frac{\pi^2}{6}\wt{\sigma}_{32}  - 2 \zeta_3\wt{\sigma}_{33}  + \frac{\pi^4}{60}\wt{\sigma}_{34}  +\left(\frac{10 \pi^2 \zeta_3}{3} - 24 \zeta_5 \right) \wt{\sigma}_{35} + \left(40 \zeta_3^2 - \frac{5 \pi^6}{168}\right)\wt{\sigma}_{36} \,. \nn
\end{align}
In obtaining these expressions we made use of the following handy identities:
\begin{align}
\cG(\Omega) \equiv \frac{e^{\gamma_E \Omega}}{\Gamma(1-\Omega)}\,,\qquad \partial_\Omega \cG\rvert_{\Omega\to 0} &= 0 \,,\qquad \partial_\Omega^2 \cG\rvert_{\Omega\to 0} = -\frac{\pi^2}{6}\,,\qquad \partial_\Omega^3 \cG\rvert_{\Omega\to 0} = -2\zeta_3\,, \\
\partial_\Omega^4 \cG\rvert_{\Omega\to 0} &= \frac{\pi^4}{60}\,,\quad \partial_\Omega^5 \cG\rvert_{\Omega\to 0} = \frac{10\pi^2}{3}\zeta_3 - 24\zeta_5 \,,\quad \partial_\Omega^6 \cG\rvert_{\Omega\to 0} = 40\zeta_3^2 - \frac{5\pi^6}{168}\,. \nn
\end{align}

%%%%%%%%%%%%%%%%%%%%%%%%%%%%%%%%%%%%%%%%%%%%%%%%%%%%%%%%%%
\section{Renormalon Cancellation Formulae}
\label{sec:INGREDIENTS}

In this appendix we list all of the renormalon-cancellation formulae required for our N$^3$LL$^\prime + \mathcal{O}(\alpha_s^2)$ thrust calculation.  
Beginning at the level of the unexpanded, renormalon-corrected cross section in \eqref{eq:Xsecexpand}-\eqref{eq:modelexpansion}, and further defining
 \begin{equation}
 \label{eq:fmodexpansion}
 f_\text{mod}^{(i)}(k - 2 \Delta) \equiv \left(\frac{\alpha_s(\mu_S)}{4\pi} \right)^i \bar{f}_\text{mod}^{(i)}(k - 2 \Delta) \,,
 \end{equation}
 the components of \eqref{eq:modelexpansion} can be written explicitly as
\begin{subequations}
\label{eq:fmodn}
\begin{align}
\bar{f}_\text{mod}^{(0)}(k - 2 \Delta) &= f_\text{mod}(k - 2 \Delta)\,,\\
\bar{f}_\text{mod}^{(1)}(k - 2 \Delta) &= -
2\,\delta^1 \left(\frac{\mu_R}{\xi}\right) f'_\text{mod}(k - 2 \Delta) \,,\\
\bar{f}_\text{mod}^{(2)}(k - 2 \Delta) &=   \Bigl[ - 2\,  \delta^2\left(\frac{\mu_R}{\xi}\right) f'_\text{mod}(k - 2 \Delta) + 2\left(\delta^1\,\frac{\mu_R}{\xi}\right)^2 \, f''_\text{mod}(k - 2 \Delta)\Bigr]\,, \\
\bar{f}_\text{mod}^{(3)}(k - 2 \Delta) &=  \left[-2 \,\delta^3 \left(\frac{\mu_R}{\xi}\right) f^\prime_\text{mod}(k - 2 \Delta) + 4 \,\delta^1\,  \delta^2\, \left(\frac{\mu_R}{\xi}\right)^2 f^{\prime\prime}_\text{mod}(k - 2 \Delta)-\frac{4}{3}\left(\delta^1\,\frac{\mu_R}{\xi}\right)^3 f^{\prime\prime\prime}_\text{mod}(k - 2 \Delta)  \right]\,,
\end{align}
\end{subequations}
where the subtraction terms $\delta^n$ and $\Delta$ have an implicit functional dependence on the soft and renormalon scales, $\lbrace \delta^n, \Delta \rbrace \equiv \lbrace \delta^n(\mu_S,\mu_R), \Delta(\mu_S,\mu_R)\rbrace$.  Indeed, the form of $\Delta(\mu_\delta,\mu_R)$ for arbitrary reference $\mu_\delta$ and subtraction $\mu_R$ scales (i.e. accounting for RGE in both scales, according to \eqref{eq:RAnomDim}) is known to the required three-loop order, and is given by
\begin{align}
\nonumber
    \Delta(\mu_\delta,\mu_R) &= \Delta(R_\Delta,R_\Delta) + 2 \left(\frac{\mu_R}{\xi} \right) \eta_\Gamma (\mu_\delta,\mu_R) \\
\nonumber
    &+\frac{R_\Delta}{2\beta_0} e^{-G\left[\alpha_s(R_\Delta)\right]} \left( \frac{2\pi}{\beta_0} e^{i\pi} \right)^{\frac{\beta_1}{2\beta_0^2}} \bigg\{ \gamma^0_R \left[\Gamma \left(-\frac{\beta_1}{2\beta_0^2},-\frac{2\pi}{\beta_0 \alpha_s(\mu_R)} \right) -\Gamma \left(-\frac{\beta_1}{2\beta_0^2},-\frac{2\pi}{\beta_0 \alpha_s(R_\Delta)} \right) \right] \\
   \nn
    &- \frac{1}{2\beta_0} \left[ \gamma_R^1 - \frac{\gamma_R^0}{\beta_0} \left( \beta_1 + \frac{B_2}{2} \right)\right]\left[\Gamma \left( -\frac{\beta_1}{2\beta_0^2}-1,-\frac{2\pi}{\beta_0 \alpha_s(\mu_R)}\right) - \Gamma \left(-\frac{\beta_1}{2\beta_0^2}-1,-\frac{2\pi}{\beta_0 \alpha_s(R_\Delta)} \right) \right] \\
   \nn
    &+ \frac{1}{4 \beta_0^2} \left[\gamma_R^2 - \frac{\gamma_R^1}{\beta_0} \left(\beta_1 + \frac{B_2}{2} \right) + \gamma_R^0 \left(B_2 + \frac{B_2 \beta_1}{2\beta_0^2} - \frac{B_3}{4\beta_0} + \frac{B_2^2}{8\beta_0^2} \right) \right] \\
      \label{eq:RGapDeltaevolve}
    &\times \left[\Gamma \left(-\frac{\beta_1}{2\beta_0^2}-2, -\frac{2\pi}{\beta_0 \alpha_s(\mu_R)} \right) - \Gamma \left(-\frac{\beta_1}{2\beta_0^2}-2, - \frac{2\pi}{\beta_0 \alpha_s(R_\Delta)} \right) \right]
    \bigg\},
\end{align}
where we take the input gap parameter to be $\Delta(R_\Delta,R_\Delta) = 0.1$ GeV at an arbitrary reference scale $R_\Delta = 1.5$ GeV (the exact value for this reference scale is not expected to be particularly consequential in the tail regions relevant to our fits \cite{Abbate:2010xh}.) and where $G$ is the anti-derivative of $1/\beta \left[ \alpha \right]$,
\begin{equation}
G \left[\alpha \right] = \frac{2\pi}{\beta_0} \left[ \frac{1}{\alpha} + \frac{\beta_1}{4\pi \beta_0} \ln \alpha - \frac{B_2}{(4\pi)^2} \alpha + \frac{B_3}{(4\pi)^3} \frac{\alpha^2}{2} \right]\,.
\end{equation}
Here $B_{2,3}$ are given as
\begin{align}
    B_2 &\equiv -\frac{\beta_2}{\beta_0} + \frac{\beta_1^2}{\beta_0^2} \,, \,\,\,\,\,\,\,\,\,\,
    B_3 \equiv - \frac{\beta_3}{\beta_0} + 2\frac{\beta_1 \beta_2}{\beta_0^2} - \frac{\beta_1^3}{\beta_0^3}\,,
\end{align}
in terms of the QCD $\beta$-function coefficients defined through
\begin{equation}
\label{eq:QCDbeta}
\mu \frac{d \alpha_s(\mu)}{d\mu} = -2 \beta_0 \frac{\alpha_s(\mu)^2}{4\pi} \left[1 + \left(\frac{\alpha_s(\mu)}{4\pi} \right) \frac{\beta_1}{\beta_0} + \left(\frac{\alpha_s(\mu)}{4\pi}\right)^2 \frac{\beta_2}{\beta_0} + \left(\frac{\alpha_s(\mu)}{4\pi}\right)^3 \frac{\beta_3}{\beta_0} \right] \,.
\end{equation}
The solution to \eqref{eq:QCDbeta} for the running coupling up to four loops is
\begin{align}
\frac{1}{\as(\mu)} &= \frac{1}{\as(m_Z)}\biggl\{ X + \frac{\as(m_Z)}{4\pi} \frac{\beta_1}{\beta_0}\ln X  + \biggl(\frac{\as(m_Z)}{4\pi}\biggr)^2 \biggl[\frac{\beta_2}{\beta_0} \Bigl(1-\frac{1}{X}\Bigr) + \frac{\beta_1^2}{\beta_0^2} \Bigl(\frac{\ln X}{X} + \frac{1}{X} - 1\Bigr) \biggr] \\ \nn
&\qquad\qquad + \biggl(\frac{\as(m_Z)}{4\pi}\biggr)^3 \frac{1}{X^2} \biggl[\frac{\beta_3}{2\beta_0} (X^2-1) + \frac{\beta_1\beta_2}{\beta_0^2}(X+\ln X - X^2) +\frac{\beta_1^3}{2\beta_0^3}[(1-X)^2-\ln^2 X]\biggr]\biggr\}\,,
\end{align}
where 
\begin{equation}
    X \equiv 1+ \frac{\as(m_Z)}{2\pi}\beta_0 \ln\frac{\mu}{m_Z}\,.
\end{equation}
The actual functional dependence of the subtraction terms $\delta^n$ in \eqref{eq:fmodn} on the reference and subtraction scales is scheme-dependent, as is the structure of the $\mu_R$-anomalous dimensions appearing in \eqref{eq:RGapDeltaevolve}.  We now present the ingredients $\delta^i$ and $\gamma_R^i$ in both the $n=0$ and $n=1$ schemes considered in the main text.
%%%%%%%%%%%%%%%%%%%%%
\subsection{$n=0$ Schemes}
\label{sec:n0}
From \eqref{eq:subtractdefine} and the fixed-order expansion of the soft function that can be constructed from \eqref{eq:Fexp}, one reads off that
\begin{equation}
\label{eq:deltaseries}
    \delta(\mu_\delta,\mu_R) = \frac{\mu_R}{2\xi} \ln \tilde{S}\Bigl(v=\frac{\xi}{\mu_R},\mu_\delta\Bigr)\equiv \frac{\mu_R}{2\xi} \sum_{n=1} \left(\frac{\alpha_s(\mu_\delta)}{4\pi}\right)^n \,\delta^n(\mu_\delta,\mu_R)
\end{equation}
for a generic $\xi$ scheme and reference scale $\mu_\delta$.  Given the known three-loop expansion of the Laplace-space soft function $\tilde{S}$, one can then easily obtain the order-by-order expressions for $\delta$ as follows:
\begin{subequations}
\label{eq:deltan0}
\begin{align}
\delta^1(\mu_\delta,\mu_R) &= \Gamma_s^0 \, L^2 + c_{\tilde{S}}^1 \,,\\
\delta^2(\mu_\delta,\mu_R) &=\frac{2}{3} \Gamma_s^0 \beta_0 \,L^3 + \Gamma_s^1 L^2 + \left(\gamma_s^1 + 2 c_{\tilde{S}}^1 \beta_0 \right)L + c_{\tilde{S}}^2 - \frac{1}{2} (c_{\tilde{S}}^1)^2 \,,\\
\delta^3(\mu_\delta,\mu_R) &= \frac{2}{3} \Gamma_s^0 \beta_0^2 \,L^4 + \frac{2}{3}\left(2\Gamma_s^1 \beta_0 + \Gamma_s^0 \beta_1 \right) \,L^3 + \left( \Gamma_s^2 +2 \gamma_s^1 \beta_0 + 4 c_{\tilde{S}}^1 \beta_0^2  \right)\,L^2 \\
\nonumber
&\quad + \left(\gamma_s^2 + 2 c_{\tilde{S}}^1 \beta_1 + 4 c_{\tilde{S}}^2 \beta_0 - 2 (c_{\tilde{S}}^1)^2 \beta_0  \right)\,L -c_{\tilde{S}}^1 c_{\tilde{S}}^2 + \frac{1}{3}(c_{\tilde{S}}^1)^3 + c_{\tilde{S}}^3 \,,
\end{align}
\end{subequations}
with $L = \ln \frac{\mu_\delta e^{\gamma_E} \xi}{\mu_R}$.  Then, from \eqref{eq:RAnomDim} and \eqref{eq:QCDbeta},
one immediately obtains the relevant $\mu_R$-anomalous dimensions in this scheme:
\begin{subequations}
\begin{align}
\gamma_R^0 &= \frac{1}{2\xi} \left(\Gamma_s^0 \,L^2 + c_{\tilde{S}}^1 \right)\,, \\
\gamma_R^1 &= \frac{1}{2\xi} \left(\frac{2}{3}\Gamma_s^0 \beta_0 \,L^3+\Gamma_s^1 \,L^2 + \left(\gamma_s^1 + 2 c_{\tilde{S}}^1 \beta_0\right)L +c_{\tilde{S}}^2 -\frac{1}{2} (c_{\tilde{S}}^1)^2  -2 \beta_0 \left( \Gamma_s^0 \, L^2 + c_{\tilde{S}}^1 \right)\right)\,,\\
\nonumber
2 \xi \, \gamma_R^2 &= \frac{2}{3}\Gamma_s^0 \beta_0^2 \,L^4 + \frac{2}{3}\left(2\Gamma_s^1 \beta_0 + \Gamma_s^0 \beta_1\right)L^3 +\left(\Gamma_s^2 + 2\gamma_s^1\beta_0 +4 c_{\tilde{S}}^1 \beta_0^2\right)L^2 +\left(\gamma_s^2 + 2 c_{\tilde{S}}^1 \beta_1 + 4 c_{\tilde{S}}^2 \beta_0 -2(c_{\tilde{S}}^1)^2 \beta_0\right)L \\
\nonumber
&- c_{\tilde{S}}^2 c_{\tilde{S}}^1 +\frac{1}{3}(c_{\tilde{S}}^1)^3 +c_{\tilde{S}}^3 \\
&- 2 \left( \frac{4}{3} \Gamma_s^0 \beta_0^2 \,L^3 + \left(\Gamma_s^0 \beta_1 + 2 \Gamma_s^1  \beta_0 \right)\,L^2 +  \beta_0 \left(4 c_{\tilde{S}}^1 \beta_0 +2 \gamma_s^1\right) \,L + 2 c_{\tilde{S}}^2  \beta_0 + c_{\tilde{S}}^1 \beta_1 - (c_{\tilde{S}}^1)^2 \beta_0\right)\,,
\end{align}    
\end{subequations}
where now $L = \ln \xi e^{\gamma_E}$ ($\mu_\delta = \mu_R$ in this calculation, see \eqref{eq:RAnomDim}). This log of course vanishes in schemes where $\xi=e^{-\gamma_E}$.

In practice, the subtraction series \eqref{eq:deltaseries} appears in \eqref{eq:ultimate} as part of a renormalon-subtracted soft function that is expanded in powers of $\as(\mu_S)$, not $\as(\mu_\delta)$, when $\mu_\delta$ and $\mu_S$ are not chosen to be equal (which is a choice we are free to make). In this case, we need to know the coefficients of the powers of $\as(\mu_S)$ when expanding \eqref{eq:deltaseries} in $\alpha_s(\mu_S)$:
\begin{equation}
\label{eq:deltaexpmuS}
    \delta(\mu_\delta,\mu_R) \equiv \frac{\mu_R}{2\xi}\sum_{n=1}^\infty\Bigl(\frac{\as(\mu_S)}{4\pi}\Bigr)^n \delta^n(\mu_S;\mu_\delta,\mu_R)\,,
\end{equation}
where, up to order $\alpha_s^3$,
\begin{subequations}
\label{eq:deltanmuS}
\begin{align}
    \delta^1(\mu_S;\mu_\delta,\mu_R) &= \delta^1(\mu_\delta,\mu_R)\,, \\
    \delta^2(\mu_S;\mu_\delta,\mu_R) &= \delta^2(\mu_\delta,\mu_R) + 2\beta_0 \ln\frac{\mu_S}{\mu_\delta} \, \delta^1(\mu_\delta,\mu_R)\,, \\
    \delta^3(\mu_S;\mu_\delta,\mu_R) &= \delta^3(\mu_\delta,\mu_R) + 2\beta_1\ln\frac{\mu_S}{\mu_\delta}\,\delta^1(\mu_\delta,\mu_R) + 4\beta_0^2 \ln^2\frac{\mu_S}{\mu_\delta}\, \delta^1(\mu_\delta,\mu_R) + 4\beta_0 \ln\frac{\mu_S}{\mu_\delta} \, \delta^2(\mu_\delta,\mu_R)\,,
\end{align}
\end{subequations}
having somewhat abused our notation by using the same symbols for $\delta^n$ on both sides---the number of arguments distinguishes them. The expansion \eqref{eq:deltaexpmuS} with coefficients \eqref{eq:deltanmuS} are the expressions we use at each order of the expansion of the renormalon-free shape function in \eqref{eq:modelexpansion}.
%%%%%%%%%%%%%%%%%%%%%%%%%%%%%%%%%%%%%%%%%%%%%%%%%%%%%%%%
\subsection{$n=1$ Schemes}
\label{sec:n1}
Setting $n=1$ in \eqref{eq:subtractdefine}, one then defines the subtraction terms as
\begin{equation}
\label{eq:deltaexpn1}
\delta(\mu_\delta,\mu_R) = \frac{\mu_R}{2\xi} \frac{d}{d \ln v} \ln \tilde{S}(v,\mu_\delta) \big \vert_{v = \xi/\mu_R} \equiv \frac{\mu_R}{2\xi} \sum_{n=1} \left(\frac{\alpha_s(\mu_\delta)}{4\pi}\right)^n \,\delta^n(\mu_\delta,\mu_R) \,.
\end{equation}
Upon solving for $\delta^i(\mu_\delta)$ and again running up to the soft scale $\mu_S$ using \eqref{eq:QCDbeta}, the generalized expressions for $\delta^i$ are finally found to be:
\begin{subequations}
\label{eq:deltan1}
\begin{align}
\delta^1(\mu_\delta,\mu_R) &= 2 \Gamma_s^0 \, L \,,
\label{eq:deltan1LO}\\
\delta^2(\mu_\delta,\mu_R) &= 2 \Gamma_s^0 \beta_0 \,L^2 +  2 \Gamma_s^1  \,L + \gamma_s^1 + 2 c_{\tilde{S}}^1 \beta_0\,, \\
\delta^3(\mu_\delta,\mu_R) &= \frac{8}{3} \Gamma_s^0 \, \beta_0^2\,L^3 + 2\left(2\Gamma_s^1 \beta_0 + \Gamma_s^0 \beta_1 \right) L^2 + 2\left(\Gamma_s^2 + 2\gamma_s^1 \beta_0 +4 c_{\tilde{S}}^1 \beta_0^2  \right) \,L \\
&\quad + \gamma_s^2 + 2 c_{\tilde{S}}^1 \beta_1 + 4 c_{\tilde{S}}^2 \beta_0 -2(c_{\tilde{S}}^1)^2 \beta_0 \,, \nn
\end{align}
\end{subequations}
with $L = \ln \frac{\mu_\delta e^{\gamma_E} \xi}{\mu_R}$.  We can again use these expressions to derive the $\mu_R$-anomalous dimensions $\gamma_R^n$, finding
\begin{subequations}
\begin{align}
\gamma_R^0 &= \frac{1}{2\xi} \left( 2 \Gamma_s^0 \, L \right)\,,\\
\gamma_R^1 &= \frac{1}{2\xi} \left( 2 \Gamma_s^0 \beta_0 \,L^2 + 2 (\Gamma_s^1 - 2 \Gamma_s^0 \beta_0)  \,L + \gamma_s^1 + 2 c_{\tilde{S}}^1 \beta_0 \right)\,,\\
\nonumber
2 \xi \,\gamma_R^2&=  \frac{8}{3} \Gamma_s^0 \, \beta_0^2\,L^3 + 2\left(2\Gamma_s^1 \beta_0 + \Gamma_s^0 \beta_1 - 4 \Gamma_s^0 \beta_0^2 \right) L^2 + 2\left(\Gamma_s^2 + 2\gamma_s^1 \beta_0 +4 c_{\tilde{S}}^1 \beta_0^2 -4 \Gamma_s^1 \beta_0 -2 \Gamma_s^0 \beta_1 \right) \,L \\
&\quad + \gamma_s^2 + 2 c_{\tilde{S}}^1 \beta_1 + 4 c_{\tilde{S}}^2 \beta_0 -2(c_{\tilde{S}}^1)^2 \beta_0 -4 \gamma_s^1 \beta_0 - 8 c_{\tilde{S}}^1 \beta_0^2 \,,
\label{eq:gammaRn1}
\end{align}
\end{subequations}
for $L = \ln (e^{\gamma_E} \xi)$ (again, $\mu_\delta=\mu_R$ here, cf. \eqref{eq:RAnomDim}.).

As above for the $n=0$ scheme, when the series \eqref{eq:deltaexpn1} is expanded in powers of $\as(\mu_S)$ rather than $\as(\mu_\delta)$, the expansion takes the same form as \eqref{eq:deltaexpmuS} with coefficients given by \eqref{eq:deltanmuS}, this time with the $n=1$ scheme subtraction terms \eqref{eq:deltan1}.

\end{widetext}

%------------------------------------------------
\begin{figure*}
\centering
\begin{tabular}{cc}
\includegraphics[width=0.95\columnwidth]{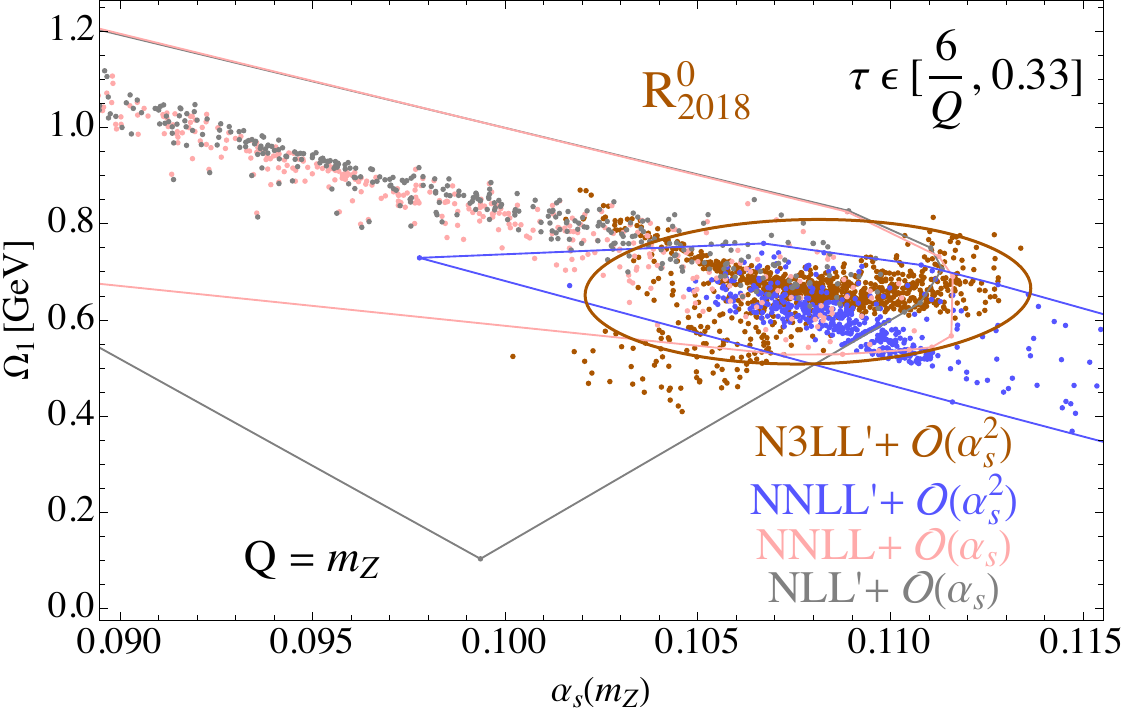} \quad
\includegraphics[width=.9555\columnwidth]{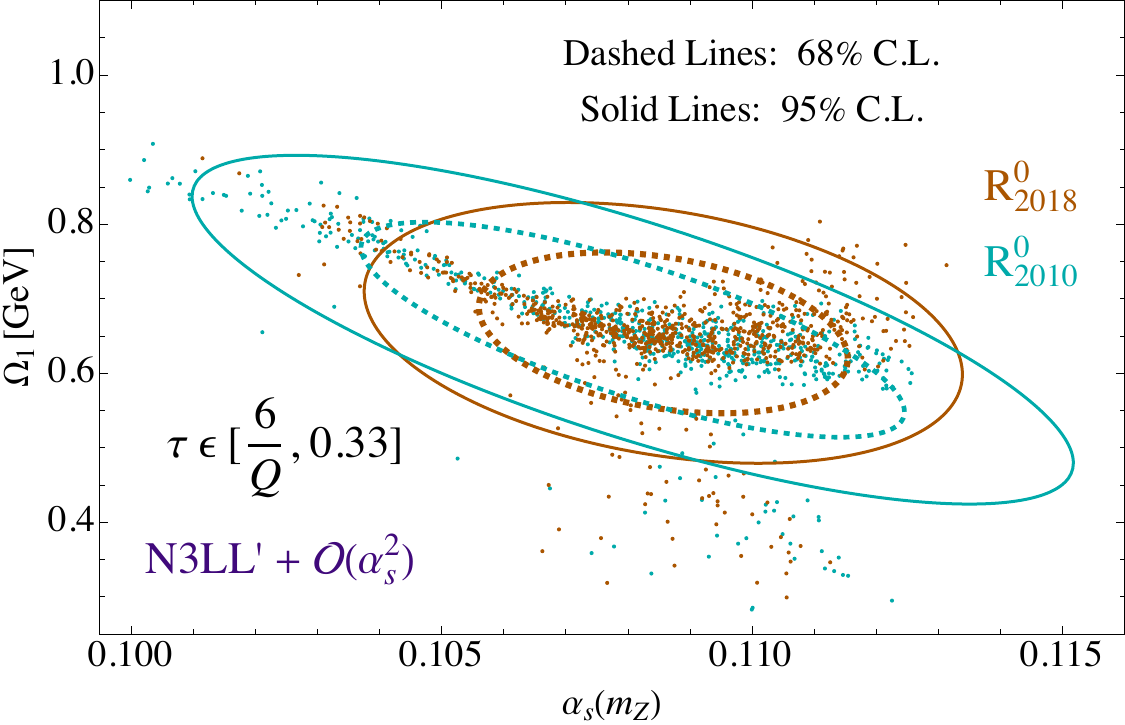}\\
\includegraphics[width=.9575\columnwidth]{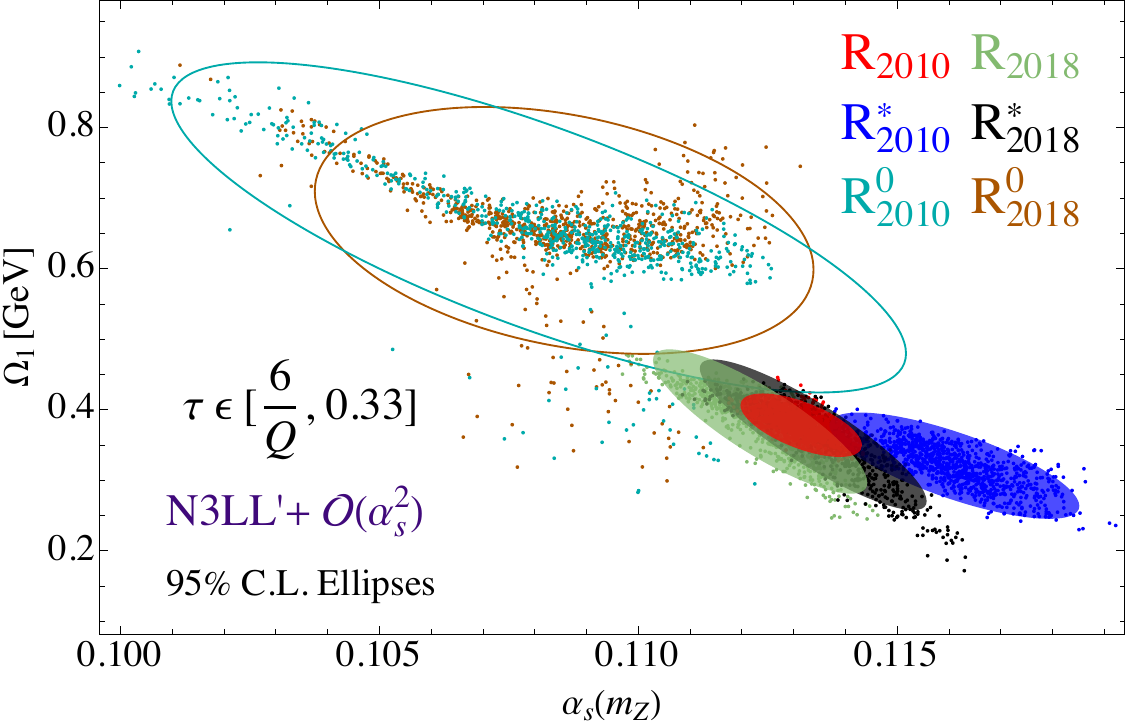} \quad
\includegraphics[width=.95\columnwidth]{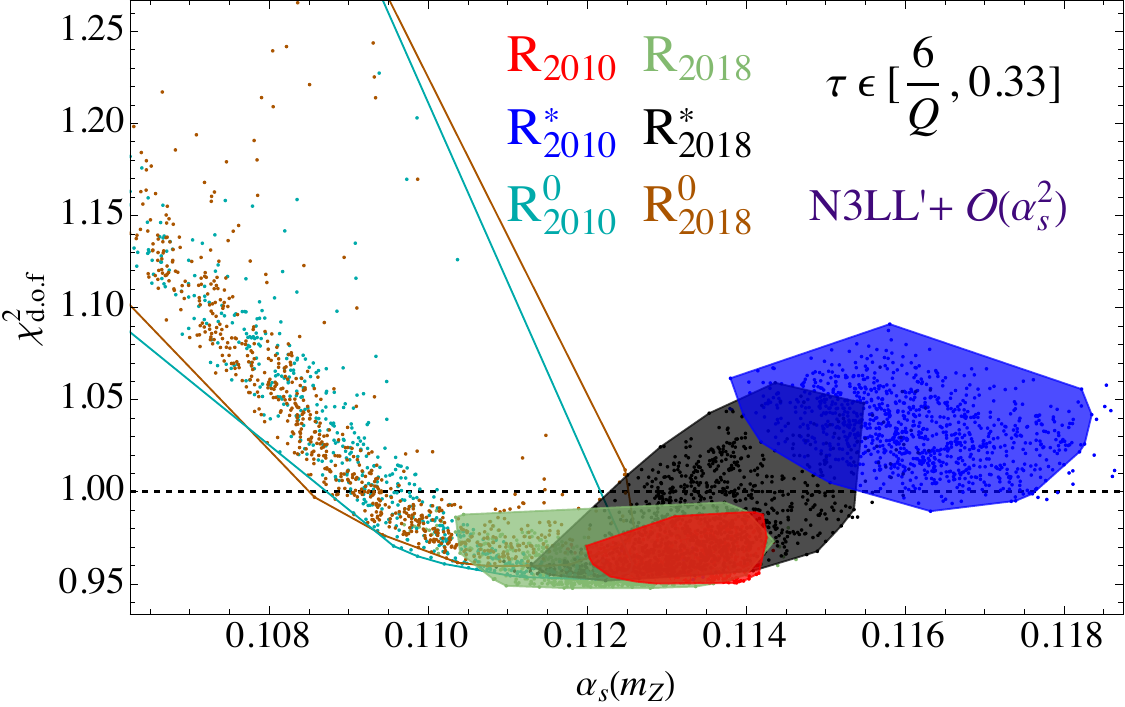}
\end{tabular}
\vspace{-1em}
\caption{Extraction results for the R$^0$ schemes, in analogue to those from Figs. \ref{fig:mZconverge}-\ref{fig:Results3}.  The {\bf{top-left}} panel represents the convergence of $\lbrace \alpha_s, \Omega_1 \rbrace$ fits at various logarithmic accuracies, and with $Q=m_Z$ data.  The {\bf{top-right}} panel represents our global $\lbrace \alpha_s, \Omega_1 \rbrace$ extractions in these schemes.  The {\bf{bottom panels}} present the R$^0$ fits relative to R$^{(\star)}$ results.  See the text for further discussion.}
\vspace{-1em}
\label{fig:R0Results1}
\end{figure*}
%------------------------------------------------
%---------------------------------------------------------------------------
\begin{table}
\centering
{\renewcommand{\arraystretch}{2.}
\begin{tabular}{|c|c|c|}
\hline
\textbf{Profiles} & \textbf{Parameters} & \boldmath{$R^0$} (Default) \\
\hline \hline
\multirow{3}{*}{2018 Profiles} & $\lbrace \alpha_s, \Omega_1 \rbrace$ &  $\lbrace 0.1086, 0.654 \rbrace$ \\
\cline{2-3} & $\lbrace \sigma_\alpha, \sigma_\Omega \rbrace$ & $\lbrace 0.0048, 0.175 \rbrace$ \\
\cline{2-3} & $\rho_{\alpha\Omega}$ & $-0.318$ \\
\hline
\multirow{3}{*}{2010 Profiles} & $\lbrace \alpha_s, \Omega_1 \rbrace$ & $\lbrace 0.1081, 0.658 \rbrace$ \\
\cline{2-3} & $\lbrace \sigma_\alpha, \sigma_\Omega \rbrace$ & $\lbrace 0.0071, 0.234 \rbrace$ \\
\cline{2-3} & $\rho_{\alpha\Omega}$ & $-0.764$ \\
\hline
\end{tabular}}
\caption{The same as Fig. \ref{tab:BestFit}, but for the $R^0$ cancellation schemes. These numbers correspond to central values of the 95\% C.L. ellipses presented in the top-right panel of Fig. \ref{fig:R0Results1}, and associated $K_{theory}$ parameters as defined in Section \ref{sec:METHOD}.}
\label{tab:R0BestFit}
\end{table}
%-----------------
%------------------------------------------------
\begin{figure}
\centering
\vspace{-1.2em}
\includegraphics[width=.9575\columnwidth]{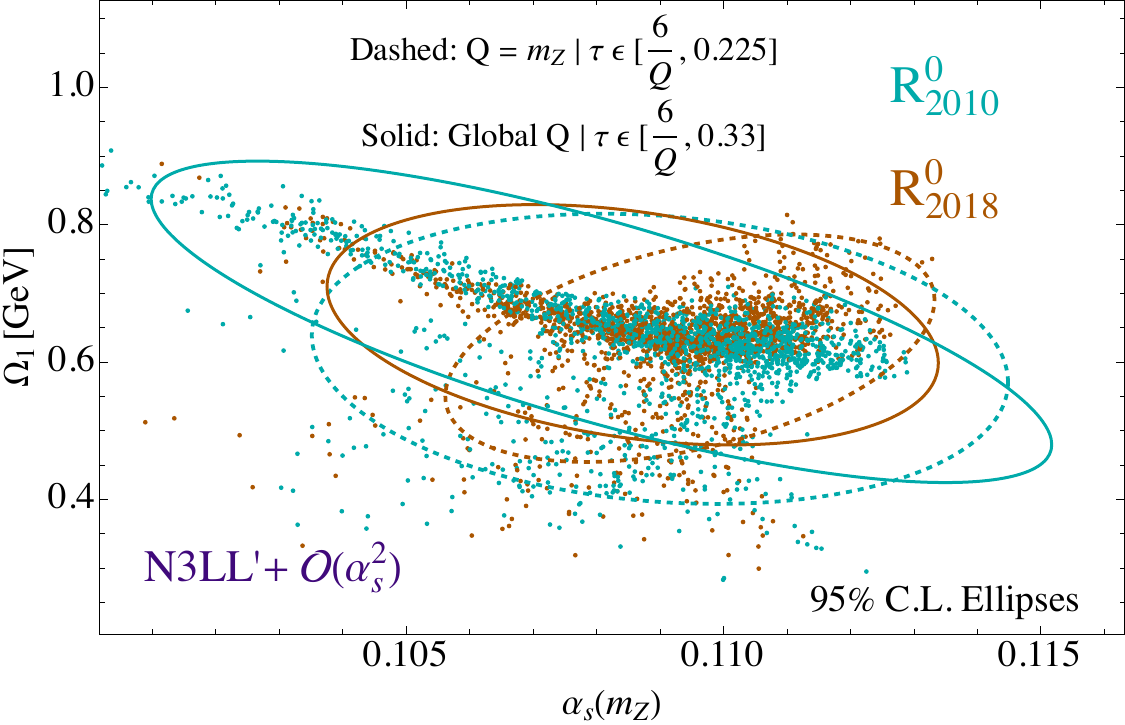}\\
\includegraphics[width=.95\columnwidth]{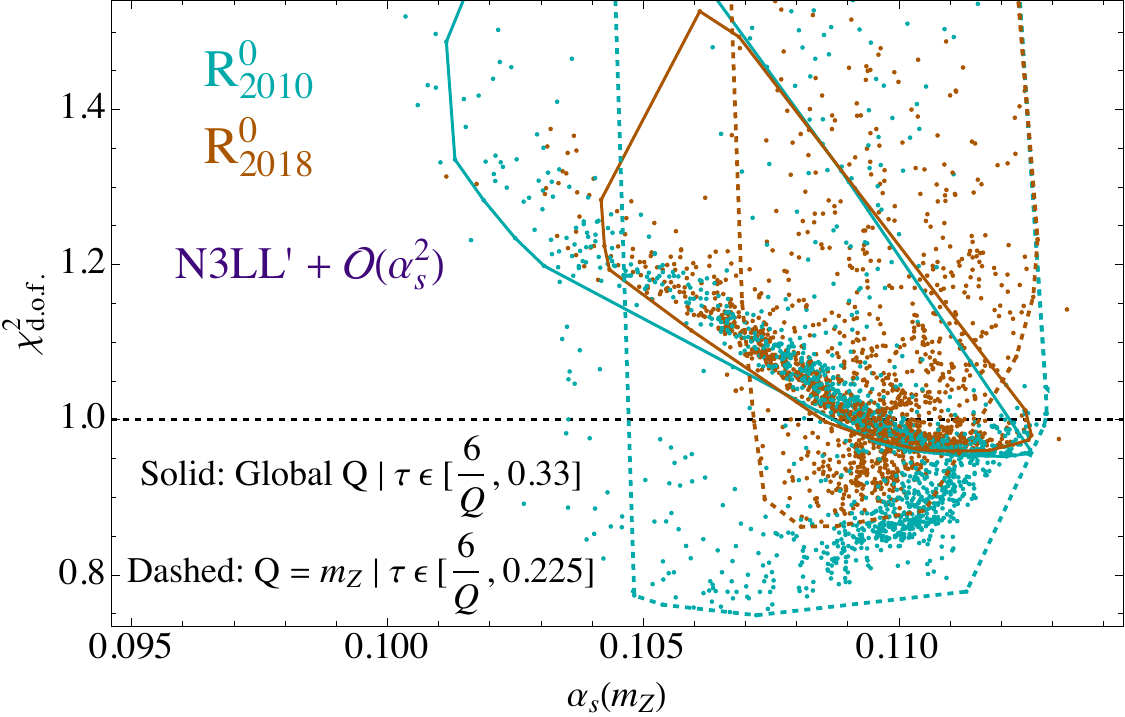}
\vspace{-1em}
\caption{$\lbrace \alpha_s, \Omega_1 \rbrace$ ({\bf{Top}}) and $\lbrace \alpha_s, \chi^2_{d.o.f} \rbrace$ ({\bf{Bottom}}) fits for the $R^0$ schemes, comparing results from global $Q$ data and default $\tau$ windows (solid lines) to those with $Q=m_Z$ data restricted to a central $\tau$ region (dashed lines).}
\vspace{-1em}
\label{fig:R0GlobalComp}
\end{figure}
%------------------------------------------------

%%%%%%%%%%%%%%%%%%%%%%%%%%%%%%%%%%%%%%%%%%%%%%
\section{Further Analysis of R$^0$ Schemes}
\label{sec:MORER0}

In this appendix we collect additional results from our study of the R$^0$ scheme defined in \eqref{eq:R0}, which as mentioned in the main text suffers from perturbative instabilities in comparison to the R$^{(\star)}$ schemes.

In Tab. \ref{tab:R0BestFit} we first collect the numerical values for the centers of the $95\%$ C.L. ellipses, their variances, and their correlation coefficients (in analogue to Tab. \ref{tab:BestFit}) obtained in our default R$^0_{2010,2018}$ fits, while in Fig. \ref{fig:R0Results1} we collect numerous plots that also serve as analogues to the R$^{(\star)}$ scheme studies in the main text.  In the top left panel we show the convergence of the R$^0_{2018}$ scheme for $Q=m_Z$ data only, where one observes a slightly erratic behavior, with final N$^{3}$LL$^\prime$ $+ \mathcal{O}(\alpha_s^2)$ results settling between the extremely low (high) $\alpha_s$ extractions found for accuracies at or below (above) N$^2$LL $+ \mathcal{O}(\alpha_s)$.\footnote{We have checked that N$^{3}$LL $+ \mathcal{O}(\alpha_s^2)$ results mimic those from the light blue N$^2$LL$^\prime$ $+ \mathcal{O}(\alpha_s^2)$ scans.} This is a manifestation of the instability presented in Fig. \ref{fig:R0convergence}, and represents a noticeable departure from the behavior shown in Fig. \ref{fig:mZconverge} for R$^{(\star)}$ schemes.  Then in the top right panel of Fig. \ref{fig:R0Results1} we show the fully global results for $\lbrace \alpha_s, \Omega_1 \rbrace$ extractions in R$^0_{2010,2018}$ schemes at $68\%$ and $95\%$ C.L., in analogue to Fig. \ref{fig:Results1} for the R$^{(*)}$ schemes.  There one sees a rather large spread of $\lbrace \alpha_s, \Omega_1 \rbrace$ values, especially for the R$^0_{2010}$ scheme.  Indeed, relative to the R$^{(\star)}$ scheme ellipses from Fig. \ref{fig:Results2}, which we also plot in the bottom left panel of Fig. \ref{fig:R0Results1}, the overall size of the R$^0$ ellipses is dramatically increased.  Furthermore, while schemes within \emph{either} the $n=1$ and $n=0$ renormalon cancellation classes are largely consistent with one another, there is only minimal overlap \emph{between} the $n=1$ and $n=0$ schemes.  Also, with respect to extracted values of $\alpha_s$ ($\Omega_1$), the qualitative impact of changing to the R$^0$ scheme is  to shift the error ellipse centers towards smaller (larger) values, regardless of the profile functions utilized. Finally, we also plot the $\alpha_s - \chi^2_{\rm{dof}}$ data from our global $Q$ extractions in the bottom-right panel of Fig. \ref{fig:R0Results1}.  There we see that a smaller proportion of $n=0$ extractions fall below the $\chi^2_{\rm{dof}}=1$ contour than do their $n=1$ counterparts, and that a number of $n=0$ fits (especially those towards extremely low $\as$) have particularly large $\chi^2_{\rm{dof}}$, and cannot yet be considered good fits. 

We examine the R$^0$ schemes closer in Fig. \ref{fig:R0GlobalComp}, where in the top (bottom) panel we have plotted 95\% C.L. regions in the $\alpha_s-\Omega_1$ ($\alpha_s-\chi^2_{\rm{dof}}$) plane obtained using global $Q$ datasets over the default fit window (solid lines) and those obtained using $Q=m_Z$ data over a reduced fit window with $6/Q \le \tau \le 0.225$ (dashed lines).  There we observe another odd feature:  including more data in the global-$Q$ scan actually \emph{widens} the $\alpha_s - \Omega_1$ ellipses and simultaneously \emph{reduces} their quality, trends which are counter-intuitive and contrary to those observed for $n=1$ schemes in Fig. \ref{fig:Results3} above.  This may indicate that R$^0$ theory comparisons to data at $Q\neq m_Z$ may not exhibit the same stability as those seen in Fig. \ref{fig:R0convergence}. 

Indeed, from the results of Figs. \ref{fig:R0Results1}-\ref{fig:R0GlobalComp} and those regarding convergence mentioned in Sec.~\ref{sec:OTHER}, it seems clear that the stability of the R$^0$ scheme is currently tenuous, and may hence be a too aggressive subtraction scheme for general $e^+e^-$ event-shape predictions. This may possibly be resolved upon a concrete, analytic determination of $c_{\tilde{S}}^3$ and a more refined set of profile variations tuned for global $Q$ values.  On the other hand, $n=0$ schemes are not uninteresting; indeed \cite{Bachu:2020nqn} finds them to be superior for top-quark-initiated event shapes, and a significant fraction of the $Q=m_Z$ fits in Fig. \ref{fig:R0GlobalComp} exhibit very good quality in terms of their $\chi^2_{\rm{dof}}$.  

%%%%%%%%%%%%%%%%%%%%%%%%%%%%%%%%%%%%%%%%%%%%%%%%%%%%%%%%
\bibliography{Alpha_Thrust}
%%%%%%%%%%%%%%%%%%%%%%%%%%%%%%%%%%%%%%%%%%%%%%%%%%%%%%%%
\end{document}